\documentclass[pre,aps]{revtex4-2}
\usepackage{amsfonts}
\usepackage{times}
\usepackage{amsmath}
\usepackage{graphicx}
\usepackage{amssymb}
\usepackage{xcolor,comment}
\usepackage{footnote}

\usepackage{placeins}

\newcommand{\figtwowidth}{0.39\textwidth}
\newcommand{\figthreewidth}{0.315\textwidth}
\newcommand{\figfourwidth}{0.235\textwidth}
\newcommand{\figonewidth}{0.58\textwidth}
\newcommand{\figpairwidth}{0.34\textwidth}

\newcommand{\e}{\mathrm{e}}

\newcommand{\ii}{\mathrm{i}}

\newcommand{\br}{\mathbf{r}}

\newcommand{\dkc}[1]{}

\begin{document}

\title{Solitary waves and vortices in a Nonlinear Schr\"odinger equation with ponderomotive nonlinearity}

\author{D.K. Campbell}
\affiliation{Department of Physics, Boston University, 590 Commonwealth Ave., Boston, MA 02215, USA}
\author{J. Cuevas-Maraver}
\affiliation{Grupo de F\'{\i}sica No Lineal, Departamento de F\'{\i}sica Aplicada I,
Universidad de Sevilla. Escuela Polit\'{e}cnica Superior, C/ Virgen de \'{A}%
frica, 7, 41011-Sevilla, Spain}
\affiliation{Instituto de Matem\'{a}ticas de la Universidad de Sevilla (IMUS). Edificio
Celestino Mutis. Avda. Reina Mercedes s/n, 41012-Sevilla, Spain}
\author{R. Goh}
\affiliation{Department of Mathematics \& Statistics, Boston University,  665 Commonwealth Ave., Boston, MA 02215, USA}
\author{Panayotis G.\ Kevrekidis}
\affiliation{Department of Mathematics and Statistics, University
of Massachusetts, Amherst, Massachusetts 01003-4515, USA}
\affiliation{Department of Physics, University
of Massachusetts, Amherst, Massachusetts 01003-4515, USA}

\begin{abstract}
In the present work we revisit a ponderomotive nonlinearity model
used to examine self-trapped laser beams in plasma. Upon briefly 
considering the exact stationary 1D solutions of the model,
we extend considerations to two spatial dimensions where we find
both solitonic and vortical structures. The solitary
waves localized in both directions are found to be spectrally stable.
However, all other structures that we consider in this model, including
line solitons ---which are homogeneous 2D extensions of 1D solitons---and
vortices of topological charge $S=1$ and $S=2$ are found to be 
spectrally unstable. The focal point of our studies then turns
to the examination of the collisions of the stable two-dimensional
solitary waves for which we map a two-parameter space of soliton 
speeds and frequencies, in terms of the potential outcomes.
While the standard scenarios of merger, inelastic collision
leading to separation, separation that leaves behind a localized
pulse are all possible, the intriguing outcome that we highlight
here is that of a longitudinal collision yielding a transverse
spliting of the solitons, either with or without a localized pulse remnant. 
\end{abstract}

\maketitle

\section{Introduction} 

The nonlinear Schr{\"o}dinger (NLS) equation~\cite{sulem,fibich} is
a prototypical dispersive nonlinear wave equation that
models topics as diverse as nonlinear optical wave
patterns of relevance to fibers and waveguides~\cite{Kivshar2003},
Bose-Einstein condensates of ultracold atomic gases~\cite{siambook},
deep water waves~\cite{ablowitz2}, and plasmas~\cite{kono} among many others. 
While the solitary waves
of the NLS and its variants have been studied very widely
in one spatial dimension, these equations have been far less studied in
two spatial dimensions. In the context of the focusing
nonlinearities which we consider in the present work,
this is not only because of the additional complexities of
higher dimensional dispersive nonlinear systems but also
because for the widely relevant
cubic nonlinearity (stemming from the Kerr effect in 
optics~\cite{Kivshar2003}, the s-wave interatomic scattering
interaction in BECs~\cite{Pitaevskii2003} etc.) the equations suffer from catastrophic
wave collapse, which makes their study both technically harder 
and also less practically and experimentally accessible.

Furthermore, in two spatial dimensions, genuinely integrable nonlinear wave models are rare~\cite{ablowitz2}, and stable solitary waves typically arise in \emph{non-integrable} systems.
Such (2+1)-dimensional solitary waves are fundamentally different from their one-dimensional integrable counterparts: instead of perfectly elastic interactions, collisions can excite internal modes, emit radiation, trigger symmetry breaking, generate bound states, or lead to fragmentation and fusion~\cite{MantonSutcliffe2004_TopologicalSolitons}. As a result, collision dynamics in non-integrable (2+1)D systems provide a sensitive probe of the internal structure, topology, and effective degrees of freedom of solitary waves, which is in stark 
contrast to the integrable structure and typically phase-shift-inducing dynamics of the integrable systems.
An example of particular recent interest in this direction
arises in nonlinear optics through the cubic-quintic NLS
model. Here the presence of 
a defocusing quintic nonlinearity counteracts collapse and enables stable two-dimensional localized states, including broad flat-top or ``droplet-like'' solitons. A very recent example of relevant
inelastic collisions within this model (in the case of 
guiding channel potentials) is given in~\cite{ZengMalomedMihalacheLiZhu2026Chaos_CQTroughs}.
The study of vortex collisions in the (single component)
cubic-quintic NLS has
been a topic of interest also in earlier work, e.g., 
~\cite{CaplanCarreteroGonzalezKevrekidisMalomed2012_MatComSim}.
In yet another very recent cubic-quintic NLS study (indeed
one involving multiple NLS components),
the instability of vortical structures led to collisions between
vortex wave states leading to the consideration of
a "solitary wave billiard"~\cite{Zezyulin2025_arXiv2512_05763}.
This has also been a topic of interest in saturable NLS models~\cite{hongkun}. Such settings have been used to 
mathematically emulate the collision of photorefractive
solitary waves that have been also monitored experimentally,
e.g., in the works of~\cite{wieslaw,Meng:97} which observed
numerous of the above scenarios including the fusion,
birth and energy exchange of solitary waves.

Beyond optics, ultracold quantum gases provide another realization of stable non-integrable solitary structures in the form of self-bound quantum droplets, stabilized by beyond-mean-field effects. Experimental observations of three-dimensional droplet collisions reveal a crossover between merging and separation regimes governed by compressibility and effective surface tension \cite{FerioliSemeghiniMasiEtAl2019_PRL}. Two-dimensional theoretical studies further classify collision outcomes into merging, separation, and evaporation or fragmentation phases, and identify post-collision collective oscillations as intrinsic dynamical signatures \cite{HuFeiChenZhang2024_arXiv2404_19295}. 
Finally, field-theoretic models provide a topologically stabilized counterpart to optical and BEC solitary waves. In the (2+1)-dimensional, so-called, ``baby Skyrme'' model~\cite{MantonSutcliffe2004_TopologicalSolitons}, localized solitons carry internal orientation and topological charge, and their interactions depend sensitively on relative phase and internal alignment. Numerical and analytical studies of baby Skyrmion scattering demonstrate orientation-dependent forces, radiation emission, and internal deformation during collisions, illustrating how topology ensures persistence of coherent structures even as non-integrable dynamics enable rich interaction scenarios \cite{PietteSchroersZakrzewski1995_NPB}.

In the present work, we revisit an apparently long
forgotten ponderomotive nonlinearity model of 
self-trapped laser beams in plasmas, examined initially
in the work of~\cite{tappert77}; see also for example ~\cite{max76, anderson79,cohen91}. Importantly, as the original
work had shown, the exponential form of the focusing nonlinearity,
decaying rapidly as the intensity increases, preserves
the structural ability of the dispersion-nonlinearity interplay
to produce stable solitary waves. 
This is in contrast to the catastrophic
collapse of the cubic NLS~\cite{sulem,fibich}). We confirm these results by identifying
these stable waves numerically (through a fixed point iteration)
in both $1$D and $2$D settings. In addition, we explore
further  structures including vortices of different
topological charge, as well as transversely homogeneous extensions
of 1D solitary waves in the form of 2D 
line solitons. For all of these additional excitations
(line and vortical solitary waves), we illustrate
their spectral transverse instability. Centering our computational
analysis around the collision of the stable solitonic building
blocks of the model that are non-topological and localized
in both directions, we retrieve all of the prototypical
phenomenologies of a non-integrable model: mergers (essentially
plastic collision) and weakly inelastic or even nearly elastic
collisions (for sufficiently large speeds). However, we also observe
some intriguing new possibilities, involving the
excitation of localized wavepackets {\it transverse} to the
direction of initial propagation. This can happen concurrently 
with the emergence of a localized wavepacket at the original
collision location, or even without such a localized pulse
remnant. Indeed, historically, based on the early computational
work of F. Tappert~\footnote{Per a private communication of F. Tappert to one of us (DKC)}, 
it was the potential of
such transverse splinters that was considered one of the most
appealing features of this model. This type of ``$\pi/2$ scattering''
is quite unusual in solitonic models (although it can happen
when suitable symmetry exists; see for a prototypical
example, e.g., the work of~\cite{RUBACK1988669}; see also ~\cite{foster15} for this behavior in Skyrmions).

Our presentation is structured as a follows. In Section II, we
present the mathematical setup of the model and
identify the prototypical coherent structures and their
stability. In section III we dive more deeply into the context
of solitary wave collisions and map out the two-dimensional
parameter space of the soliton speed ($c$) vs. the soliton
oscillation frequency ($\omega$). In so doing we uncover
all the different regimes, including ones releasing 
transverse splinters. Finally, in section IV we summarize
our findings and present our conclusions, as well as 
discussing a number of directions for future study.

\section{Model and Mathematical Setup}

In our study, we consider a NLS equation with a ponderomotive nonlinearity in the form
\begin{equation}\label{eq:NLS}
\ii\partial_t\psi=-\frac{1}{2}\nabla^2\psi-g(1-\e^{-\beta|\psi|^2})\psi
\end{equation}

Without loss of generality we choose $g=\beta=1$, as
these factors can be absorbed in a rescaling of independent
and dependent variables. We seek stationary solutions $\phi(\br)$ in the form
\begin{equation}
\psi(\mathbf{r},t)=\e^{\ii\omega t}\phi(\br)
\end{equation}

Here $\phi(\br)$ will represent the stationary wavefunction for both 1D and 2D solitons. In the former case, $\phi(\br)=\phi(x)$ In the Among the 2D setting, we distinguish three cases: namely 1) line solitons, 2) bright radial solitons and 3) bright vortex solitons. Line solitons are described by $\phi(\br)\equiv\phi(x,y)=\phi(x)$, whereas radial and vortex solitons' wavefunction can be written as $\phi(\br)=\phi(r)\exp(\ii S\theta)\,\, S\in \mathbb{N}$, with $S=0$ corresponding to the radial soliton without vorticity (i.e., the non-topological case). In those cases, the stationary form of 
Eq.~(\ref{eq:NLS}) can be written as:
\begin{equation}\label{eq:NLS_1D}
-\omega\phi+\frac{1}{2}\partial_{xx}\phi+g(1-\e^{-\beta\phi^2})\phi=0
\end{equation}
for 1D and line solitons, and
\begin{equation}\label{eq:NLS_2D}
-\omega\phi+\frac{1}{2}\left[\frac{1}{r}\partial_{r}\left(r\partial_r\phi\right)-\frac{S^2}{r^2}\phi\right]+g(1-\e^{-\beta\phi^2})\phi=0
\end{equation}
for 2D radial and vortex solitons.

\begin{figure}[!htbp]
\begin{tabular}{cc}
\includegraphics[width=\figtwowidth]{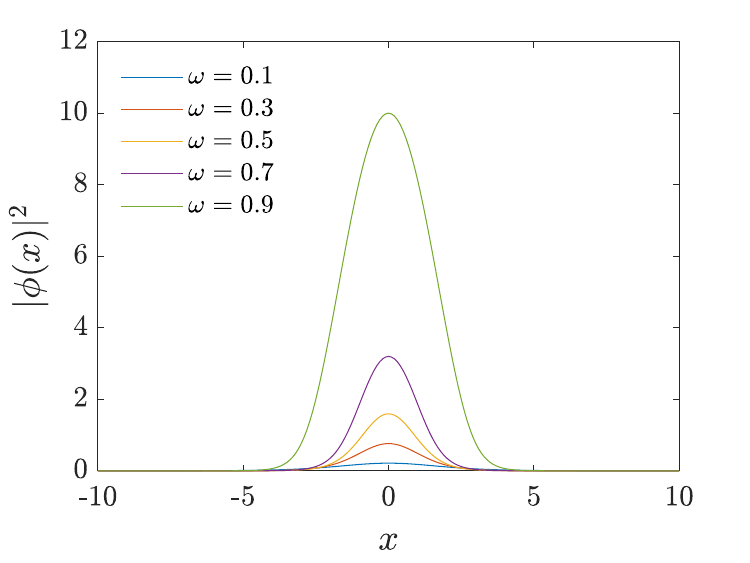} &
\includegraphics[width=\figtwowidth]{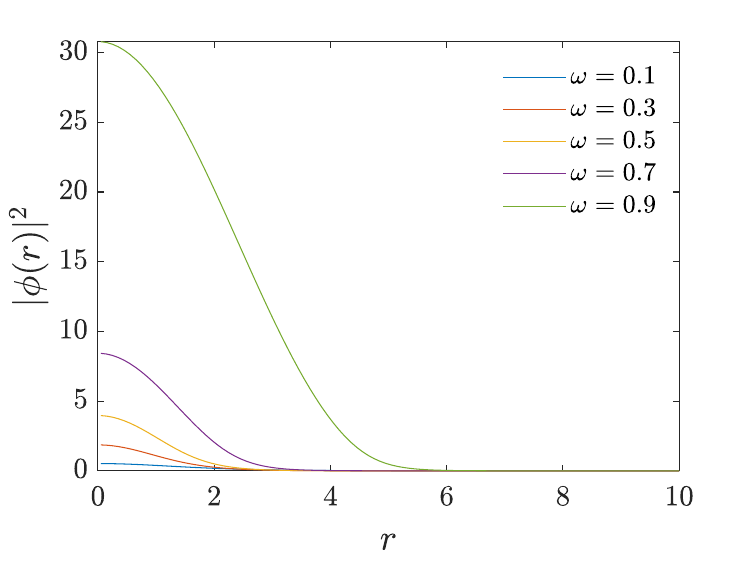} \\
\includegraphics[width=\figtwowidth]{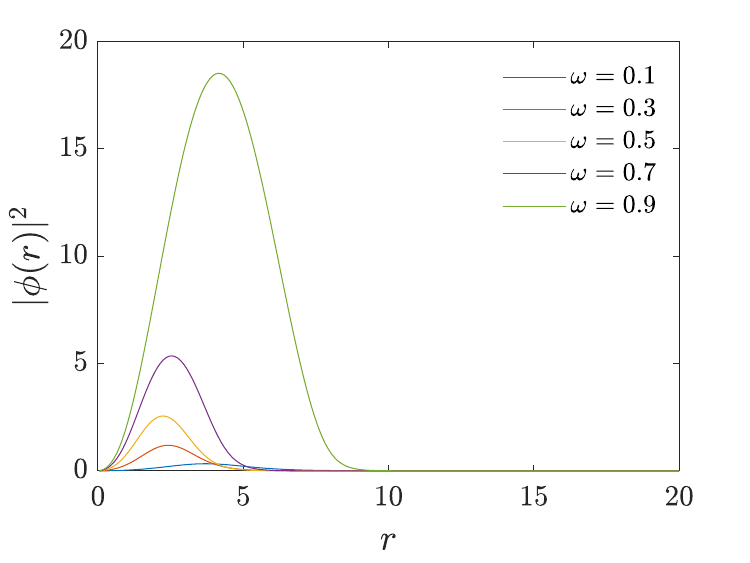} &
\includegraphics[width=\figtwowidth]{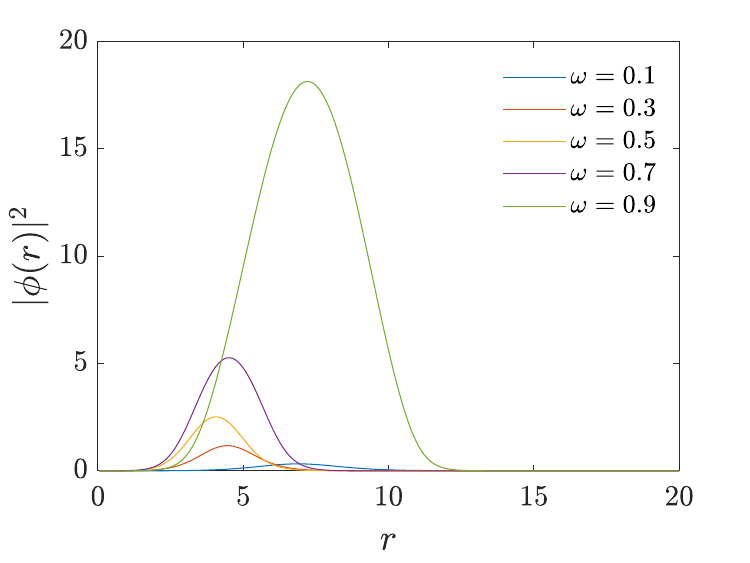} \\
\end{tabular}%
\caption{Profiles, for some selected frequencies, for 1D solitons (top left), radial 2D solitons (top right) and vortex solitons with $S=1$ (bottom left) and $S=2$ (bottom right).}
\label{fig:profiles}
\end{figure}

{In order to find stationary soliton solutions to these equations, we make use of Newton-Raphson method upon choosing a suitable seed. Finite difference methods are used to discretize the spatial derivative (in our case, with a discretization spacing $h=0.1$). For the 1D case (\ref{eq:NLS_1D}), periodic boundary conditions are implemented with the lattice nodes taken at $x_n=nh$ (with $n=-N/2\ldots N/2$), whereas for the 2D case (\ref{eq:NLS_2D}) free end boundary conditions are used, with the nodes taken at $r_n=(1/2+n)h$ (with $n=0\ldots N$); this choice is made in order to avoid the singularities of the Laplacian in polar coordinates at $r=0$. In both cases, we have taken $N=1000$, so the domain is $[-50,50]$ and $[0,100]$ in, respectively, 1D and 2D settings.}

Instability for those cases can be studied by means of introducing a perturbation $\xi(\mathbf{r},t)$ to the wavefunction $\psi(\br,t)$ in (\ref{eq:NLS}). In the co-rotating frame, one can write
\begin{equation}
\psi(\br,t)=\e^{\ii\omega t}\left[\phi(\br)+\xi(\br,t)\right].
\end{equation}

When studying the transverse (in)stability of line solitons, the perturbation is expressed as
\begin{equation}
\xi(\br,t)=a(x)\e^{\lambda t}\e^{\ii k_y y}+b^*(x)\e^{\lambda^* t}\e^{-\ii k_y y}
\end{equation}
recovering the spectral stability for the 1D soliton case when $k_y=0$, yet
extending considerations to transverse perturbations by decomposing the $y$-independent eigenvalue
problem into transverse Fourier modes. On the other hand, in order to study the azimuthal stability of radial and vortex solitons, one has to write~{\cite{Kollar}}:
\begin{equation}
{\xi(\br,t)=a(r)\e^{\lambda t}\e^{\ii (k_\theta+S)\theta}+b^*(r)\e^{\lambda^* t}\e^{-\ii (k_\theta-S)\theta}}
\end{equation}
and, similarly to the 1D case, the spectral stability for radial solitons is obtained when $k_\theta=0$.
I.e., here the azimuthal portion of the perturbation (for such waveforms) is decomposed into
Fourier modes.

Then, the spectral problem can be written in the form:

\begin{equation}
\lambda \left(\begin{array}{c} a(z)\\  b(z)\end{array}\right)=-\ii
\left(\begin{array}{cc}L_0^++L_1 & L_2 \\-L_2^* & -(L_0^-+L_1)^* \end{array}\right)
\left(\begin{array}{c}a(z)\\b(z)\end{array}\right)
\end{equation}
with $z\equiv x$ in the quasi-1D case, and $z\equiv r$ in the radial 2D case. Block operators $L$ are of the form
\begin{align*}
{L_0^+=L_0^-} &= -\frac{1}{2}\left(\partial_{xx}-k_y^2\right) && \textrm{(for the quasi-1D case of line
solitons)} \\
{L_0^{\pm}} &= {-\frac{1}{2}\left[\frac{1}{r}\partial_{r}\left(r\partial_r\phi\right)-\frac{(S\pm k_\theta)^2}{r^2}\phi\right]} && \textrm{(for the radial 2D case)} \\
L_1 &= \omega-g\left[1-(1-\beta |\phi|^2)\e^{-\beta|\phi|^2}\right]  & \\
L_2 &= -g\beta\phi^2\e^{-\beta|\phi|^2}. & 
\end{align*}

\begin{figure}[!tbp]
\begin{tabular}{cc}
\includegraphics[width=\figtwowidth]{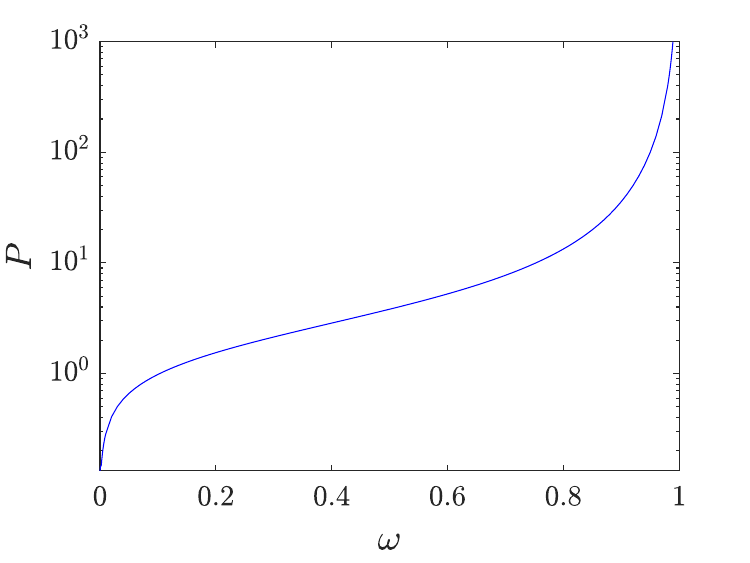} &
\includegraphics[width=\figtwowidth]{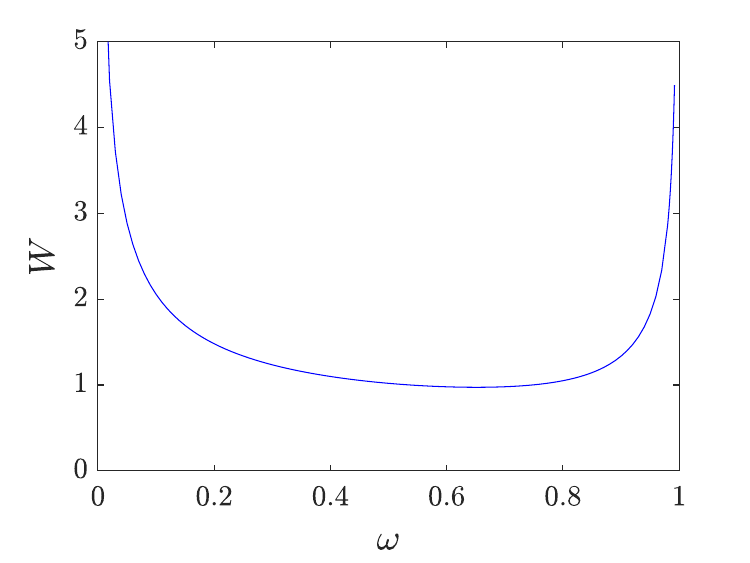} \\
\includegraphics[width=\figtwowidth]{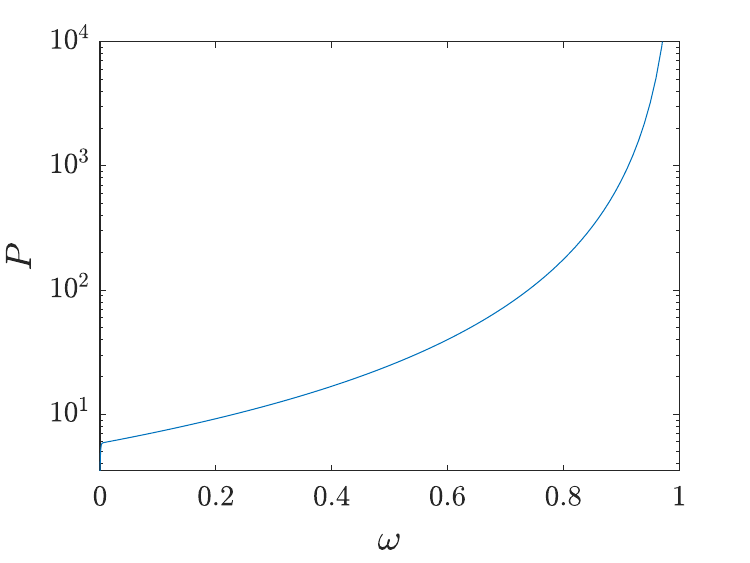} &
\includegraphics[width=\figtwowidth]{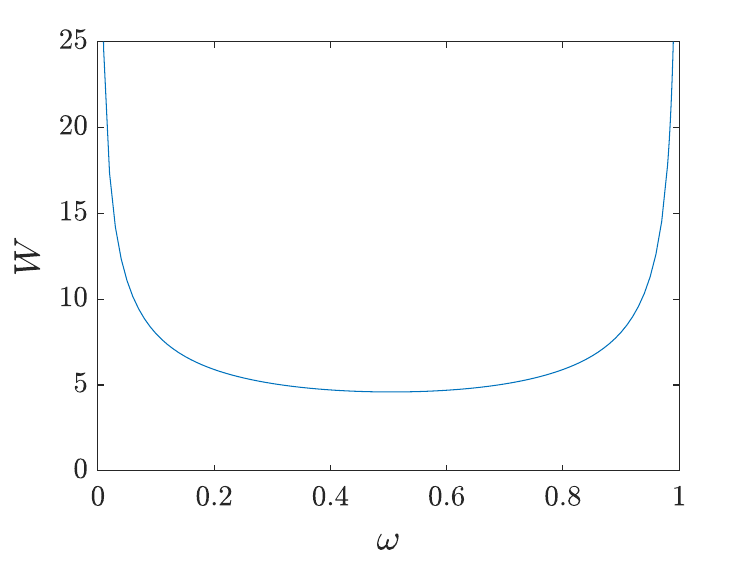} \\
\end{tabular}%
\caption{Dependence of the norm (left panels) and width (right panels) with the frequency for 1D (top panels) and 2D radial (bottom panels) solitons}
\label{fig:norm}
\end{figure}

In Fig.~\ref{fig:profiles} we show some selected profiles for all the various solutions. We can observe that the maxima of the vortex solitons are displaced from $r=0$, although this location has a non-monotonic behaviour. This is natural to expect as in these waves, the single-valuedness
of the wavefunction enforces that $\phi(0)=0$.
We also depict, in Fig.~\ref{fig:norm} the norm and the width for the 1D and 2D radial solitons. These magnitudes are defined as

\begin{equation}
P=\int_\Omega |\phi(\br)|^2\mathrm{d}\br,\qquad W=\frac{1}{P}\int_\Omega r^2|\phi(\br)|^2\mathrm{d}\br
\end{equation}
with $\Omega$ being the integration domain. 
It can be seen here that the norm is monotonically growing from the linear limit 
(of the extended but small amplitude state) all the way to
the extended but highly nonlinear state of $\omega \rightarrow 1$. 

\begin{figure}[!tbp]
\centering
\begin{tabular}{cc}
\multicolumn{2}{c}{\includegraphics[width=\figpairwidth]{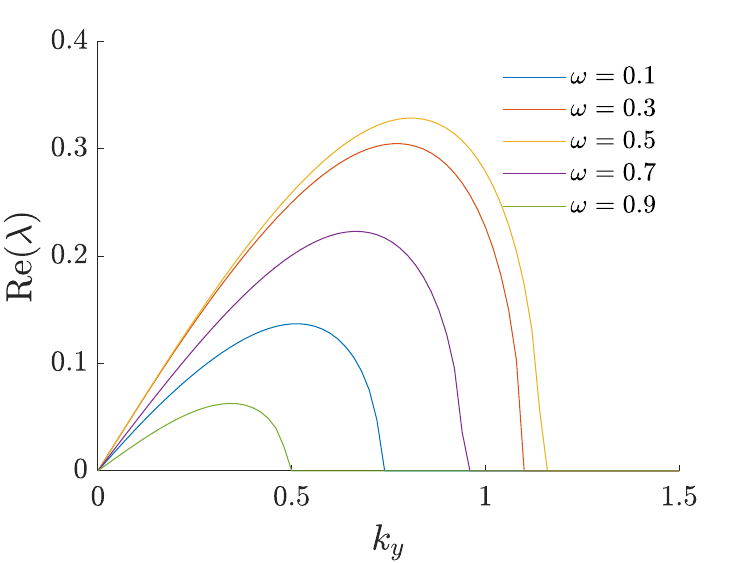}} \\
\includegraphics[width=\figpairwidth]{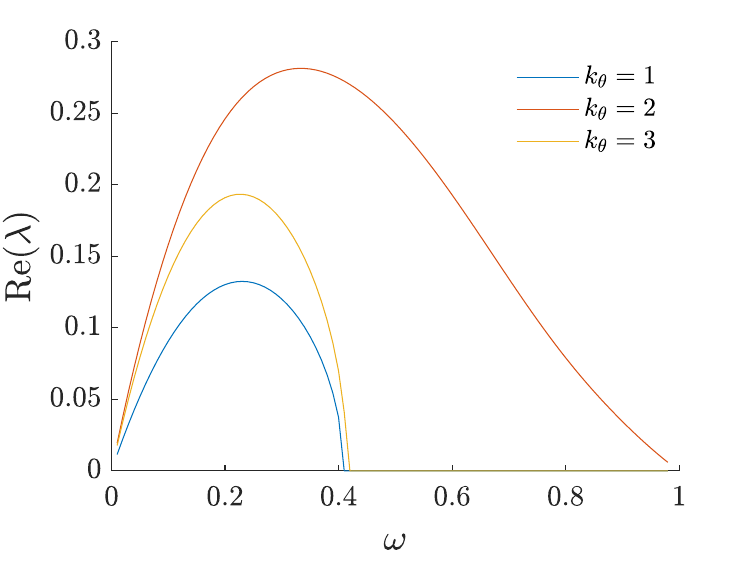} &
\includegraphics[width=\figpairwidth]{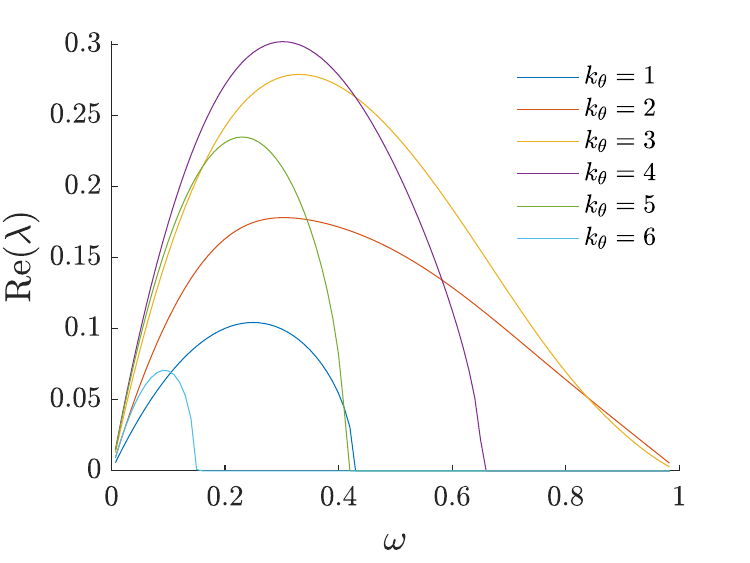} \\
\end{tabular}%
\caption{Dependence of the real part of the right-most eigenvalues for (top) line solitons and (bottom) vortex solitons with $S=1$ (left) and $S=2$ (right). Notice that in the latter,  only the values of $k_\theta$ leading to instabilities have been depicted. In all cases, unstable eigenmodes for each frequency indicates spectral instability for all such solitary waves.}
\label{fig:stability}
\end{figure}

From the Vakhitov--Kolokolov criterion~\cite{VakhitovKolokolov1973_VKCriterion} ---which, incidentally, was originally developed
in a saturable nonlinearity model---, both 1D and 2D radial solitons are expected to be stable, as $P'(\omega)>0$ and both solutions are the ground states for the respective NLS equation. For this reason, we will only show the stability results for line and vortex solitons in Fig.~\ref{fig:stability}.
For line solitons, we show the transverse instability results for some selected, yet
representative, values of the frequency. 
Nevertheless, all the frequencies considered are identified as possessing a long
wavelegth transverse instability.
Notice that the dependence of the growth rate with $k_y$ does
not vary monotonically with $\omega$; the growth rate vanishes as $k_y$ vanishes,
and it also vanishes for sufficiently large wavenumbers $k_y$. For vortex solitons, we consider the azimuthal instability for different values of $k_\theta$. Again, the growth rate vanishes as
$k_\theta \rightarrow 0$ and not all wavenumbers considered are unstable for each frequency.
However, there are always (i.e., for each frequency considered) wavenumbers that are found
to be transversely unstable. Accordingly, we find that this model possesses no stable line
or vortical solitary waves. Thus, we will henceforth focus on solitary wave interactions
only for the spectrally stable states, namely for the non-topological solitons.

\begin{figure}[!tbp]
\begin{tabular}{cc}
\includegraphics[width=\figpairwidth]{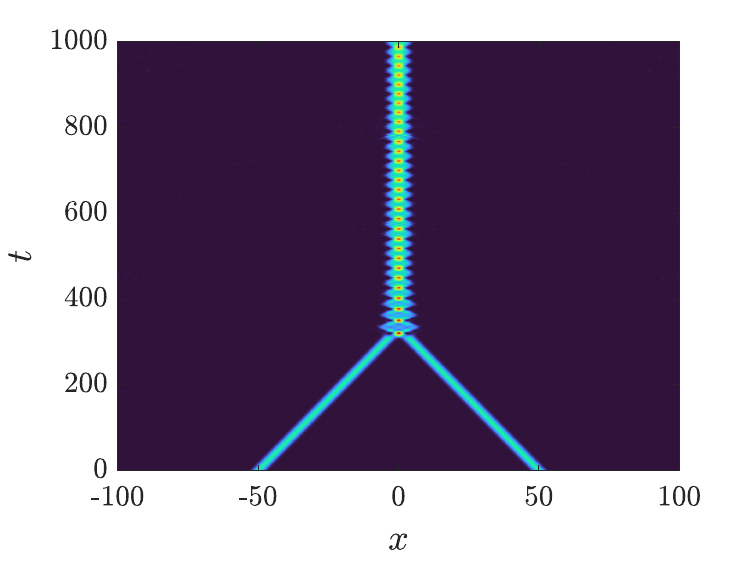} &
\includegraphics[width=\figpairwidth]{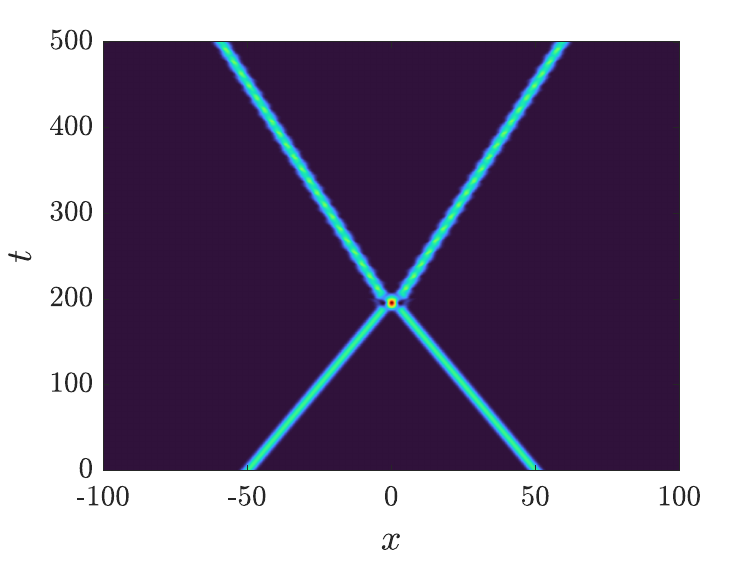} \\
\includegraphics[width=\figpairwidth]{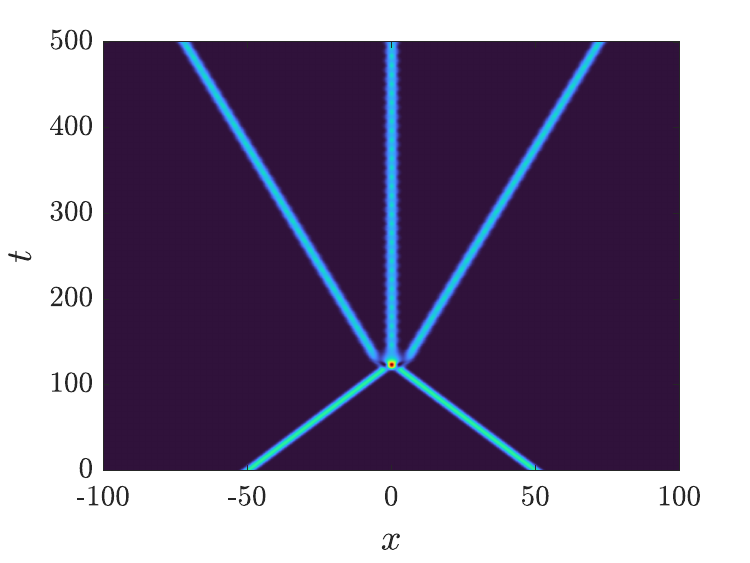} &
\includegraphics[width=\figpairwidth]{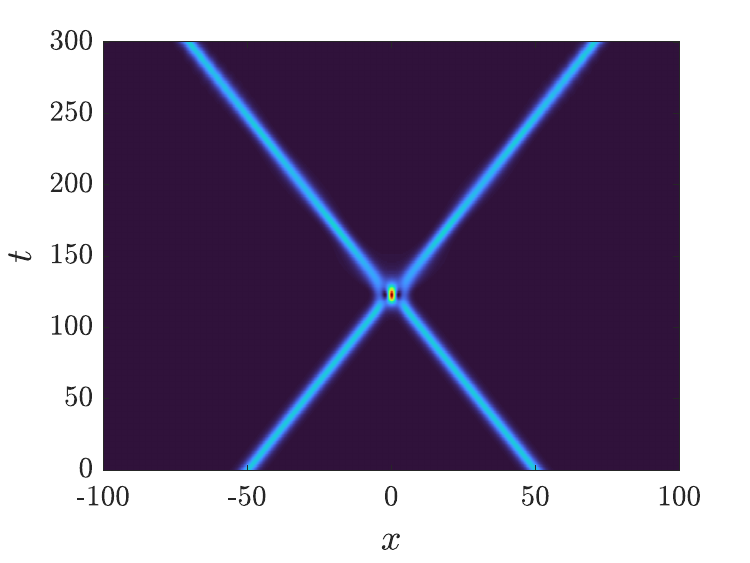} \\
\end{tabular}%
\caption{Power density plots showing the evolution of the head-on collisions of 1D solitons with the same speed, produced via a Galilean boost. Speeds and frequencies at each of the panels are: (top left) $\omega=0.9$, $c=0.15$; (top right) $\omega=0.9$, $c=0.25$; (bottom left) $\omega=0.9$, $c=0.4$ (bottom right) $\omega=0.1$, $c=0.4$.}
\label{fig:collisions1D}
\end{figure}

\section{Collisions of stable non-topological solitons}

We consider symmetric head-on collisions between two solitons moving in opposite directions with the same speed $c$. They are generated by means of a Galilean boost from a stationary soliton $\phi(x)$. {To this aim, we have made use of 4th order Runge-Kutta integrators. For the simulations of 1D solitons, we have kept the lattice parameter $h=0.1$, and the time step has been taken as $\delta=0.01$. However, for simulations of 2D solitons, we have increased the lattice parameter to $h=0.2$, and the time step was $\delta=0.005$; in addition, to perform such simulations, we had to make use of GPU computing.} 
 
In the one-dimensional case, the observed outcome is very similar to that of colliding solitons in the saturable NLS equation \cite{chris,snyder}. Fig.~\ref{fig:collisions1D} shows the typical behaviors. For fixed $\omega=0.9$, we have changed the speed, so that for low speed the solitons get trapped
in a fusion event; when the speed is increased, the solitons collide nearly elastically and, for high enough speed, there is a mixed trapping and reflection of the solitons, i.e., they still
separate but leave behind a localized wave at the collision point. The critical value of the speed
separating those behaviors depends on the frequency. However, for small enough frequency, the trapping-and-reflection regime is not observed and, in addition, the solitons collide nearly elastically. 
 Although, as we showed in the previous section, solitary waves exist for arbitrarily small
 powers, there exists a finite power threshold before a localized pulse can be ``shed'' as the result
 of a collision event.

\begin{figure}[!tbp]
\begin{center}
\begin{tabular}{cc}
\includegraphics[width=\figtwowidth]{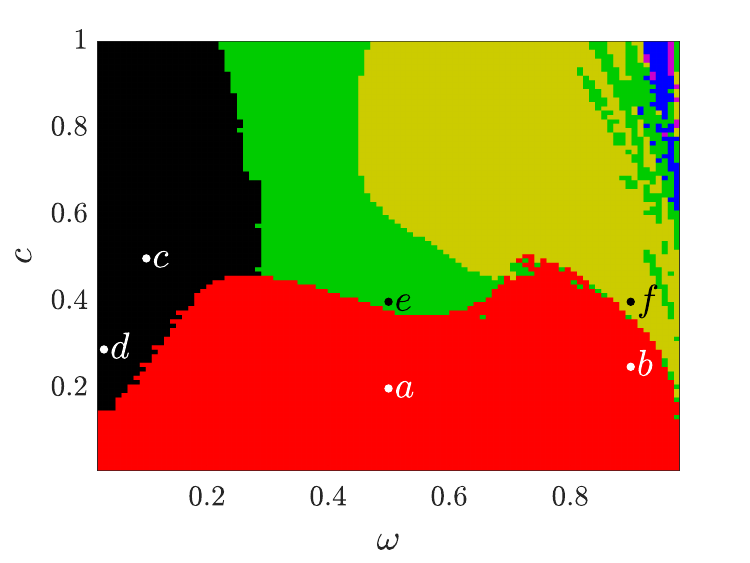} &
\includegraphics[width=\figtwowidth]{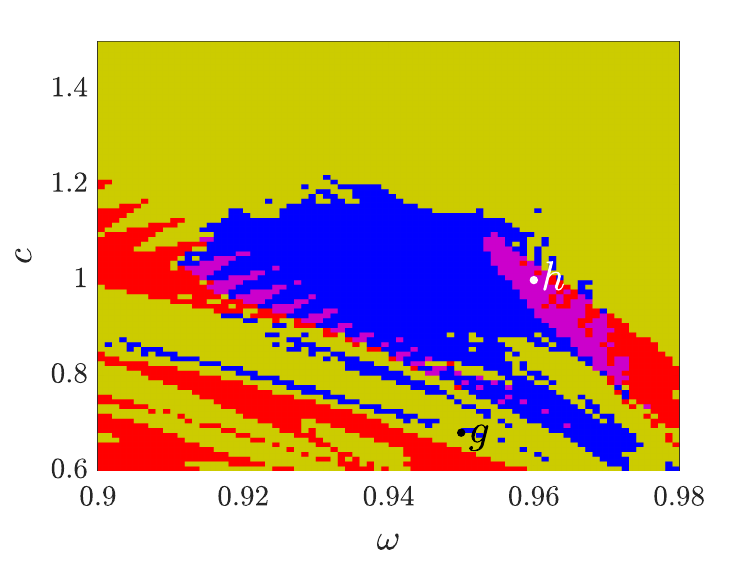}
\end{tabular}
\end{center}%
\caption{Diagram showing, in the $(\omega,c)$ plane, the different regimes for head-on collisions. The colors have the following interpretation. Red: trapping; black: dispersion; green: reflection without bound states; yellow: reflection with bound state; blue: transverse breaking without bound states; pink: transverse breaking with bound states. The right panel focuses on the region with transverse breaking. {Letters correspond to the isosurfaces in Fig.~\ref{fig:isosurfaces}.}}
\label{fig:carpet}
\end{figure}

We have analyzed in greater detail the collisions between solitons in a 2D setting. The different regimes are summarized in the $\omega$-$c$ plane of Fig.~\ref{fig:carpet}.

\begin{figure}[!tbp]
\begin{tabular}{cccc}
\includegraphics[width=\figfourwidth]{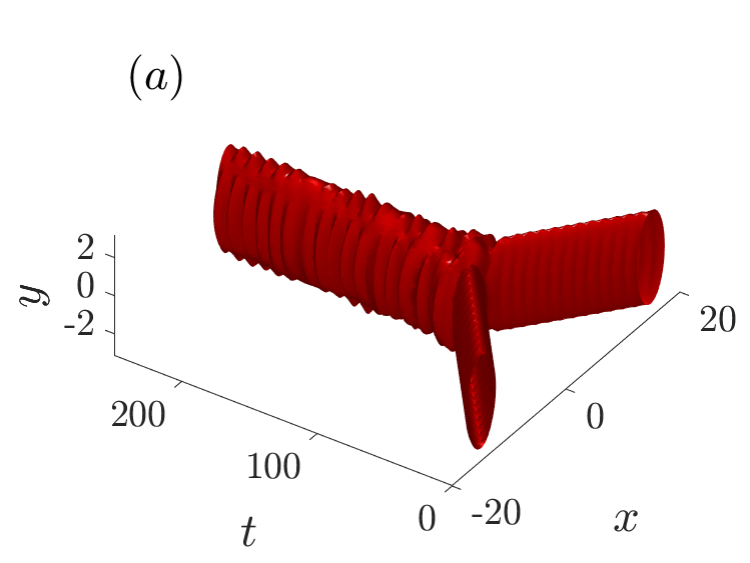} &
\includegraphics[width=\figfourwidth]{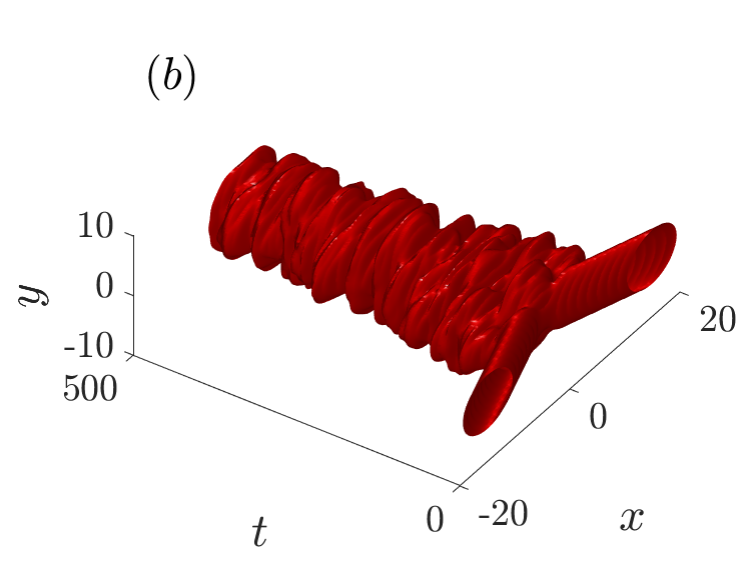} &
\includegraphics[width=\figfourwidth]{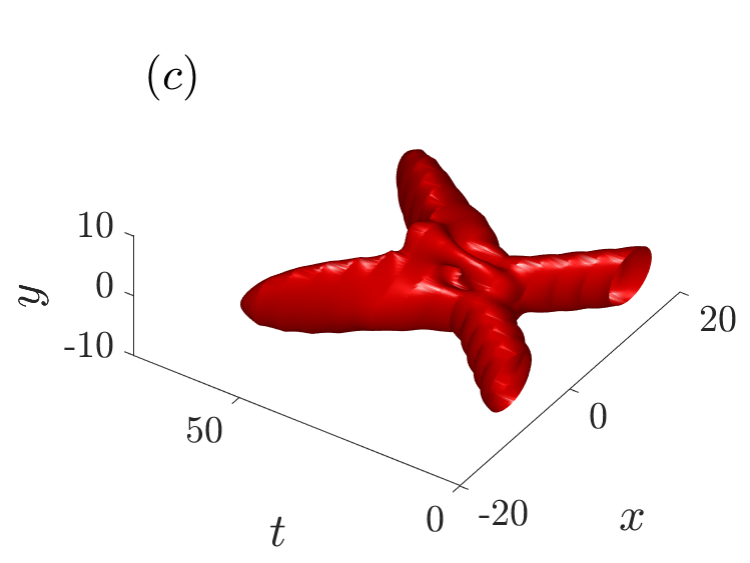} & 
\includegraphics[width=\figfourwidth]{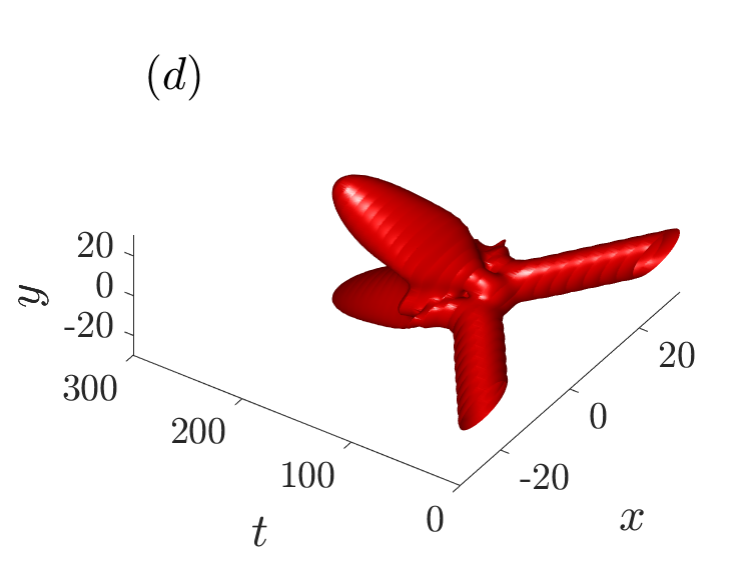} \\
\includegraphics[width=\figfourwidth]{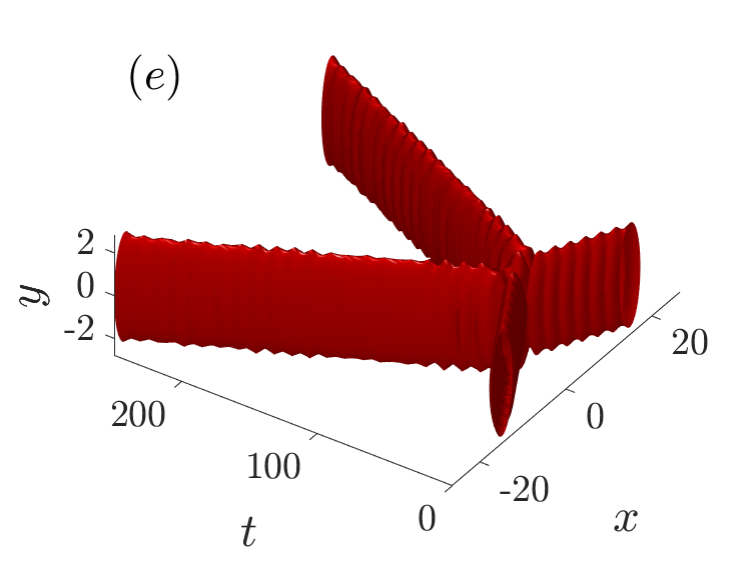} & 
\includegraphics[width=\figfourwidth]{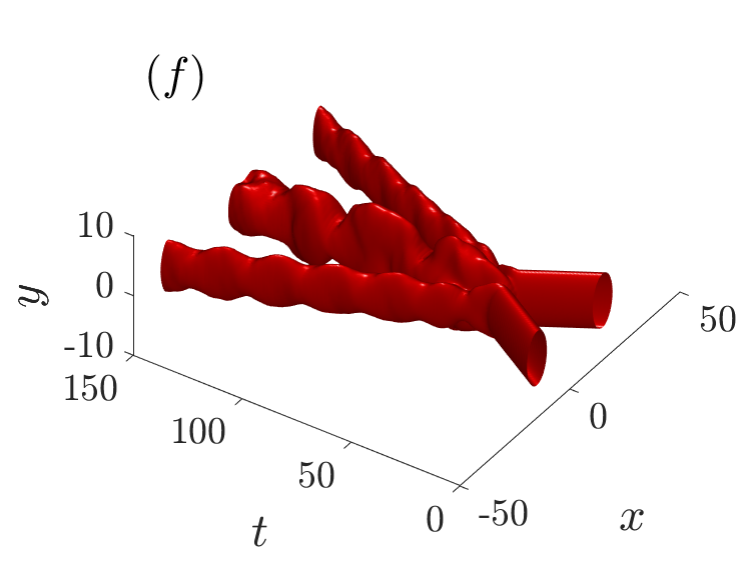} &
\includegraphics[width=\figfourwidth]{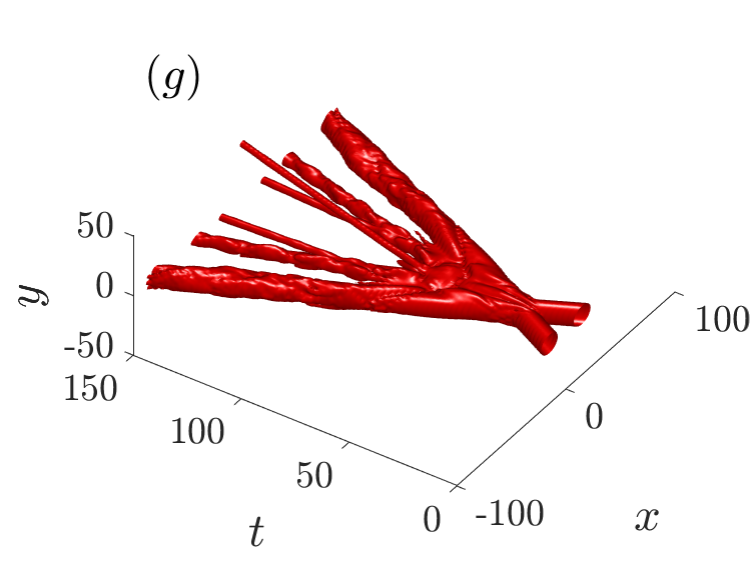} & 
\includegraphics[width=\figfourwidth]{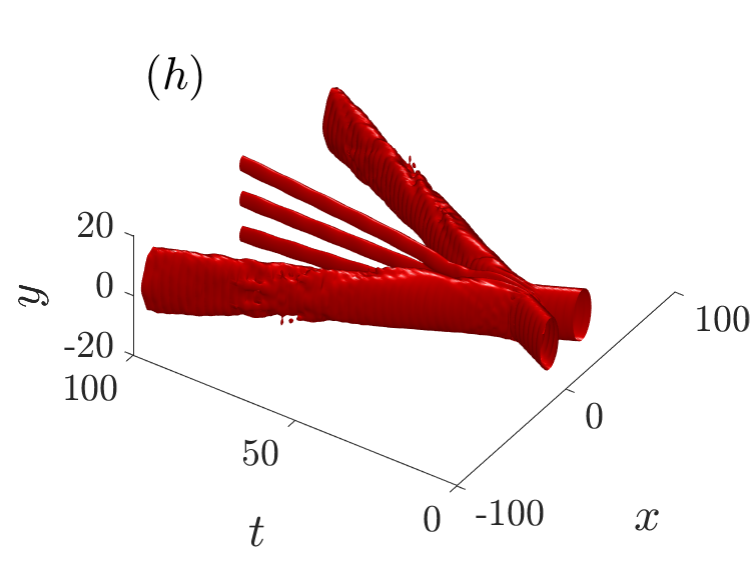} \\
\end{tabular}%
\caption{Isosurfaces of the power density $|\psi(x,y,t)|^2$ when it is equal to 1/20th of the maximum density of the corresponding stationary soliton, except for the two last one where the square root of the power density $|\psi(x,y,t)|$ is used in order to clearly illustrate the transverse
field excitation. {Letters correspond to the dots in Fig.~\ref{fig:carpet}.}}
\label{fig:isosurfaces}
\end{figure}

We explain the regimes below, in addition to showing several typical outcomes; to this aim, an isosurface (see Fig.~\ref{fig:isosurfaces}) and some snapshots  for each simulation are included in the text and some movies are included as supplemental material.

\begin{figure}[!htbp]
\begin{tabular}{ccc}
\includegraphics[width=\figthreewidth]{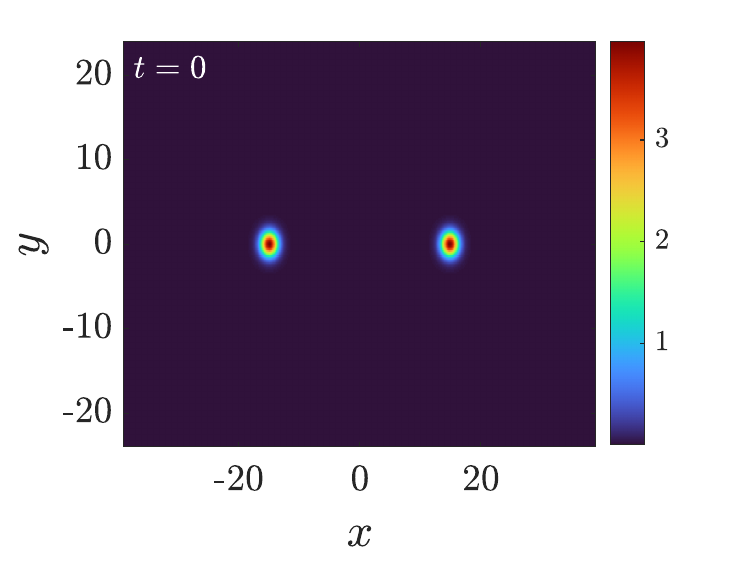} &
\includegraphics[width=\figthreewidth]{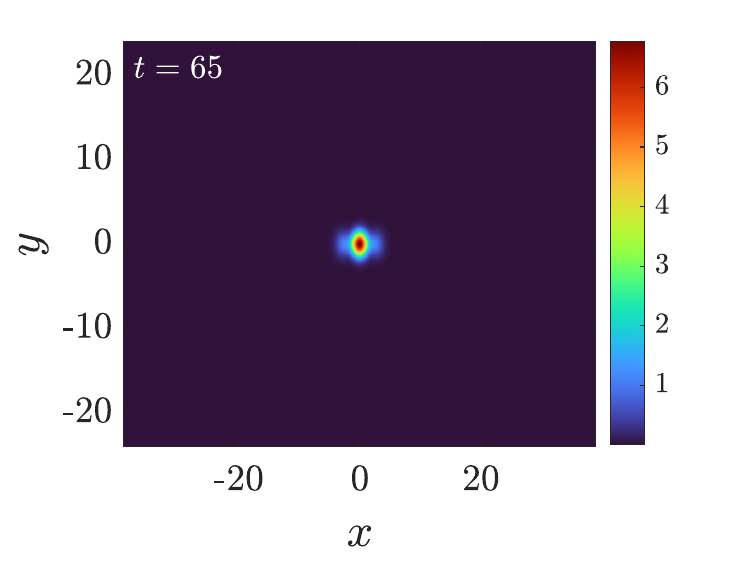} &
\includegraphics[width=\figthreewidth]{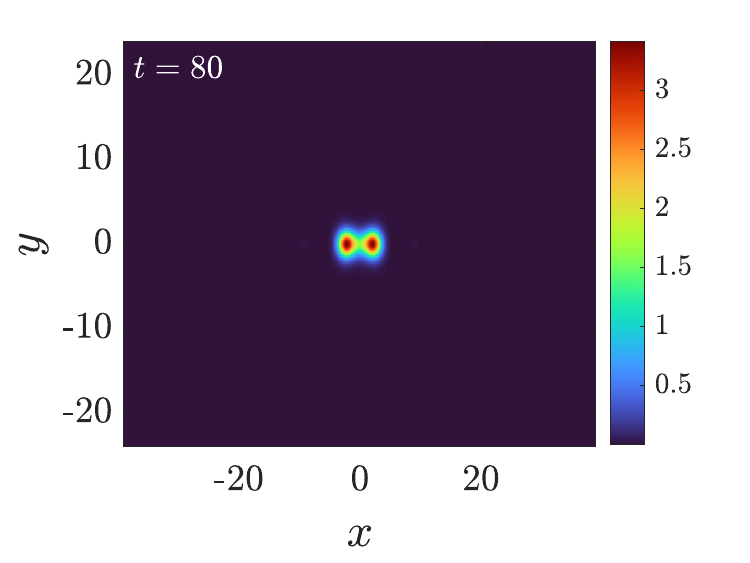} \\
\includegraphics[width=\figthreewidth]{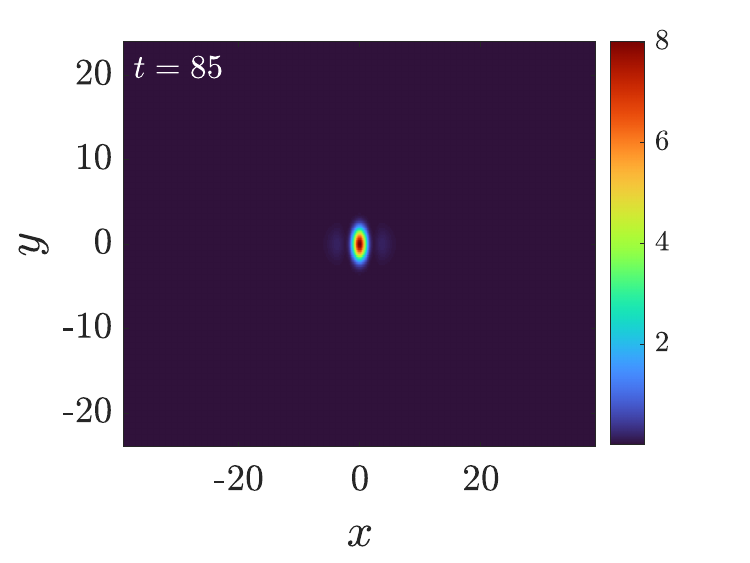} &
\includegraphics[width=\figthreewidth]{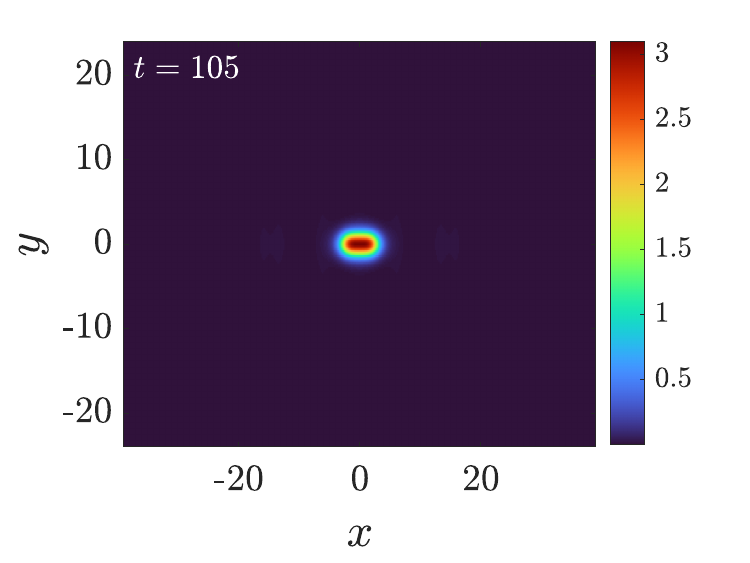} &
\includegraphics[width=\figthreewidth]{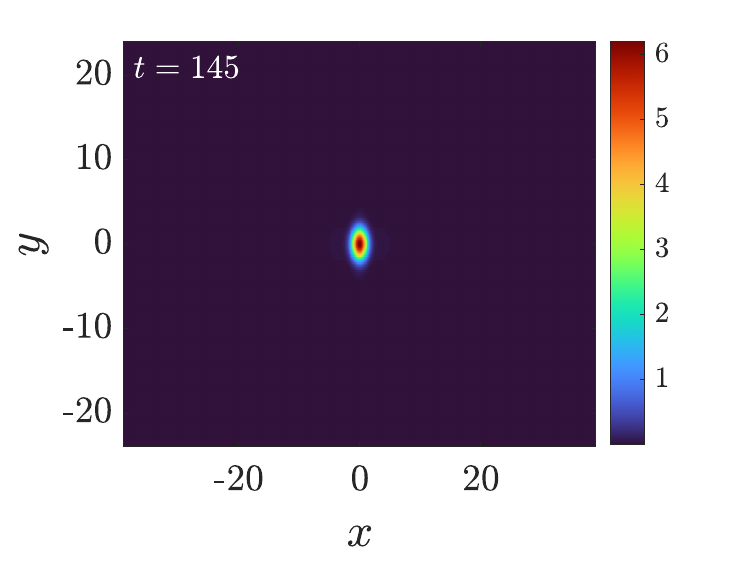} \\
\end{tabular}%
\caption{Power density plots showing the evolution of the head-on collisions of 2D solitons at different values of time for $\omega=0.5$ and $c=0.2$. This is a case leading to the fusion of the two
solitary waves. See also companion movie \texttt{movie\_01.gif}.}
\label{fig:simul2D_trapping1}
\end{figure}

\begin{figure}[!htbp]
\begin{tabular}{cccc}
\includegraphics[width=\figfourwidth]{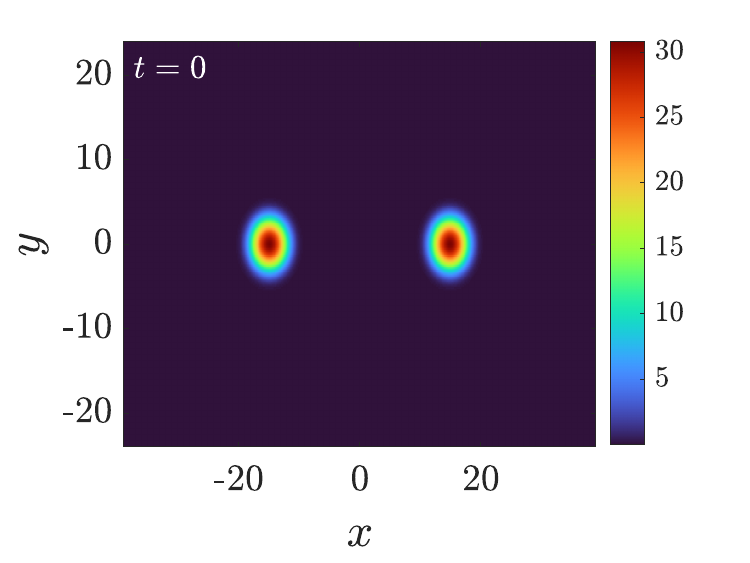} &
\includegraphics[width=\figfourwidth]{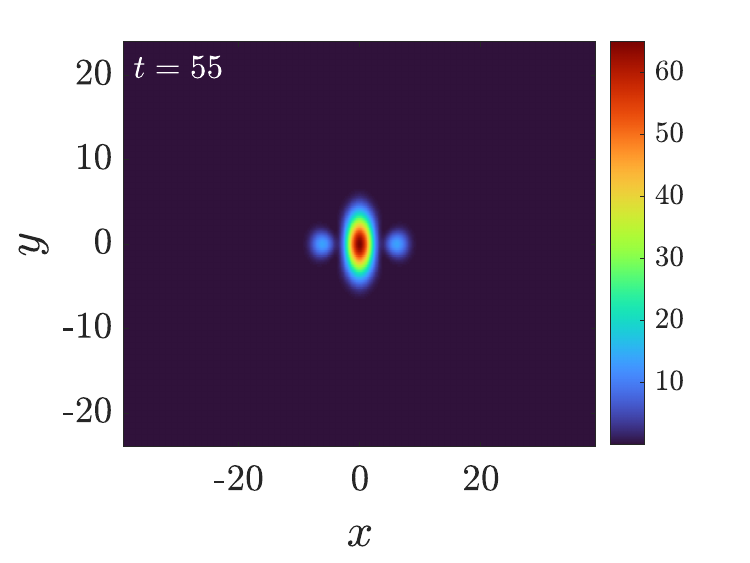} &
\includegraphics[width=\figfourwidth]{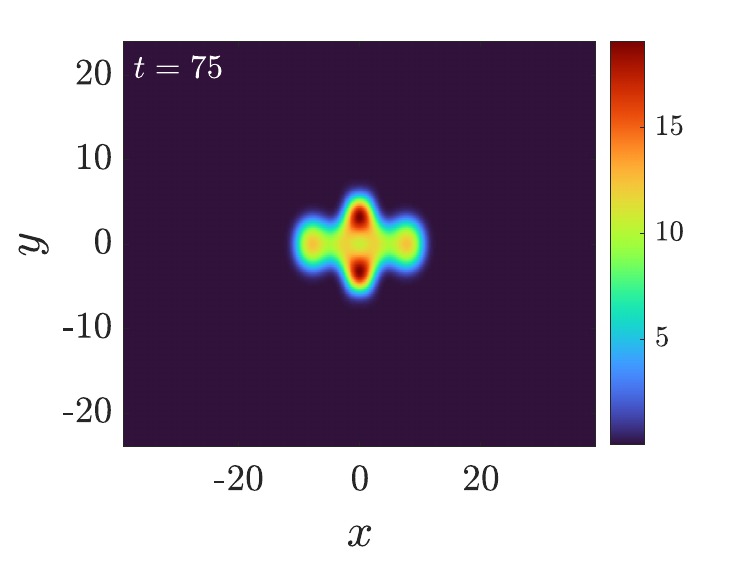} &
\includegraphics[width=\figfourwidth]{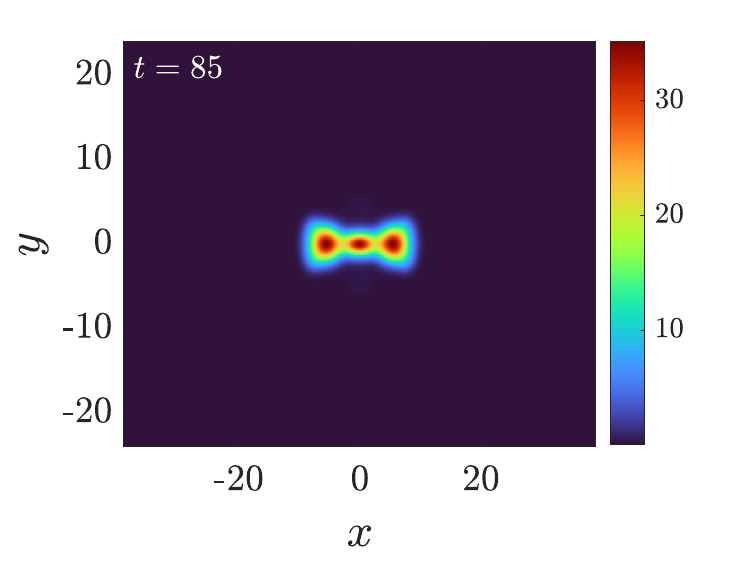} \\
\includegraphics[width=\figfourwidth]{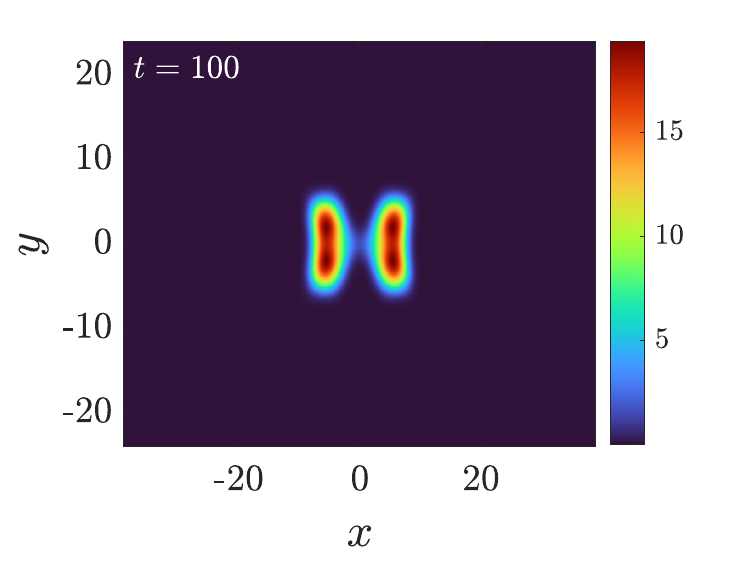} &
\includegraphics[width=\figfourwidth]{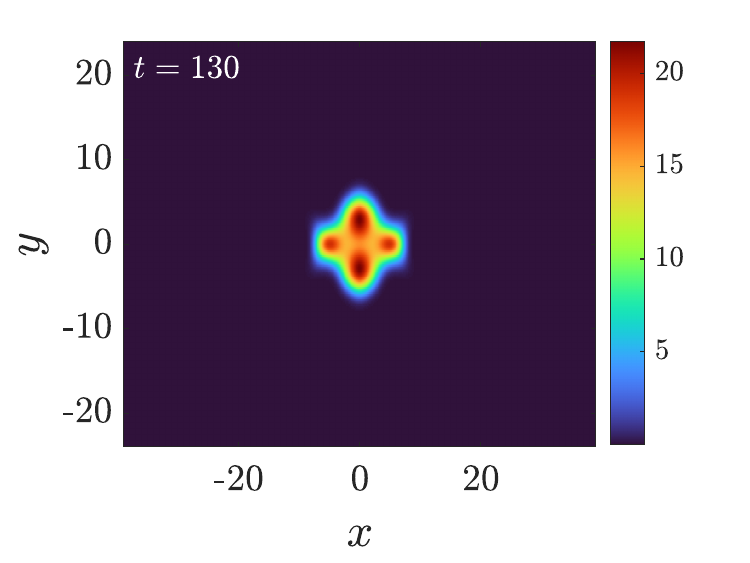} &
\includegraphics[width=\figfourwidth]{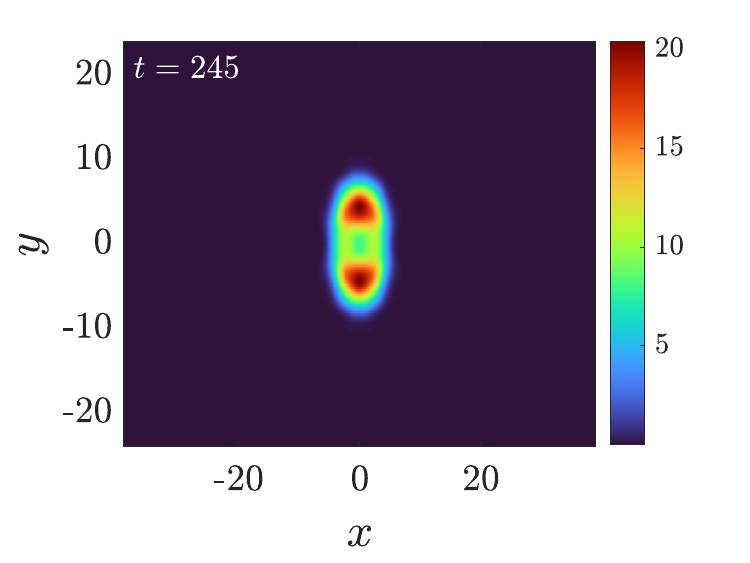} &
\includegraphics[width=\figfourwidth]{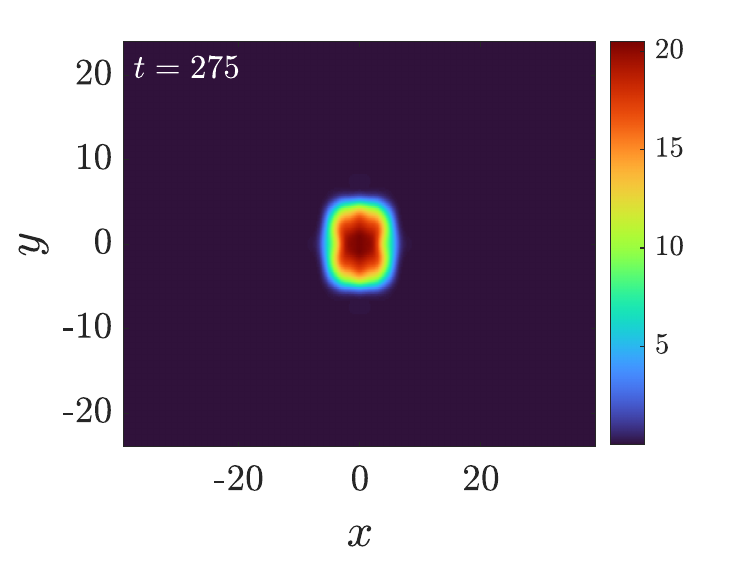} \\
\end{tabular}%
\caption{Power density plots showing the evolution of the head-on collisions of 2D solitons at different values of time for $\omega=0.9$ and $c=0.25$. Here the waves feature a complex oscillation 
(and occasional transverse excitation), however, they are never able to escape each other's attraction. See also companion movie \texttt{movie\_02.gif}.}
\label{fig:simul2D_trapping2}
\end{figure}

\begin{figure}[!htbp]
\begin{tabular}{ccc}
\includegraphics[width=\figthreewidth]{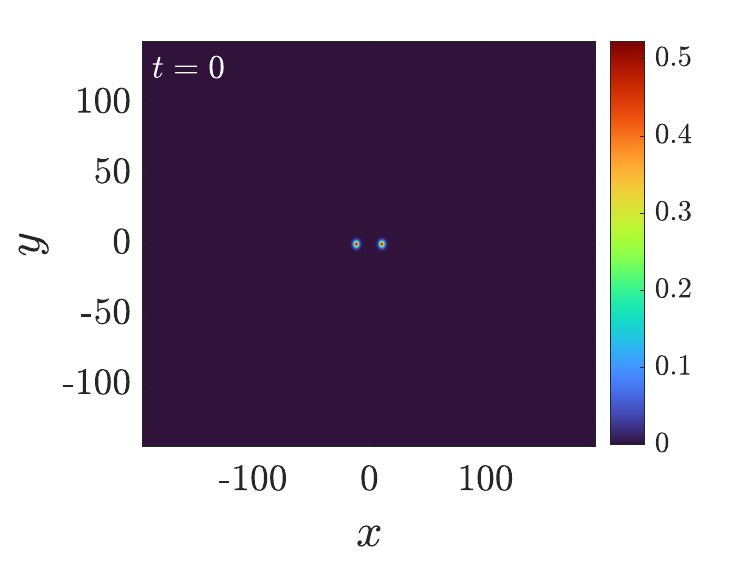} &
\includegraphics[width=\figthreewidth]{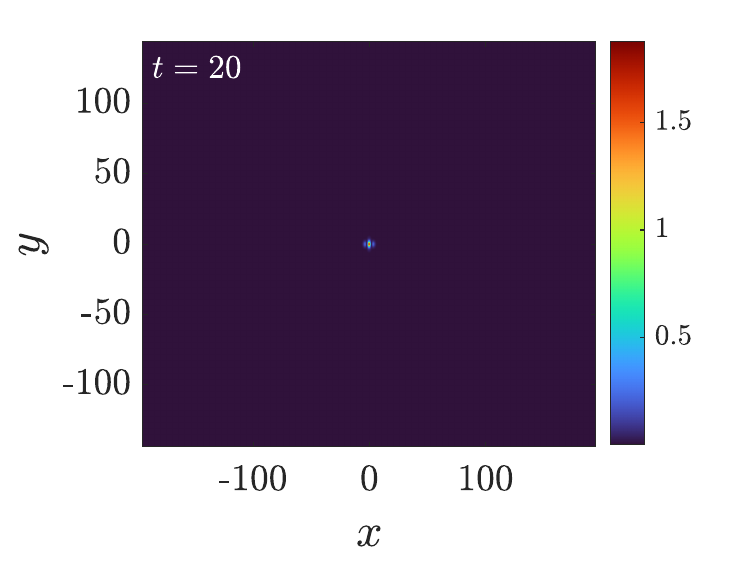} &
\includegraphics[width=\figthreewidth]{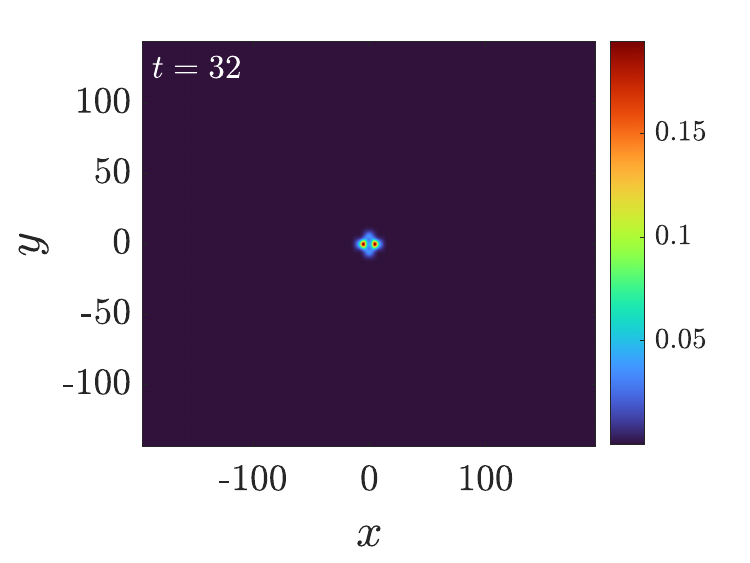} \\
\includegraphics[width=\figthreewidth]{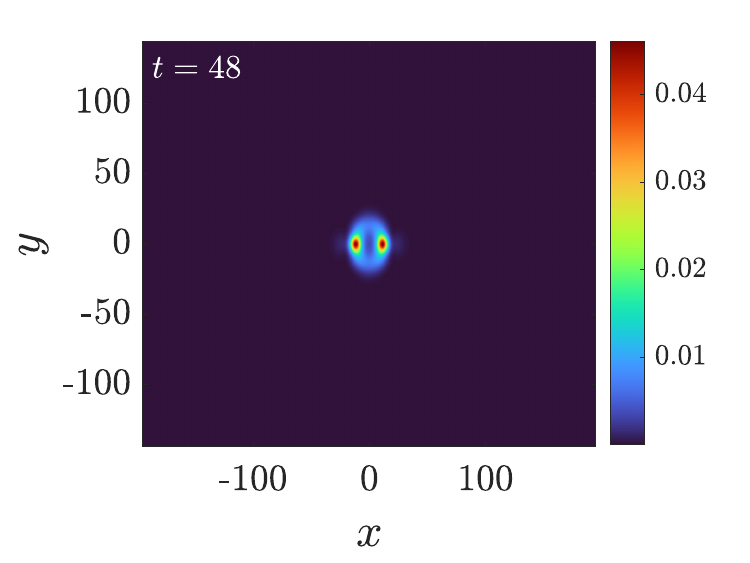} &
\includegraphics[width=\figthreewidth]{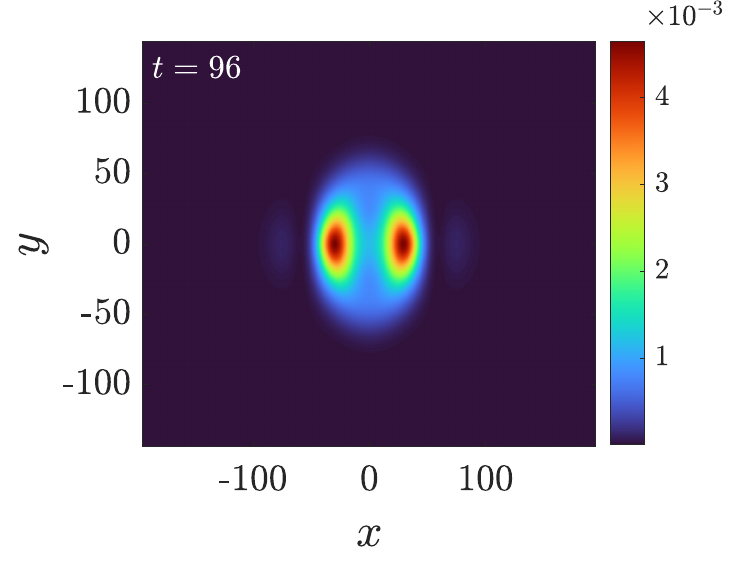} &
\includegraphics[width=\figthreewidth]{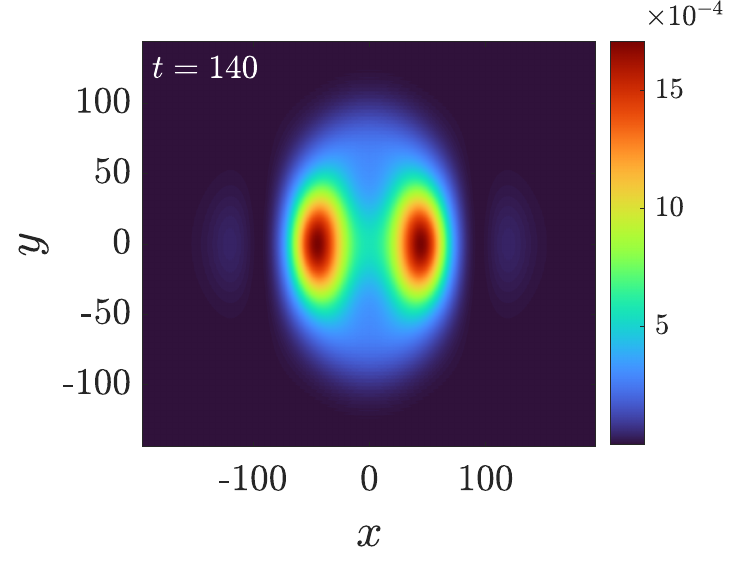} \\
\end{tabular}%
\caption{Power density plots showing the evolution of the head-on collisions of 2D solitons at different values of time for $\omega=0.1$ and $c=0.5$. Here, the system is led to eventual dispersion. See also companion movie \texttt{movie\_03.gif}.}
\label{fig:simul2D_dispersion}
\end{figure}

\begin{figure}[!htbp]
\begin{tabular}{ccc}
\includegraphics[width=\figthreewidth]{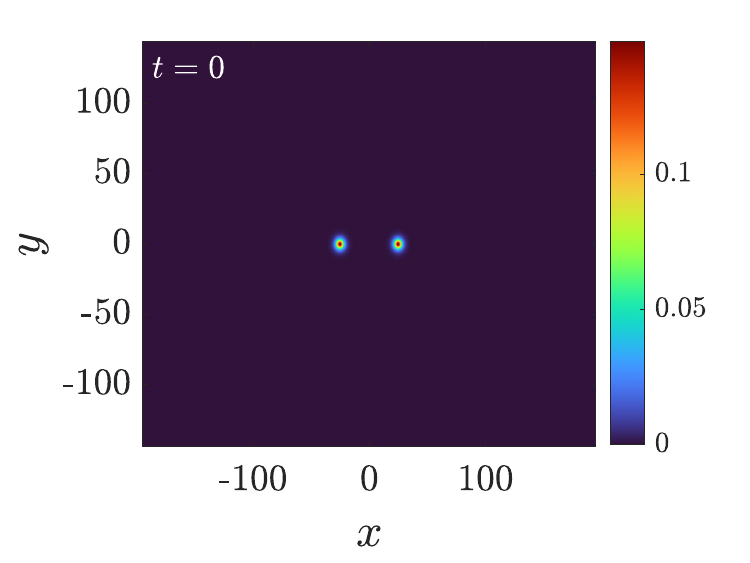} &
\includegraphics[width=\figthreewidth]{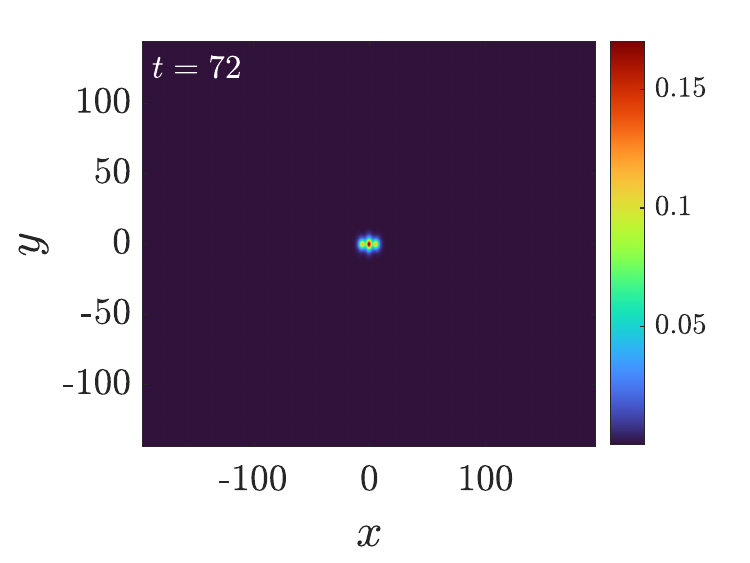} &
\includegraphics[width=\figthreewidth]{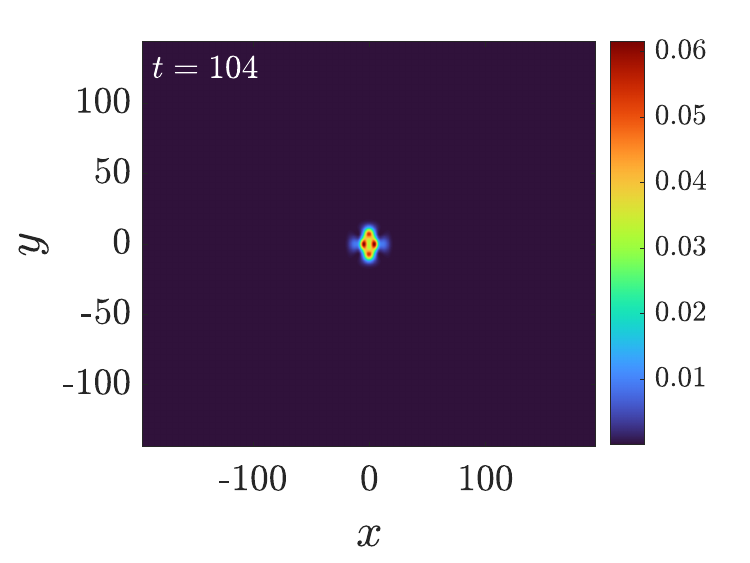} \\
\includegraphics[width=\figthreewidth]{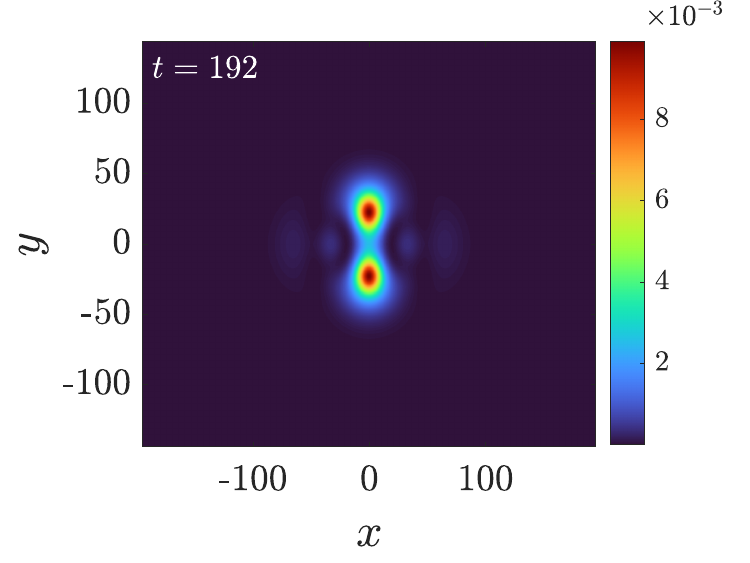} &
\includegraphics[width=\figthreewidth]{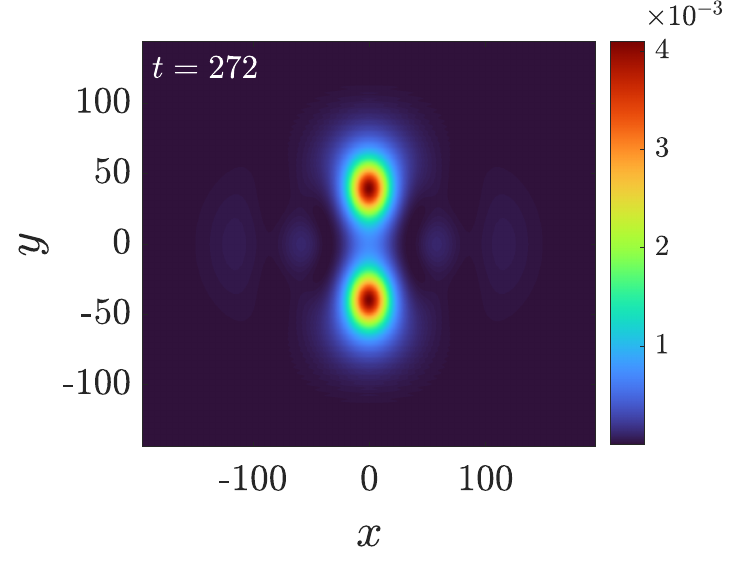} &
\includegraphics[width=\figthreewidth]{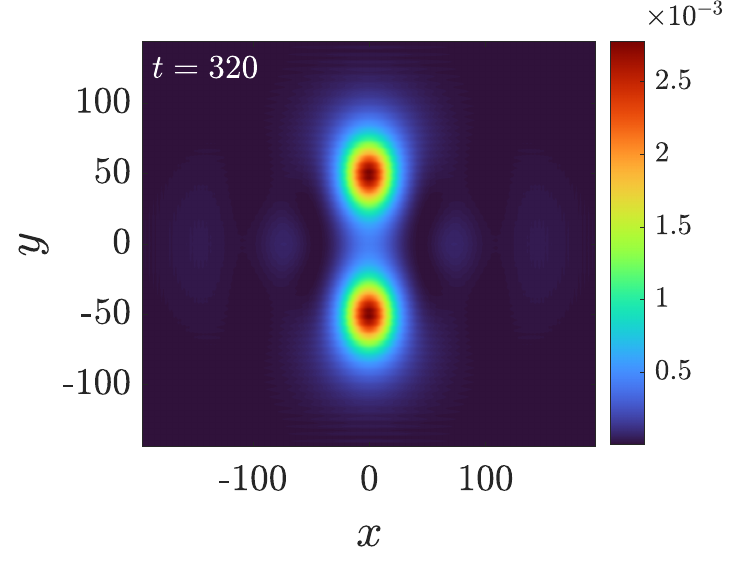} \\
\end{tabular}%
\caption{Power density plots showing the evolution of the head-on collisions of 2D solitons at different values of time for $\omega=0.03$ and $c=0.29$. Here, the system is led to eventual dispersion in the transversal direction. See also companion movie \texttt{movie\_04.gif}.}
\label{fig:simul2D_dispersiontrans}
\end{figure}

\begin{figure}[!htbp]
\begin{tabular}{ccc}
\includegraphics[width=\figthreewidth]{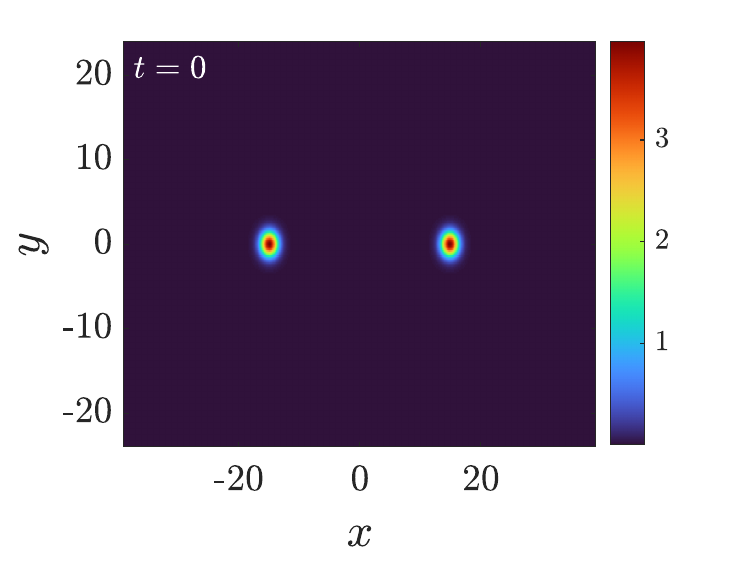} &
\includegraphics[width=\figthreewidth]{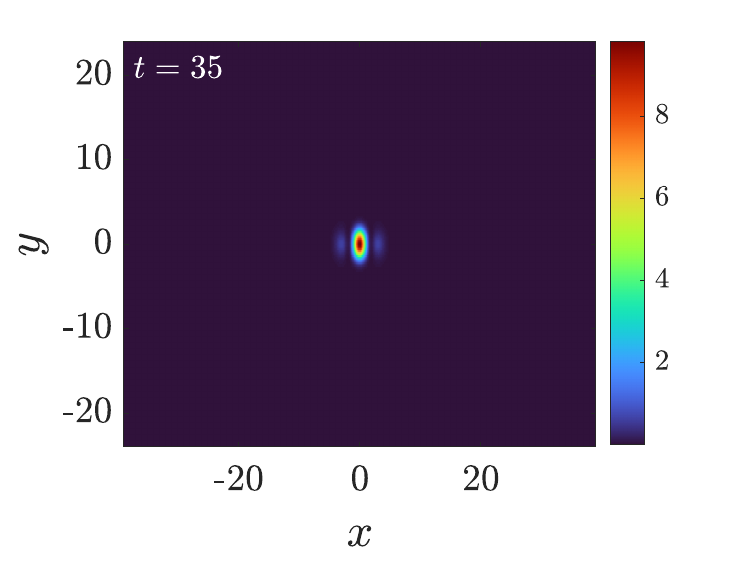} &
\includegraphics[width=\figthreewidth]{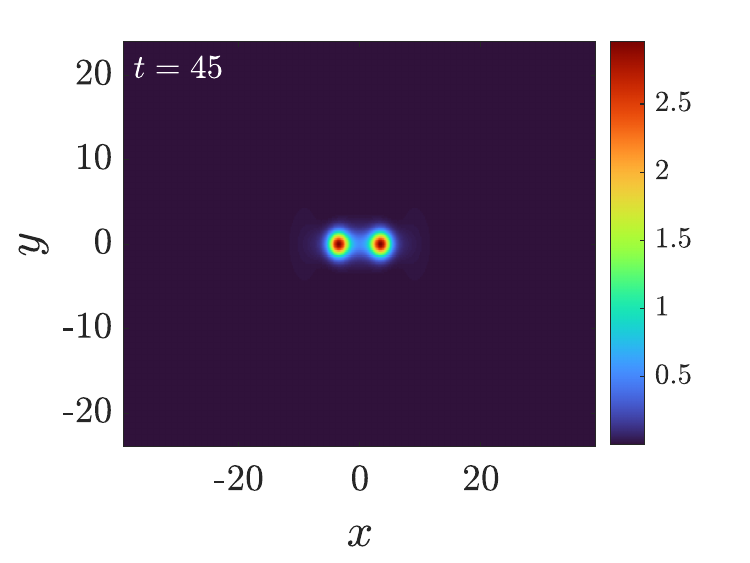} \\
\includegraphics[width=\figthreewidth]{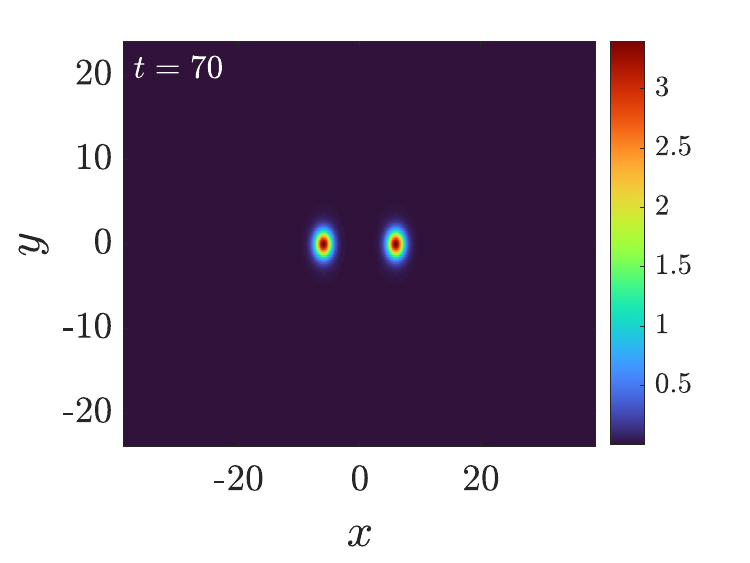} &
\includegraphics[width=\figthreewidth]{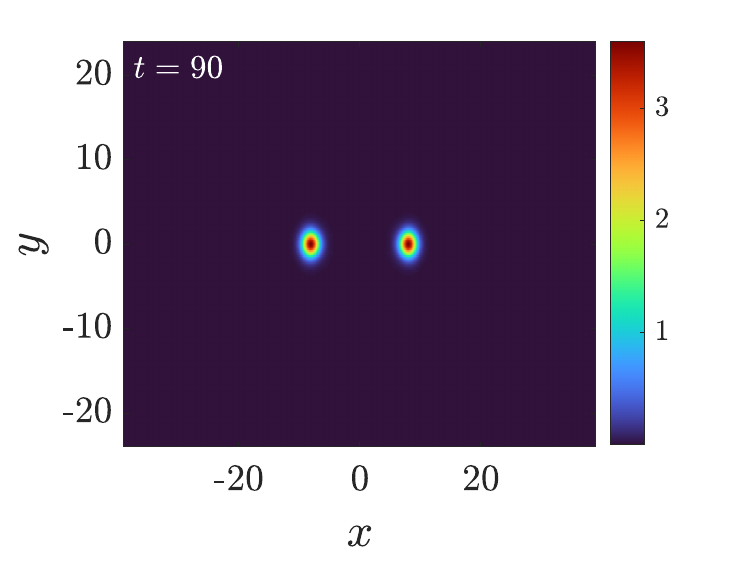} &
\includegraphics[width=\figthreewidth]{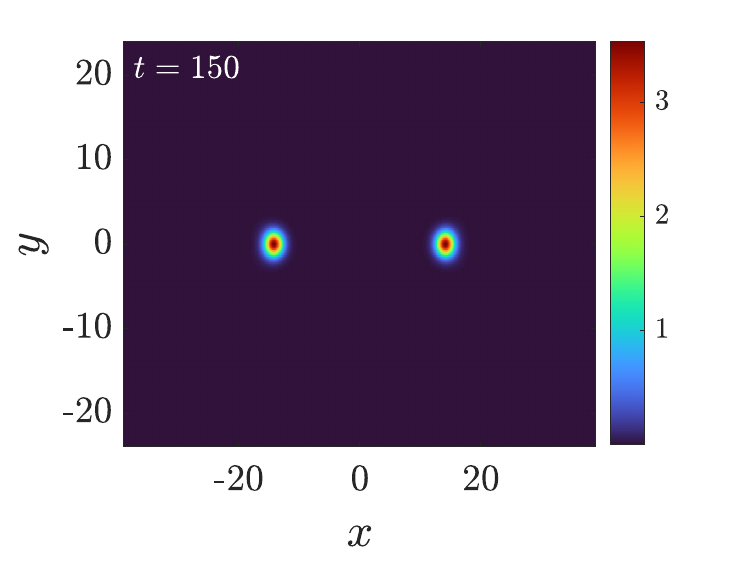} \\
\end{tabular}%
\caption{Power density plots showing the evolution of the head-on collisions of 2D solitons at different values of time for $\omega=0.5$ and $c=0.4$. The solitary waves separate (are reflected) without leaving
behind a localized pulse. See also companion movie \texttt{movie\_05.gif}.}
\label{fig:simul2D_reflection1}
\end{figure}

\begin{figure}[!htbp]
\begin{tabular}{cccc}
\includegraphics[width=\figfourwidth]{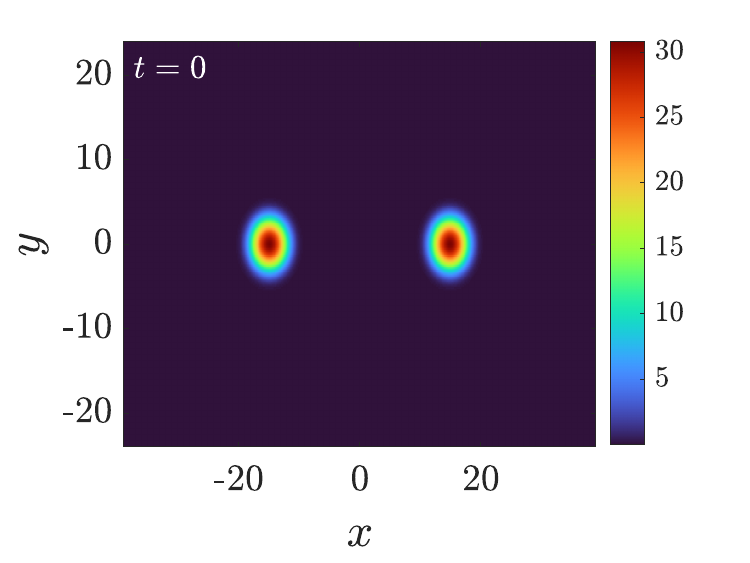} &
\includegraphics[width=\figfourwidth]{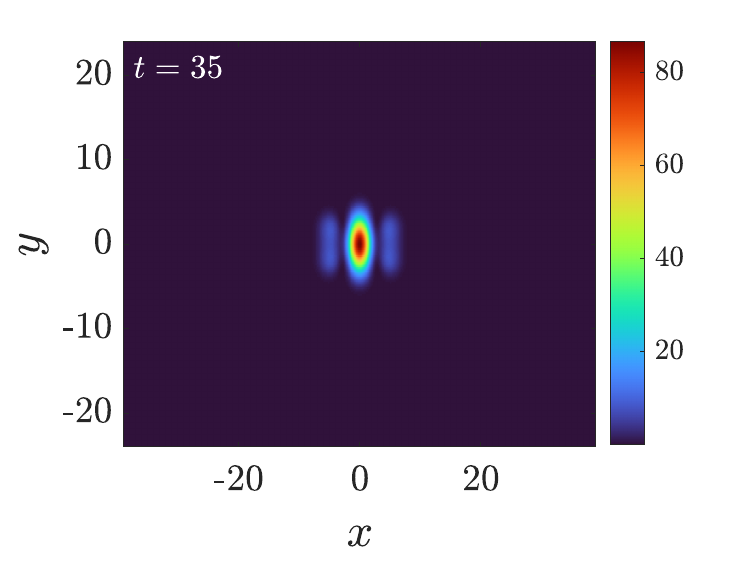} &
\includegraphics[width=\figfourwidth]{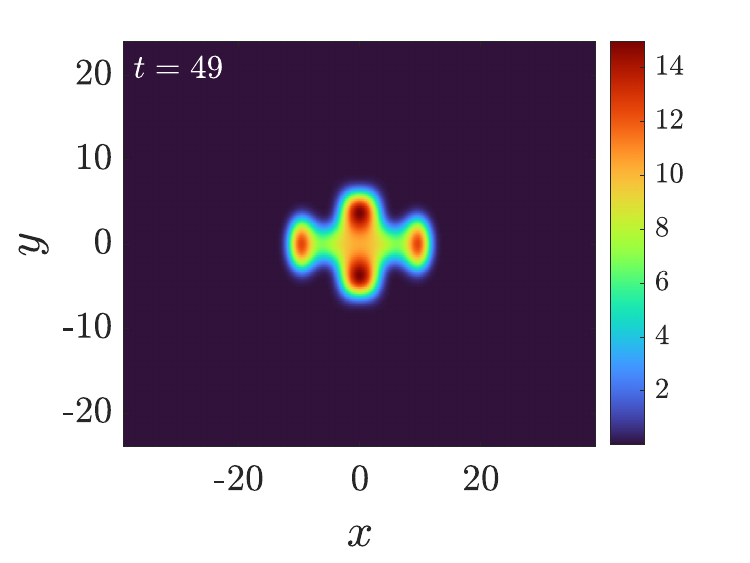} &
\includegraphics[width=\figfourwidth]{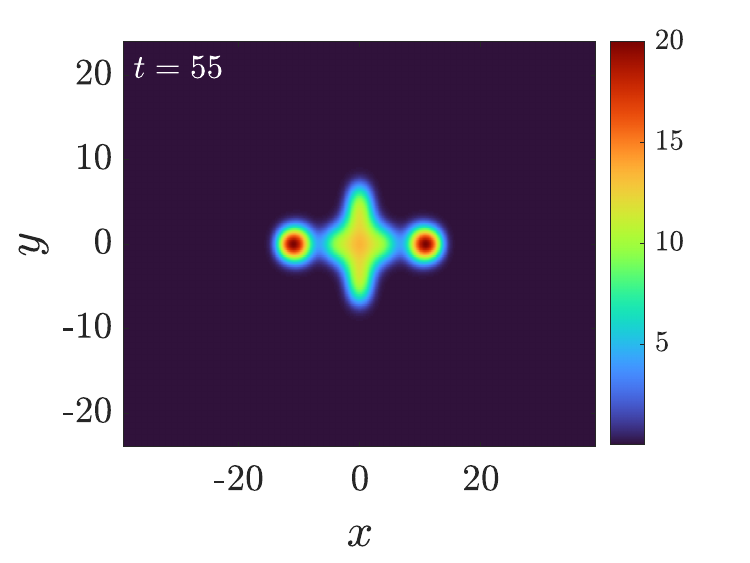} \\
\includegraphics[width=\figfourwidth]{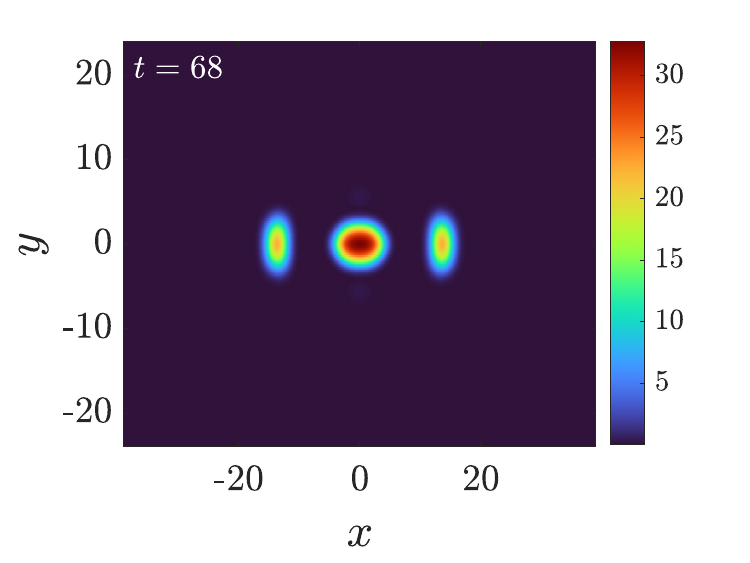} &
\includegraphics[width=\figfourwidth]{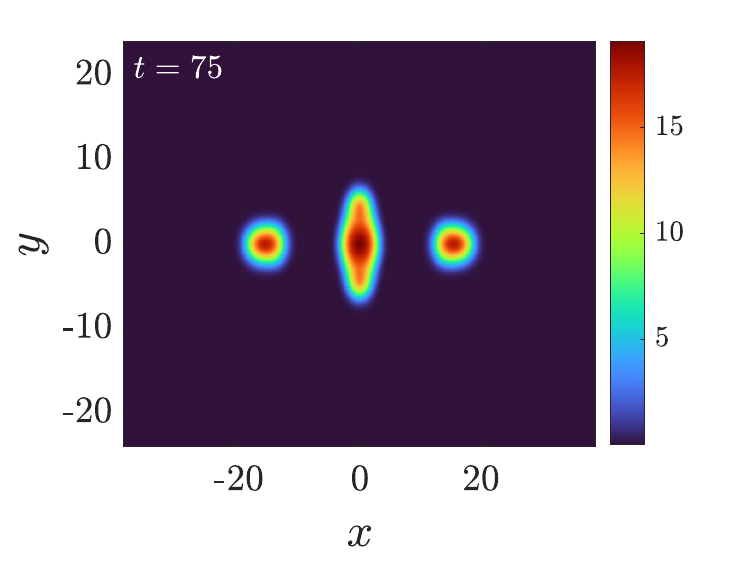} &
\includegraphics[width=\figfourwidth]{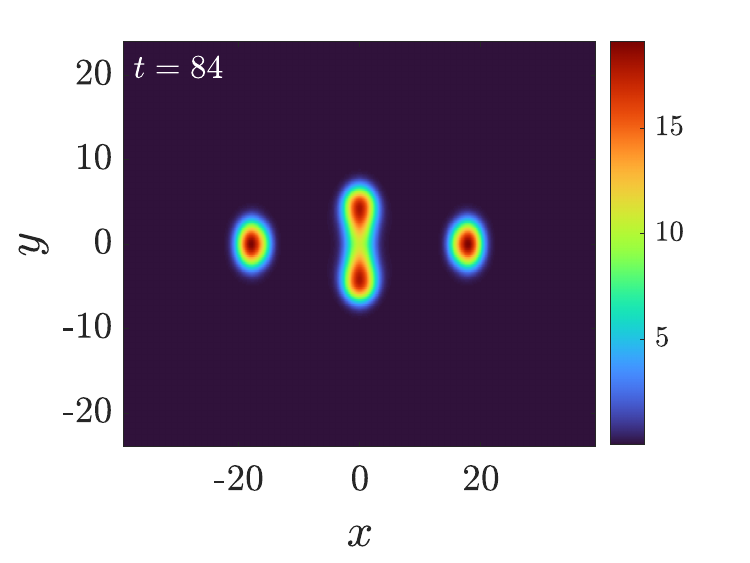} &
\includegraphics[width=\figfourwidth]{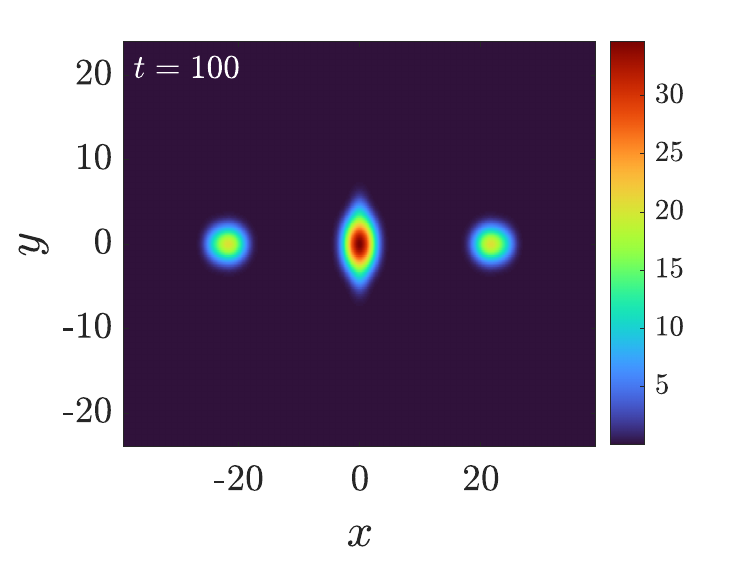} \\
\end{tabular}%
\caption{Power density plots showing the evolution of the head-on collisions of 2D solitons at different values of time for $\omega=0.9$ and $c=0.4$. The splitting process ---in addition to partial reflection--- creates a remnant solitary
wave at the original collision location. See also companion movie \texttt{movie\_06.gif}.}
\label{fig:simul2D_reflection2}
\end{figure}

\begin{figure}[!htbp]
\begin{tabular}{cccc}
\includegraphics[width=\figfourwidth]{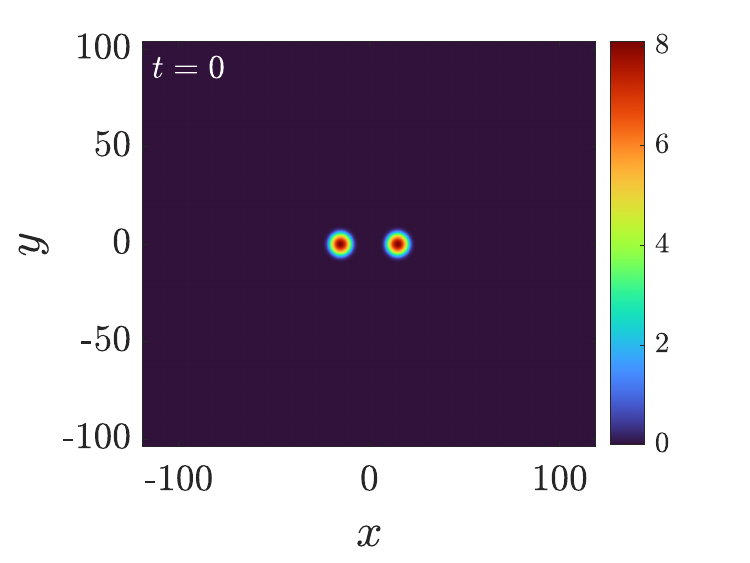} &
\includegraphics[width=\figfourwidth]{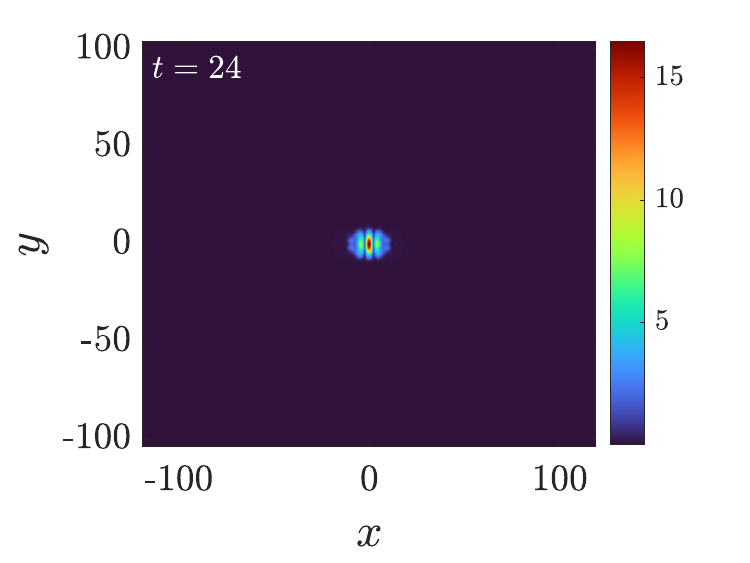} &
\includegraphics[width=\figfourwidth]{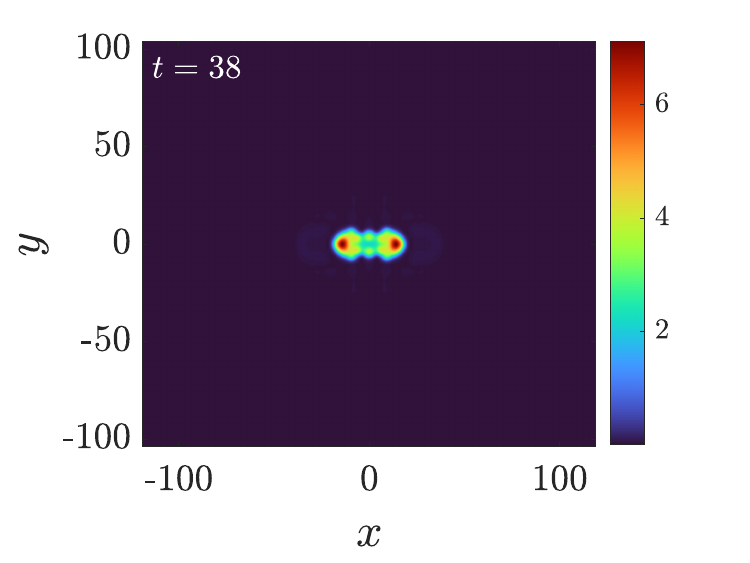} &
\includegraphics[width=\figfourwidth]{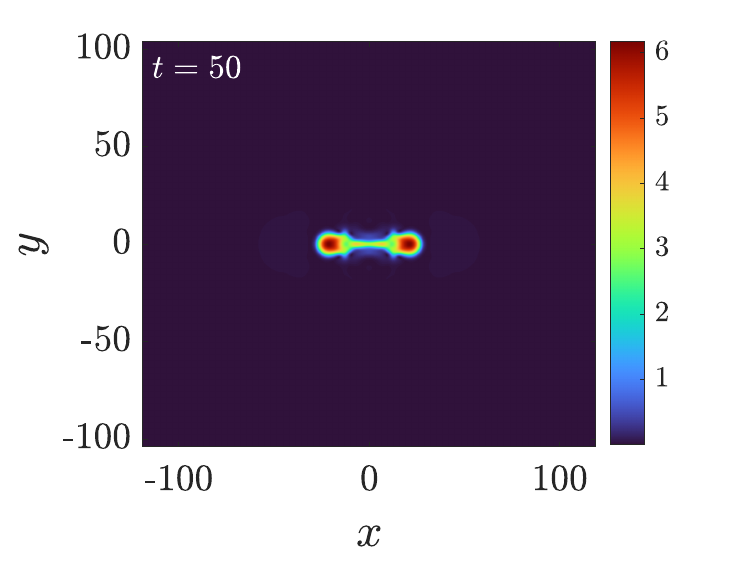} \\
\includegraphics[width=\figfourwidth]{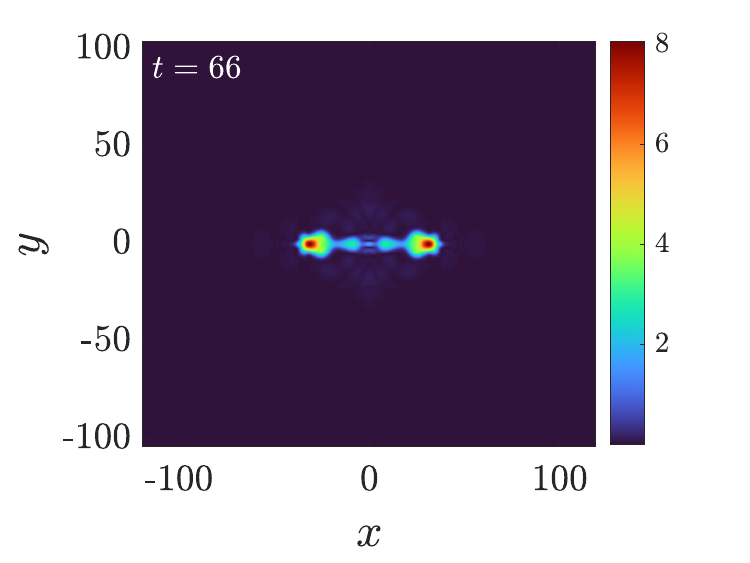} &
\includegraphics[width=\figfourwidth]{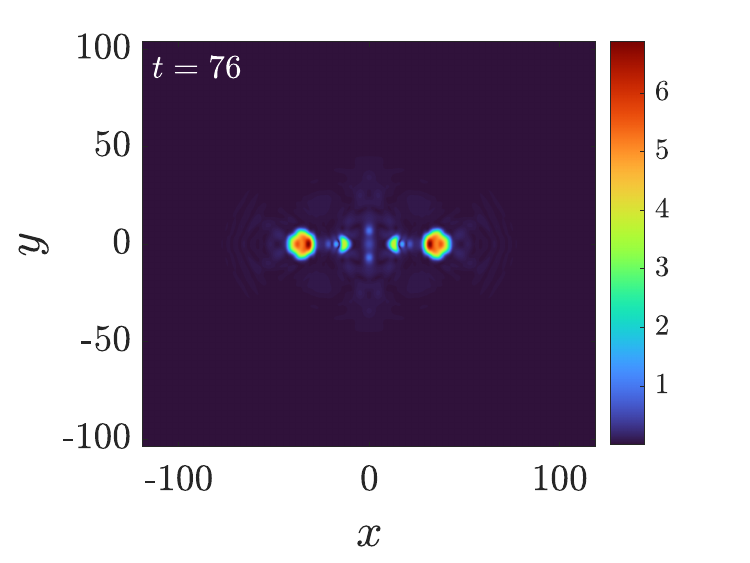} &
\includegraphics[width=\figfourwidth]{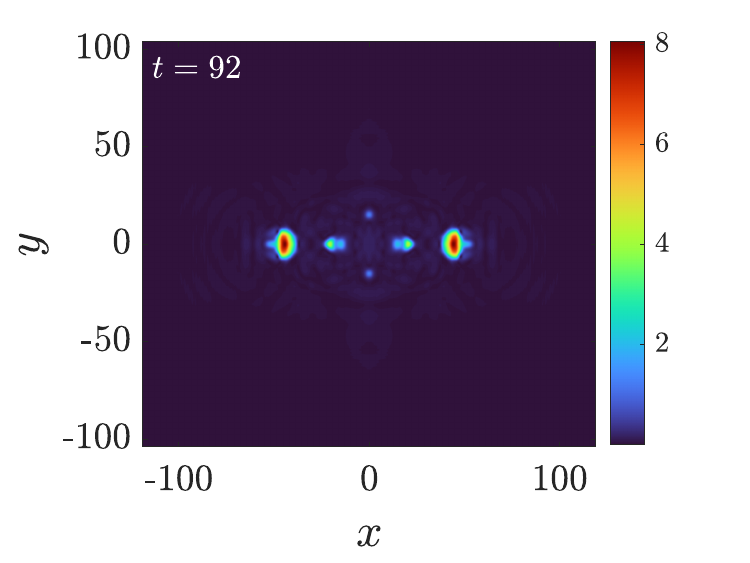} &
\includegraphics[width=\figfourwidth]{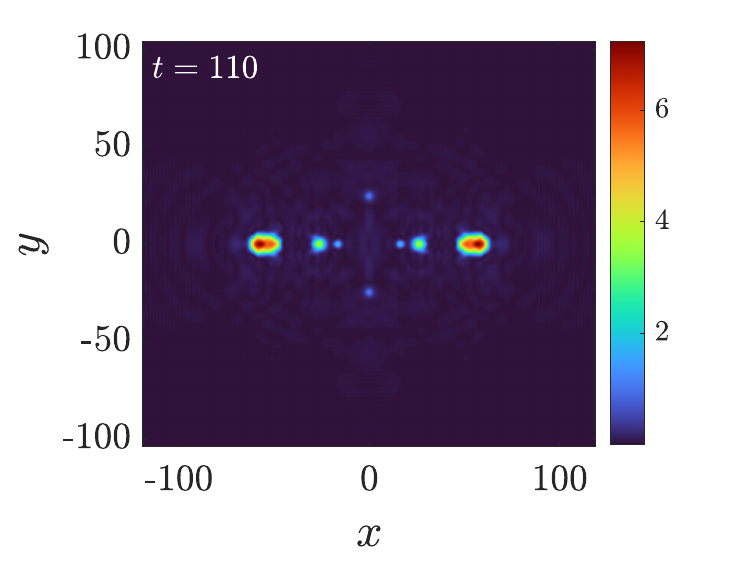} \\
\end{tabular}%
\caption{Square root of the power density plots showing the evolution of the head-on collisions of 2D solitons at different values of time for $\omega=0.95$ and $c=0.68$. Multiple splinters
along the original collision direction are accompanied by a small portion of the mass emitted
along the transverse direction. See also companion movie \texttt{movie\_07.gif}.}
\label{fig:simul2D_transverse1}
\end{figure}

\begin{figure}[!htbp]
\begin{tabular}{cccc}
\includegraphics[width=\figfourwidth]{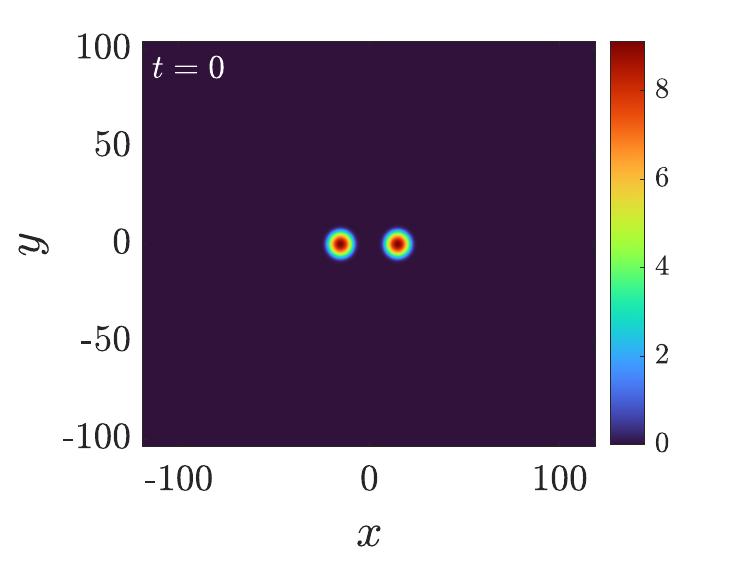} &
\includegraphics[width=\figfourwidth]{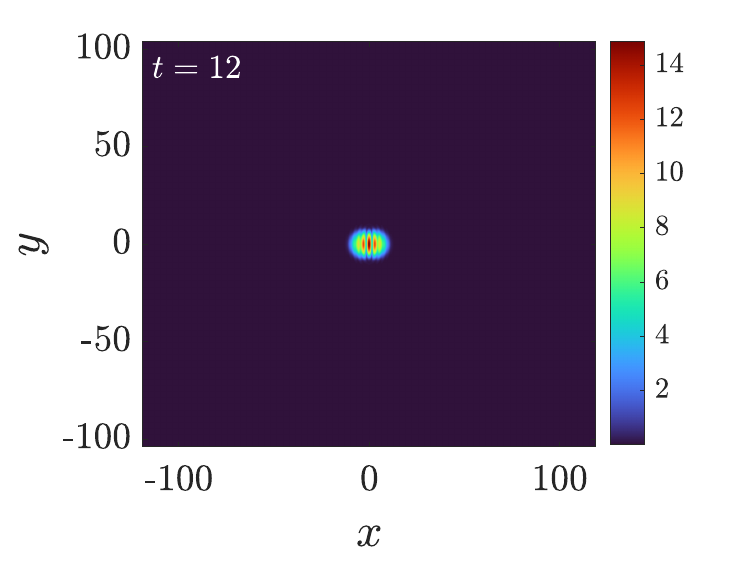} &
\includegraphics[width=\figfourwidth]{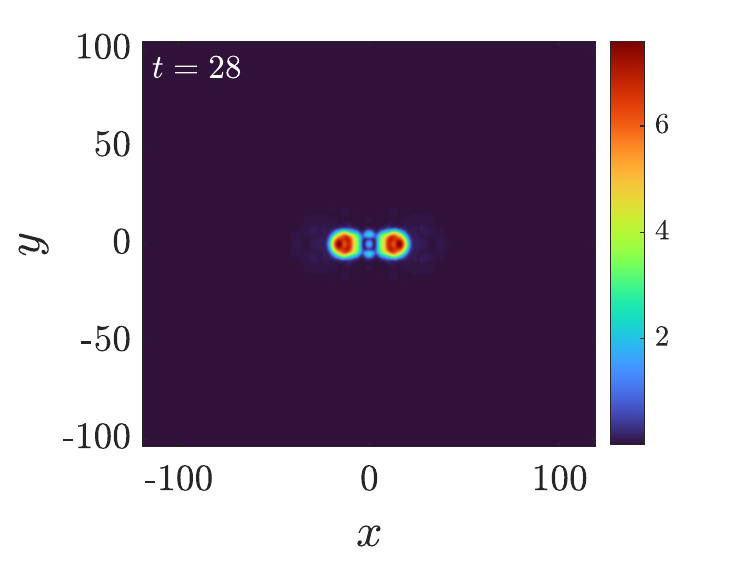} &
\includegraphics[width=\figfourwidth]{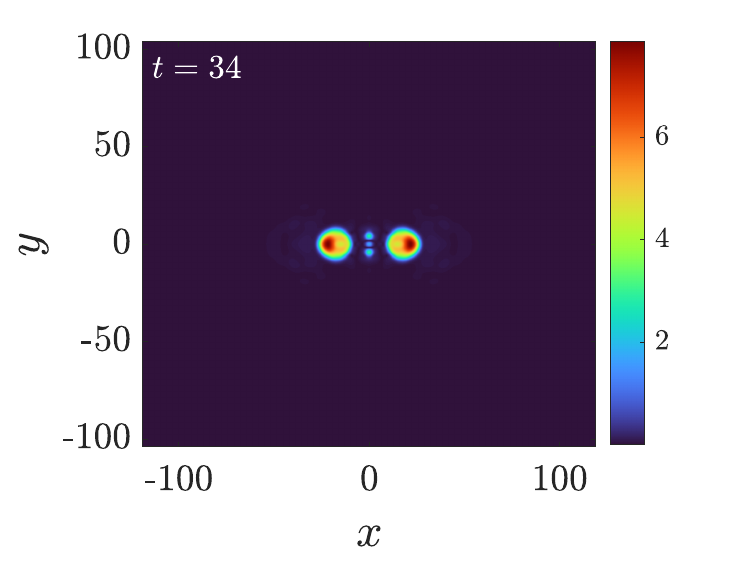} \\
\includegraphics[width=\figfourwidth]{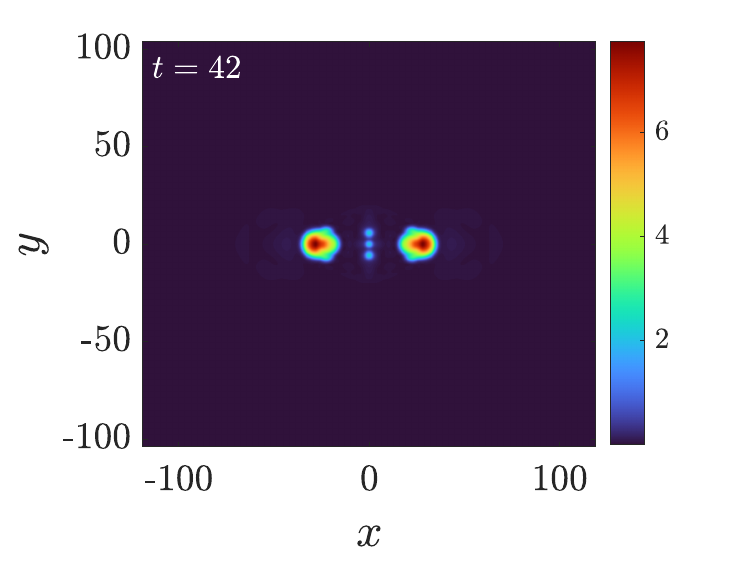} &
\includegraphics[width=\figfourwidth]{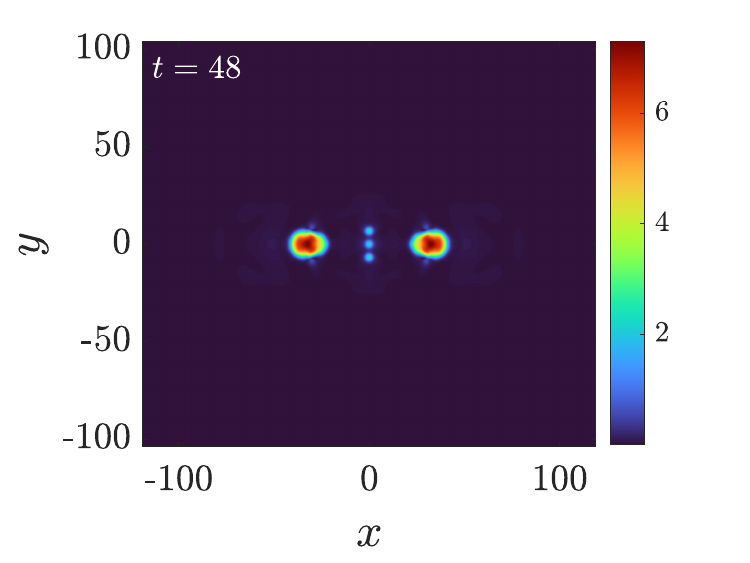} &
\includegraphics[width=\figfourwidth]{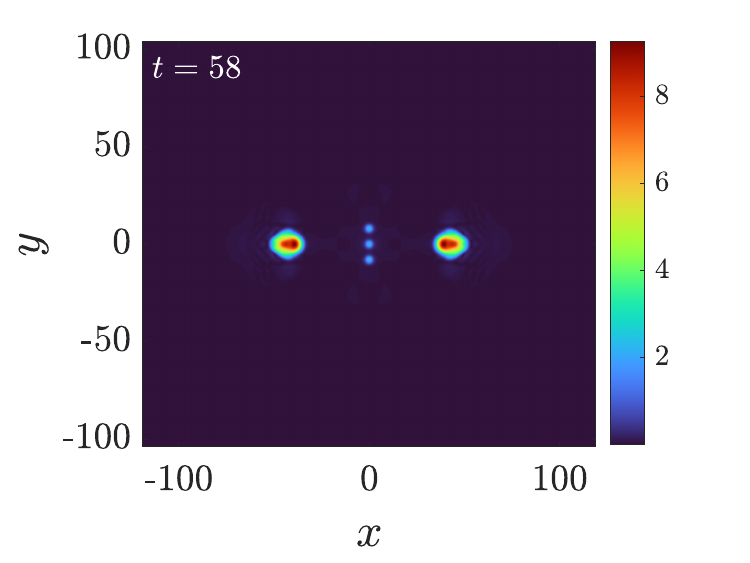} &
\includegraphics[width=\figfourwidth]{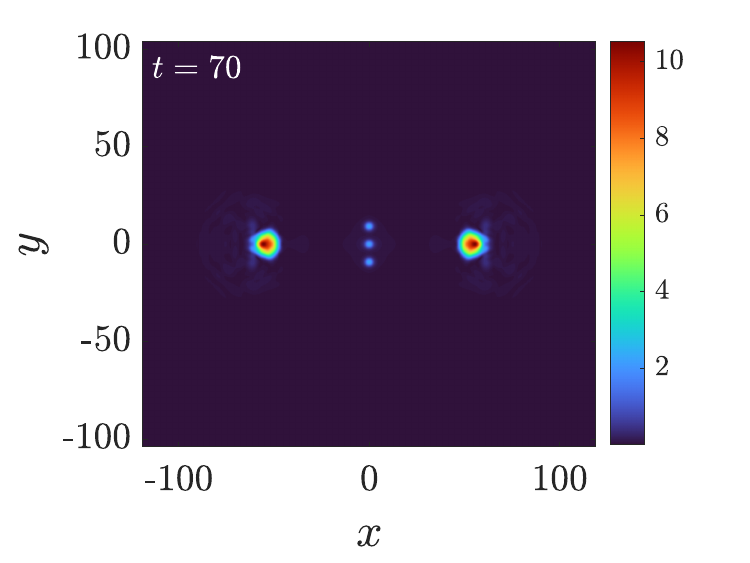} \\
\end{tabular}%
\caption{Square root of the power density plots showing the evolution of the head-on collisions of 2D solitons at different values of time for $\omega=0.96$ and $c=1$. Similar to the previous figure,
as concerns the transverse emission, but now a portion of the mass stays localized at the collision
location as well. See also companion movie \texttt{movie\_08.gif}.}
\label{fig:simul2D_transverse2}
\end{figure}

\begin{figure}[!htbp]
\begin{center}
\includegraphics[width=\figonewidth]{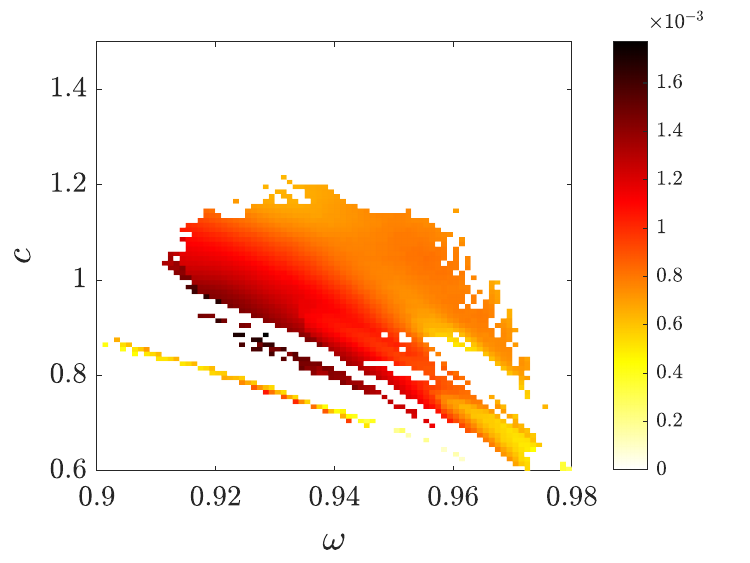}
\end{center}%
\caption{Quantity measuring the splitting rate $R$, given in (\ref{eq:split}), versus $\omega$ and $c$. This
shows the small portion of the mass emitted in the form of localized wavepackets along the
$y$-direction.}
\label{fig:splitting_rate}
\end{figure}

\begin{itemize}

\item {\em Trapping: } for low collision velocities $c$, the solitons get trapped and form a bound state. When the frequency is increased, the resulting bound state gets excited featuring
breathing oscillations. Here, the kinetic energy of the incoming solitary waves gets
transformed into internal oscillation (breathing) energy of the resulting fused, practically
stationary solitary structure. Figure~\ref{fig:simul2D_trapping1} shows the outcome for ($\omega=0.5$, $c=0.2$), which is supplemented by the movie of \texttt{movie\_01.gif}, whereas Fig.~\ref{fig:simul2D_trapping2} and movie \texttt{movie\_02.gif} do the same for ($\omega=0.9$, $c=0.25$). {See also points  (a) and (b) in Fig.~\ref{fig:carpet} and the corresponding panels in Fig.~\ref{fig:isosurfaces}.}
    
\item {\em Dispersion: } when $\omega$ is small ($\lesssim0.2$) and the velocity is high enough, the bound state increases its width and, at the same time, its amplitude decreases; see Fig.~\ref{fig:simul2D_dispersion} and \texttt{movie\_03.gif} for $\omega=0.1$ and $c=0.5$, and Fig.~\ref{fig:simul2D_dispersiontrans} and \texttt{movie\_04.gif} for $\omega=0.03$ and $c=0.29$. In the former case, the solitons are reflected in the incoming direction, but in the latter case, the solitons are deflected in the transversal direction. 
In both cases, the short time of the interaction, in conjunction with the low mass of the incoming solitary waves does not accomplish the formation of a localized bound state and eventually leads to dispersion. {See also points (c) and (d) in Fig.~\ref{fig:carpet} and the corresponding panels in Fig.~\ref{fig:isosurfaces}.}
    
\item {\em Reflection: } if $\omega$ is not small and the speed is large enough, the solitons are reflected after the collision. This reflection could be accompanied by a remaining bound state if the frequency is large enough, i.e., if the system has sufficient mass to shed some of it in the
form of a localized excitation. Notice that there is no definitive critical speed separating both behaviors, as can be observed in Fig.~\ref{fig:carpet} especially for large enough $\omega$. Figure~\ref{fig:simul2D_reflection1} and \texttt{movie\_05.gif} show a typical reflection without a remaining bound state and Fig.~\ref{fig:simul2D_reflection2} and \texttt{movie\_06.gif} do the same for a typical reflection involving a remaining bound state. {See also points (e) and (f) in Fig.~\ref{fig:carpet} and the corresponding panels in Fig.~\ref{fig:isosurfaces}.}
    
\item {\em Transverse breaking: } the most intrguing of behaviors of the ponderomotive
system arises for large enough $\omega$ ($\gtrsim0.9$). In this case, 
the soliton is wide enough and transports enough power density so that, if the speed is also large, the solitons can be broken into smaller ones. This can lead to the appearance of splinters at a direction {\it transversal} to the original one, a feature generally quite uncommon (although not 
unprecedented~\cite{RUBACK1988669}) in nonlinear wave systems. These new splinters have an amplitude very much smaller than the original ones. In addition, they appear at very narrow regions in the $(\omega,c)$ plane. Figure~\ref{fig:simul2D_transverse1} and \texttt{movie\_07.gif} show a case where the original soliton pair breaks into three pairs moving in the $x$-direction and a new pair emerges in the $y$-direction (without a localized state emerging). On the other hand, Fig.~\ref{fig:simul2D_transverse2} and \texttt{movie\_08.gif} correspond to a case where only a bound state and a pair of solitons in the $y$-direction emerge, apart from the reflected solitons in the $x$ direction. {See also points (g) and (h) in Fig.~\ref{fig:carpet} and the corresponding panels in Fig.~\ref{fig:isosurfaces}}. From Fig.~\ref{fig:carpet} one can observe that the transverse breaking region disappears when $c\gtrsim1.2$, i.e., for sufficiently fast-moving solitons. Figure~\ref{fig:splitting_rate} shows the dependence with respect to $\omega$ and $c$ of the following quantity    
    \begin{equation}\label{eq:split}
    R=\frac{2}{P}\lim_{t\rightarrow\infty}\int_{-W}^{W}\mathrm{d}x\int_{W}^{\infty}\mathrm{d}y|\psi(\mathbf{r},t)|^2
    \end{equation}
    with $W$ being the width of each splinter and $P$ being the soliton's norm. This quantity is a measure of the density split in the $y$ direction. As can be observed, this is a small fraction of the systems original power,
    yet a finite, observable one. 

\end{itemize}

\FloatBarrier

\section{Conclusions and Future Challenges}

In the  present work we have revisited an intriguing system modeling the dynamics of
ponderomotive focusing of electromagnetic waves in plasmas. These types of effects
are of relevance in plasma compression or in plasma column heating among others. 
The effective model put forth in the work of~\cite{tappert77} is that of an 
exponentially decaying (with the intensity of the beam) nonlinearity, which, in 
addition to its physical interest for plasmas, bears also significant mathematical
value in the realm of dispersive nonlinear wave equations of the nonlinear Schr{\"o}dinger
type. This is because the well-known cubic focusing effect leads to catastrophic collapse
in higher dimensions, whereas the present exponential decay of the nonlinearity with intensity
---similarly to what is known, e.g., for saturable or cubic-quintic nonlinearities which
have been argued to be of relevance to optics, among other fields--- is sufficient to mitigate
the collapse features and to, instead, generate in higher dimensions, as also in one spatial
dimension, stable non-topological solitary waves. In our study we systematically brought to bear
a spectral stability analysis of such waves in one- and two-dimensions, while revealing
that other structures, including vortical and line soliton ones are dynamically unstable
in the present model. This prompted us to explore a topic of considerable interest 
---even experimentally~\cite{Meng:97,wieslaw}--- for nonlinear wave models bearing stable
non-topological solitary waves in higher dimensions: namely. the collision of two wave structures. 
We have showcased a systematic analysis of such collisional outcomes as a function of the
frequency and the speed of the colliding waves. We found that, in addition to the
fusion (for low speeds) or dispersion (for high speeds, but low frequencies), there are 
further effects involving reflection (intermediate frequencies and higher speeds),
reflection with a remnant wave and finally some intriguing events (for high speeds and
frequencies) including the emission of wavepackets in the transverse direction, a feature
quite unusual (although not unprecedented) in nonlinear wave settings. 
{We have also performed several numerical experiments on the collisions of out-of-phase solitons. For a phase difference of $\pi$, both 1D and 2D solitons are repelled after collisions. A phase difference different from $0$ or $\pi$ implies a spatial symmetry breaking of the system and the outcome becomes unpredictable and extremely complex.}

The present work naturally paves the way for numerous further studies of the
ponderomotive nonlinearity, which is of particular interest in its own right both
from a mathematical point of view (e.g., establishing the well-posedness features of
the relevant model), but also from a physical perspective. The original work of~\cite{tappert77}
obtained (in addition to invariants) moment-like equations for the evolution of
the solitary waves. Generalizing such considerations to a systematic variational
characterization of the solitary waves~\cite{MalomedV} would be of particular interest
in its own right. Lastly, we have limited our study to considerations of 1- and 2-spatial
dimensions, an analysis of 3D solitary waves and their dynamics/interactions could be
of interest in its own right. Such studies are currently in progress and will be
described in future publications.

\appendix

\section{Unstable solitons}

We briefly also discuss the evolution of unstable vortices. In all the cases shown below, instabilities are oscillatory, and the dynamics conserve the angular momentum, given by

\begin{equation}
    L_z=\frac{1}{P}\int \psi^*(\br)\left(\br\times\nabla\right)\psi(\br) \mathrm{d}\br
\end{equation}

\begin{figure}[!htbp]
\begin{tabular}{cccc}
\includegraphics[width=\figfourwidth]{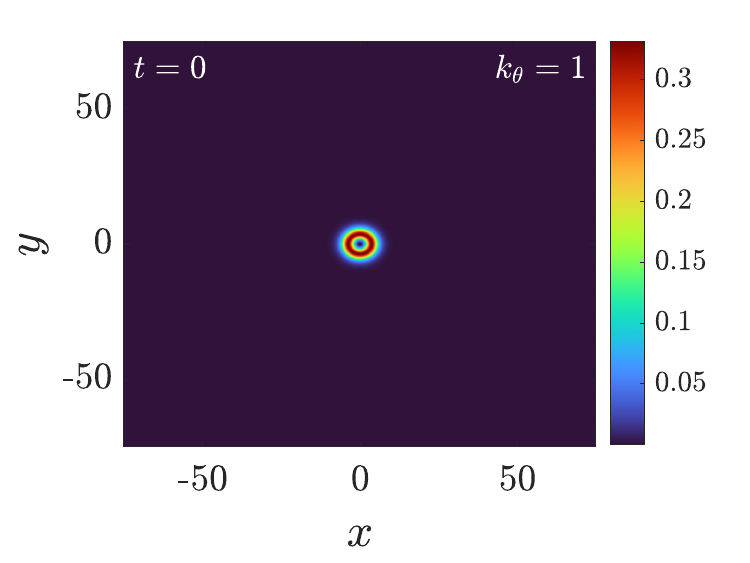} &
\includegraphics[width=\figfourwidth]{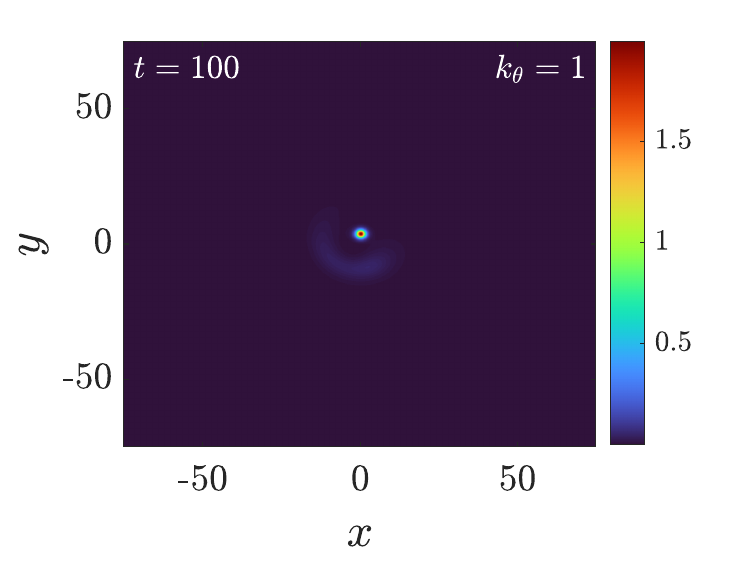} &
\includegraphics[width=\figfourwidth]{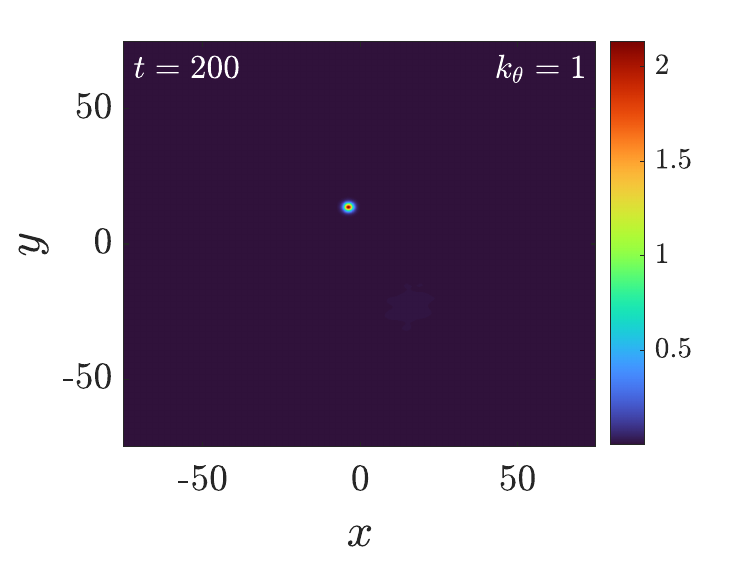} &
\includegraphics[width=\figfourwidth]{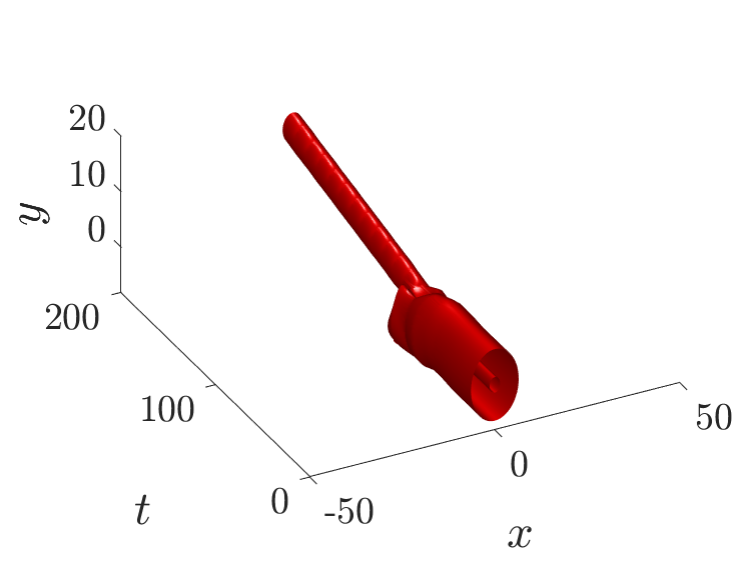} \\
\includegraphics[width=\figfourwidth]{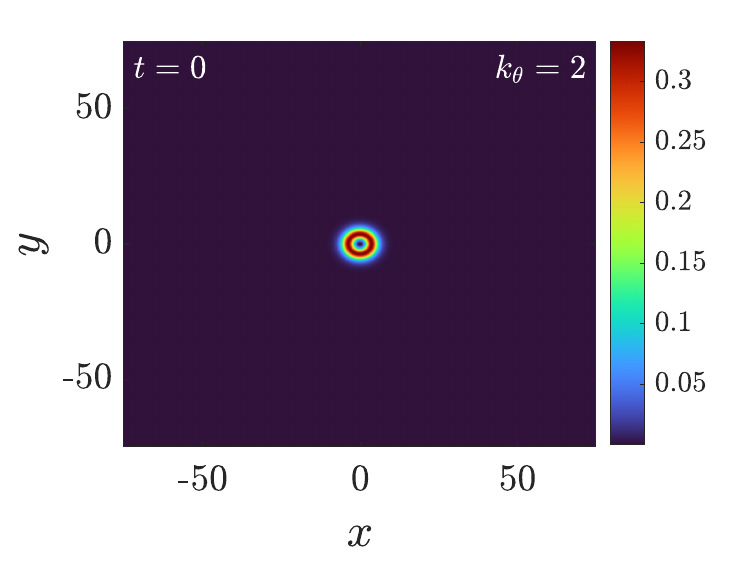} &
\includegraphics[width=\figfourwidth]{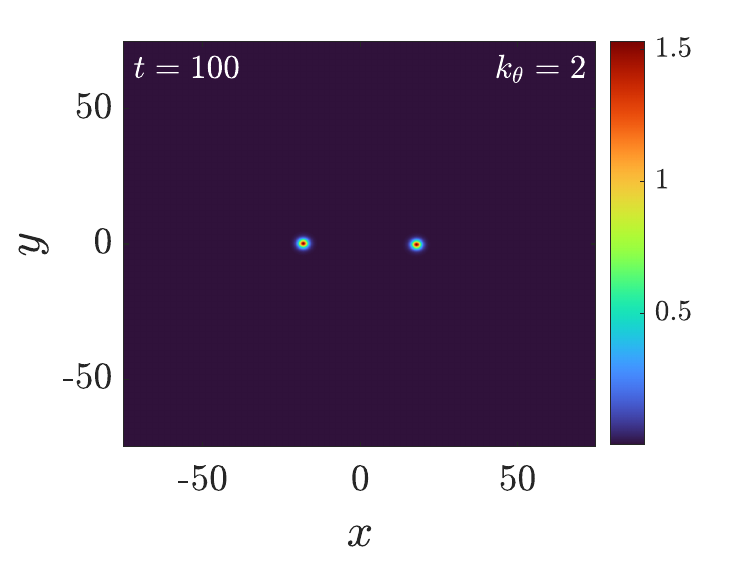} &
\includegraphics[width=\figfourwidth]{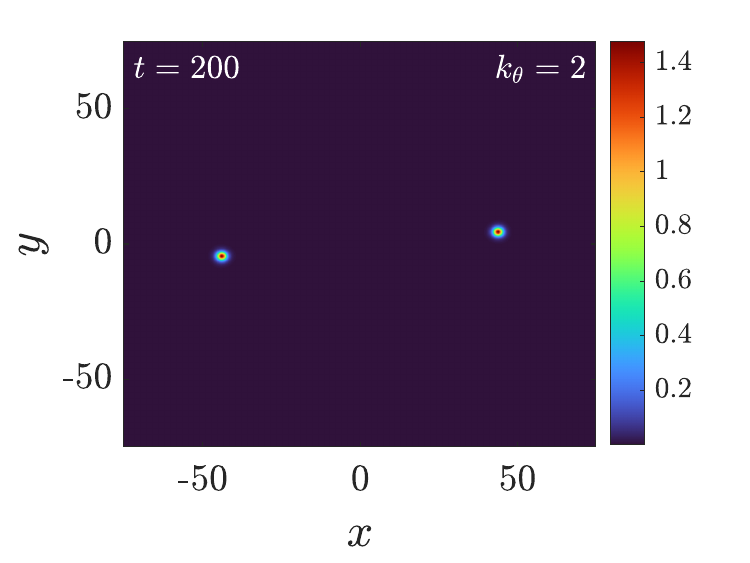} &
\includegraphics[width=\figfourwidth]{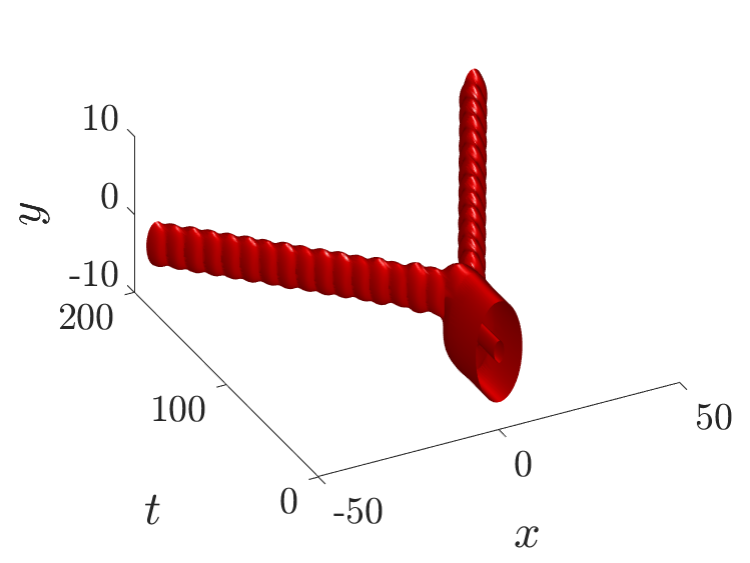} \\
\includegraphics[width=\figfourwidth]{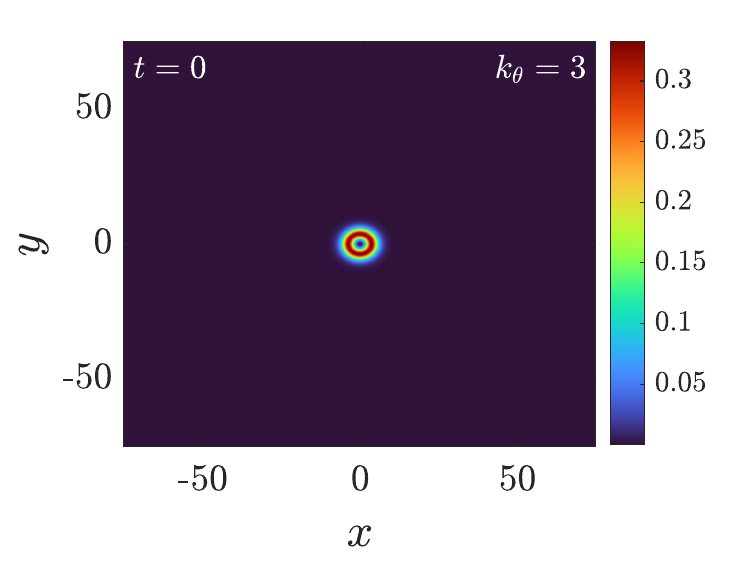} &
\includegraphics[width=\figfourwidth]{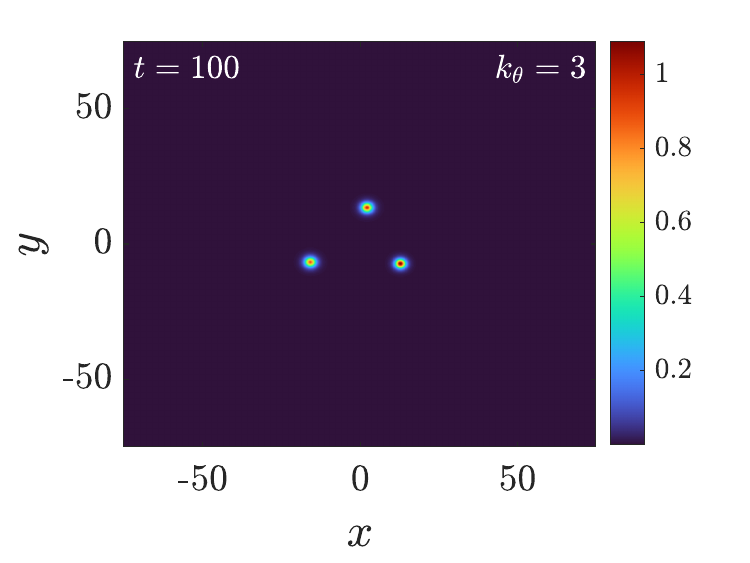} &
\includegraphics[width=\figfourwidth]{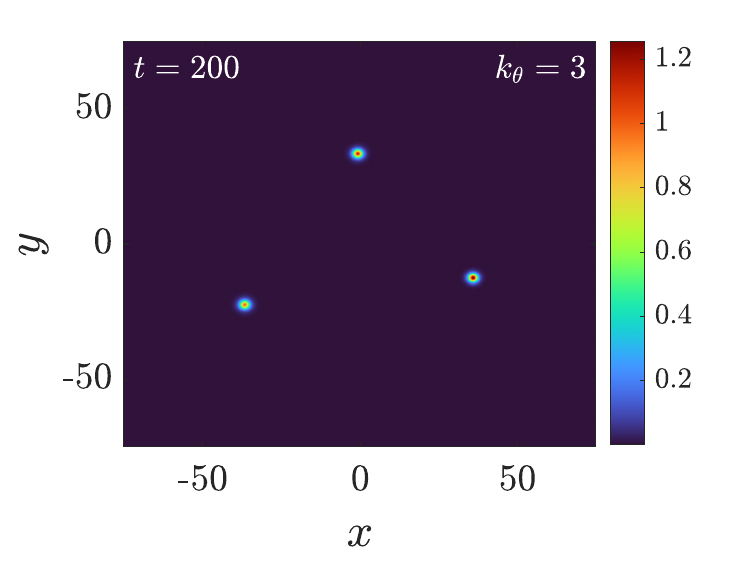} &
\includegraphics[width=\figfourwidth]{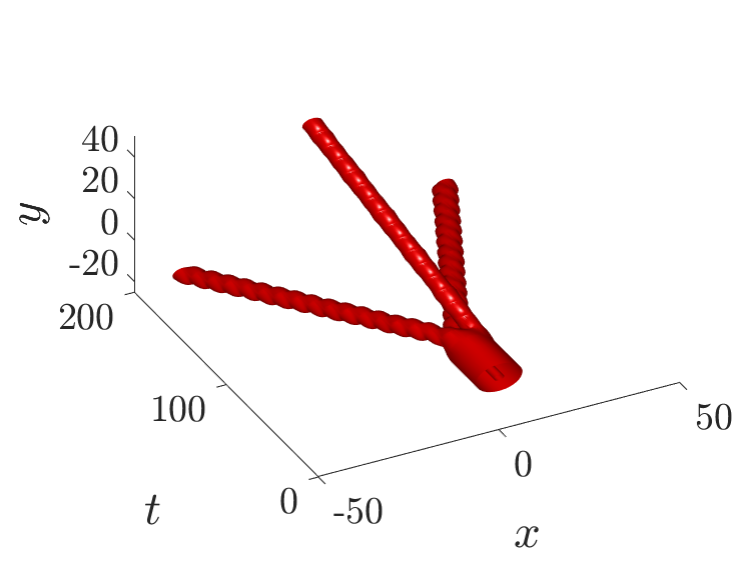} \\
\end{tabular}%
\caption{Power density plots showing the evolution of unstable topological solitons with $S=1$ and $\omega=0.1$ when they are perturbed along the direction of the azimuthal instability eigenmode with $k_\theta=1$ (top panels), $k_\theta=2$ (middle panels) and $k_\theta=3$ (bottom panels), which can be followed in more detail at \texttt{movie\_09.gif}, \texttt{movie\_10.gif} and \texttt{movie\_11.gif} respectively. Rightmost panels show the isosurfaces of the power density $|\psi(x,y,t)|^2$ when it is equal to 1/5th of the maximum density of the corresponding stationary topological soliton.}
\label{fig:simulS1}
\end{figure}

For $S=1$ vortices, there are azimuthal instabilities caused by eigenmodes with $k_{\theta}=2$ for every value of $\omega$, whereas for $\omega\leq0.41$ ($\omega\leq0.42$), there are also instabilities by $k_{\theta}=1$ ($k_{\theta}=3$) eigenmodes. We show in Fig.~\ref{fig:simulS1} (and the companion files \texttt{movie\_09.gif}, \texttt{movie\_10.gif} and \texttt{movie\_11.gif}) the evolution of the $S=1$ vortex with $\omega=0.1$ when is perturbed by the mode with $k_\theta=1$, $k_\theta=2$ and $k_\theta=3$, respectively. Notice that the vortex transforms in $k_\theta$ travelling 
single hump solitary waves which are dynamically
robust in this setting.

\begin{figure}[!htbp]
\begin{tabular}{cccc}
\includegraphics[width=\figfourwidth]{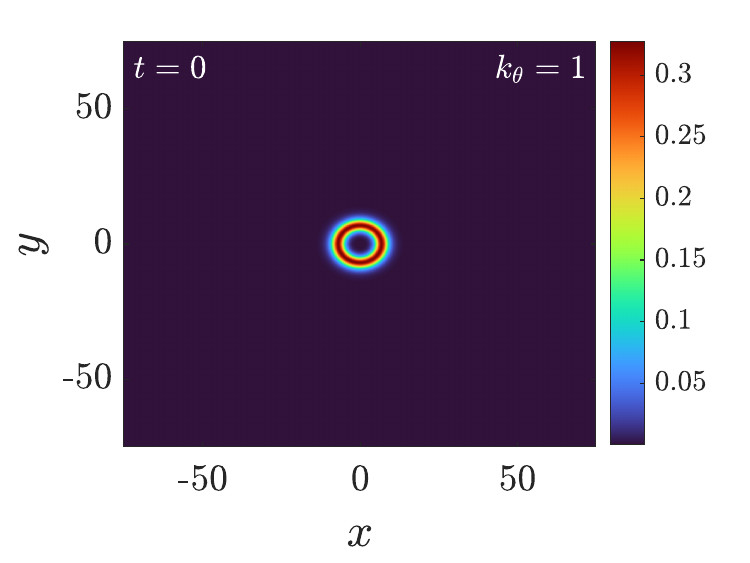} &
\includegraphics[width=\figfourwidth]{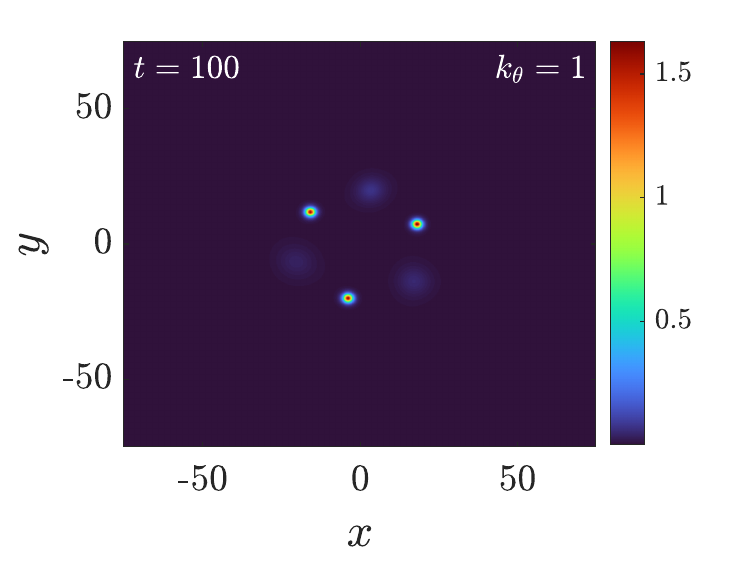} &
\includegraphics[width=\figfourwidth]{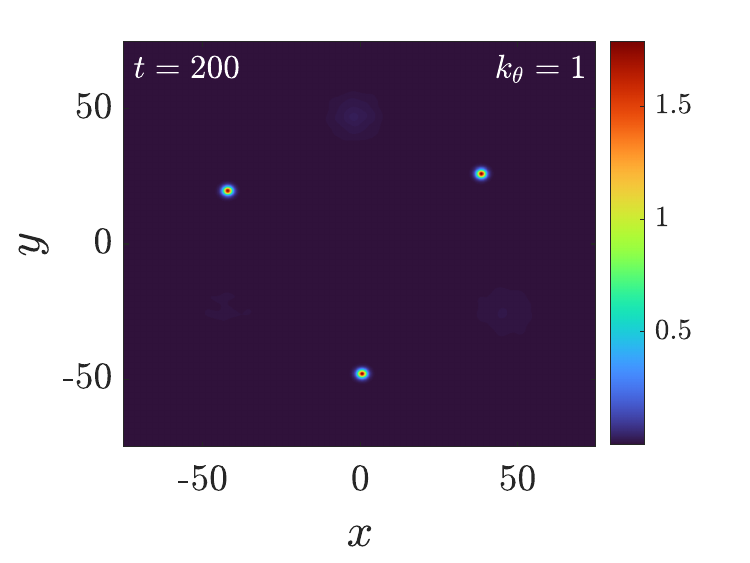} &
\includegraphics[width=\figfourwidth]{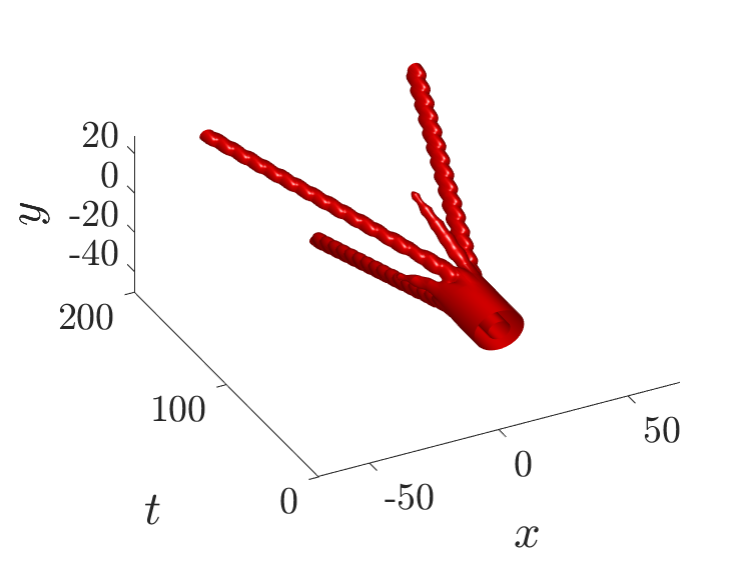} \\
\includegraphics[width=\figfourwidth]{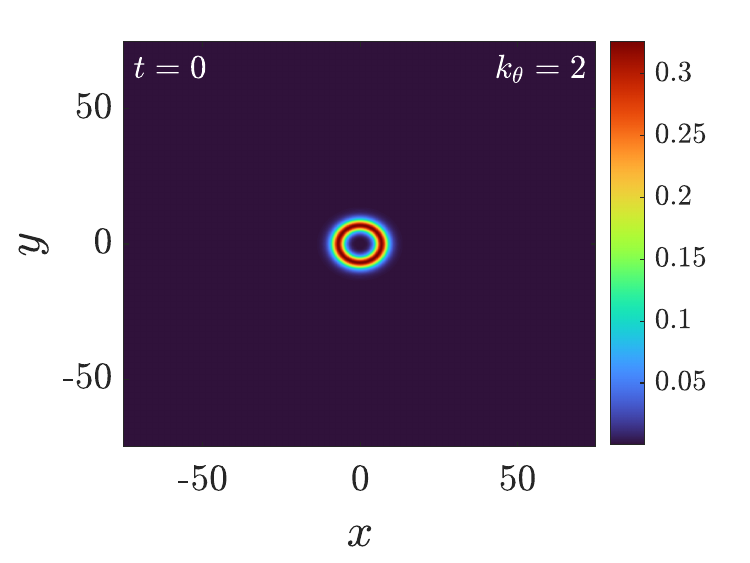} &
\includegraphics[width=\figfourwidth]{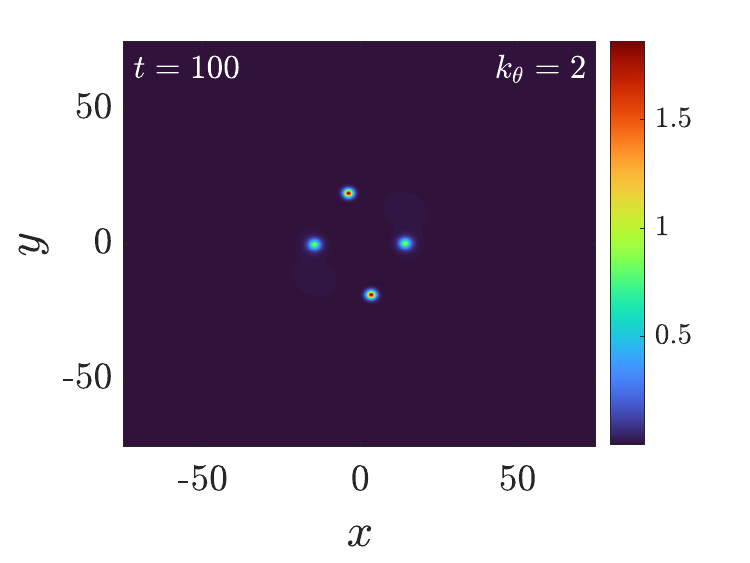} &
\includegraphics[width=\figfourwidth]{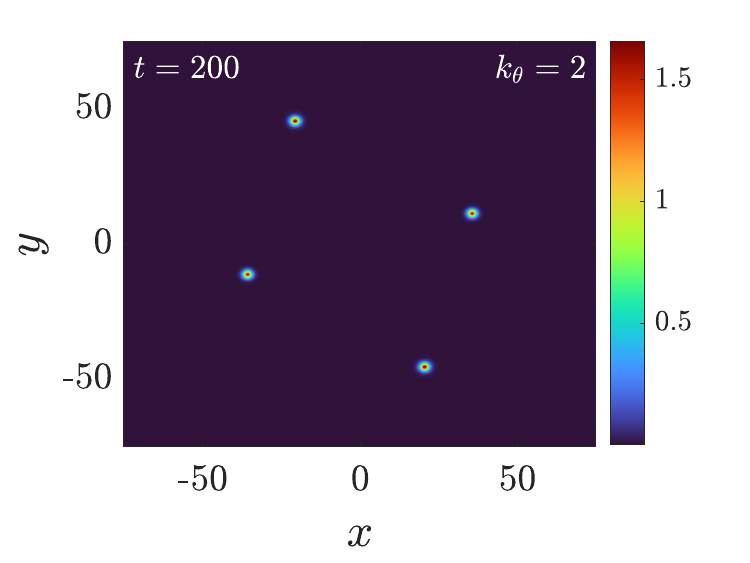} &
\includegraphics[width=\figfourwidth]{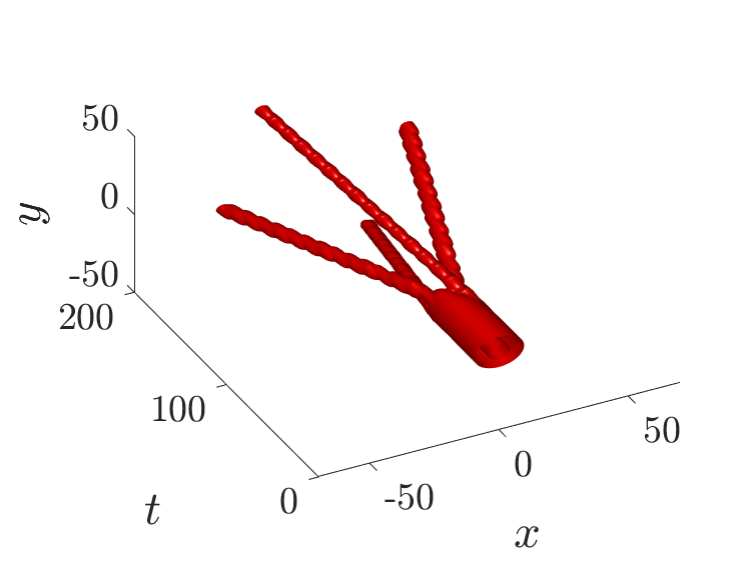} \\
\includegraphics[width=\figfourwidth]{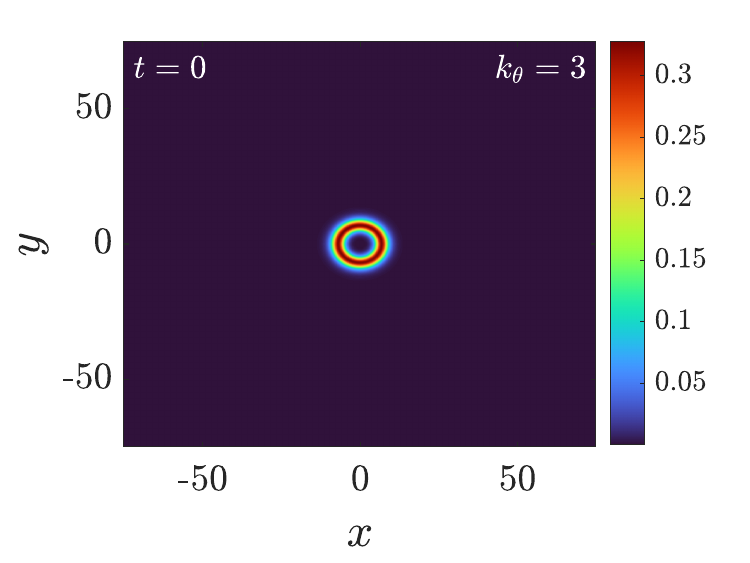} &
\includegraphics[width=\figfourwidth]{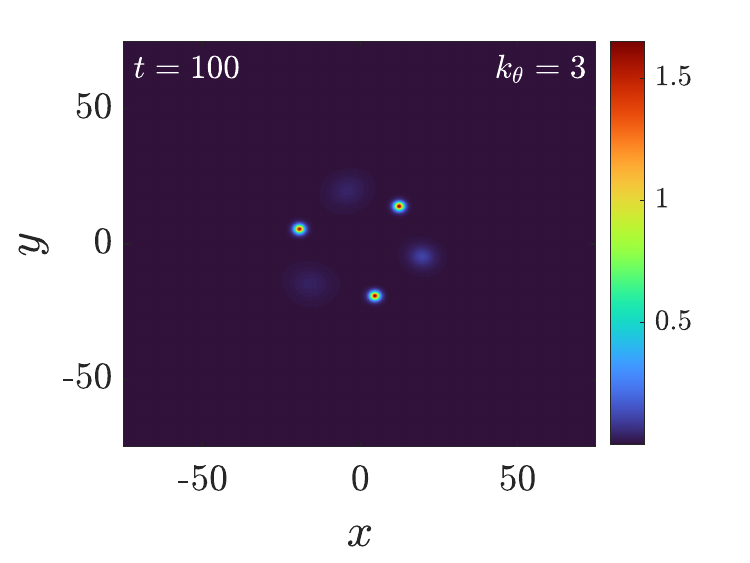} &
\includegraphics[width=\figfourwidth]{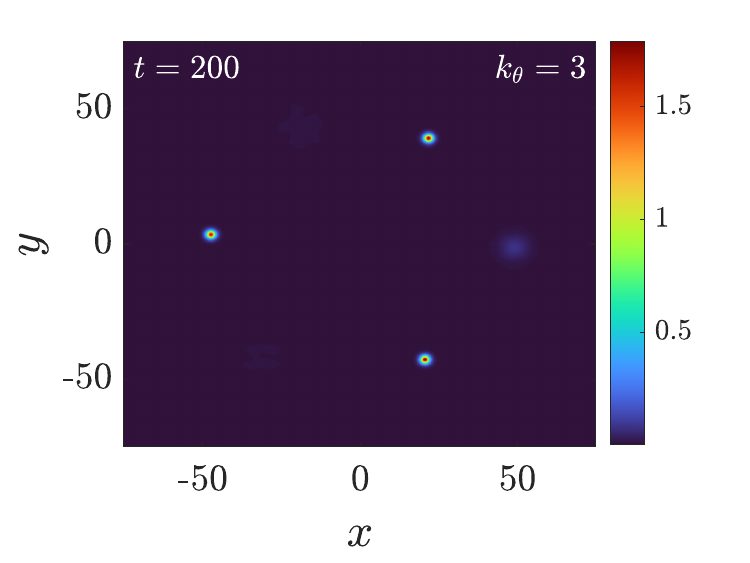} &
\includegraphics[width=\figfourwidth]{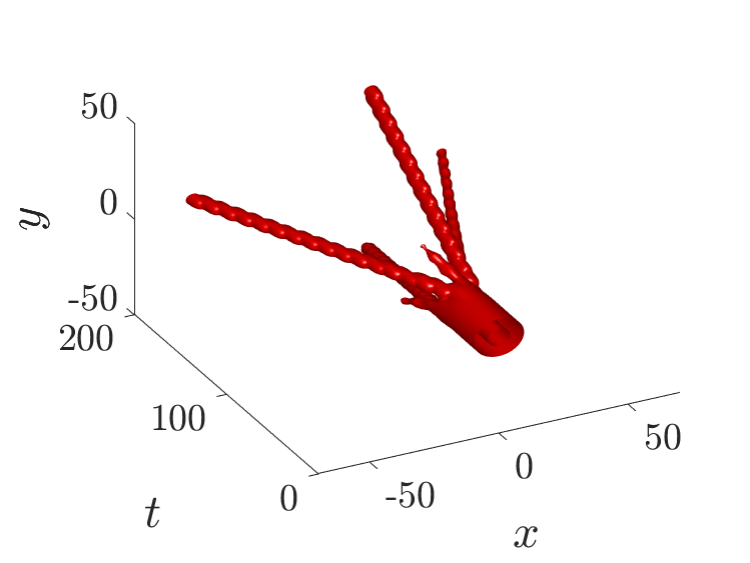} \\
\end{tabular}%
\caption{Power density plots showing the evolution of unstable topological solitons with $S=2$ and $\omega=0.1$ when they are perturbed along the direction of the azimuthal instability eigenmode with $k_\theta=1$ (top panels), $k_\theta=2$ (middle panels) and $k_\theta=3$ (bottom panels), which can be followed in more detail at \texttt{movie\_12.gif}, \texttt{movie\_13.gif} and \texttt{movie\_14.gif} respectively. Rightmost panels show the isosurfaces of the power density $|\psi(x,y,t)|^2$ when it is equal to 1/5th of the maximum density of the corresponding stationary topological soliton.}
\label{fig:simulS2a}
\end{figure}

\begin{figure}[!htbp]
\begin{tabular}{cccc}
\includegraphics[width=\figfourwidth]{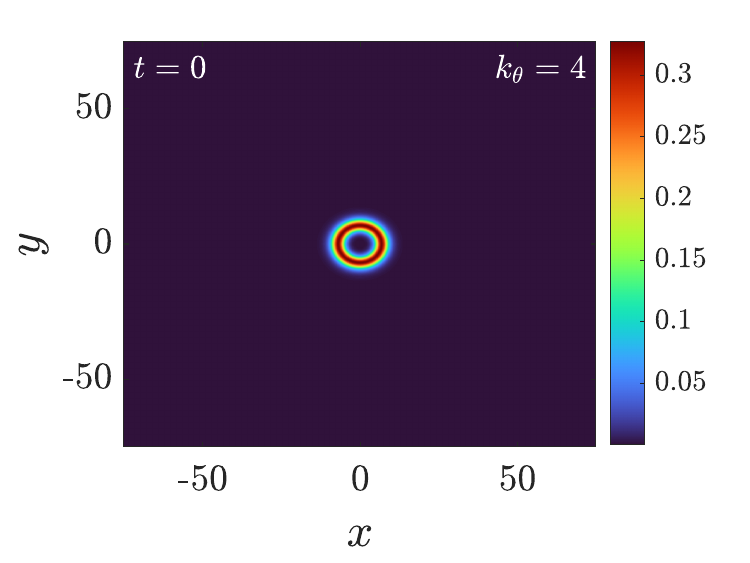} &
\includegraphics[width=\figfourwidth]{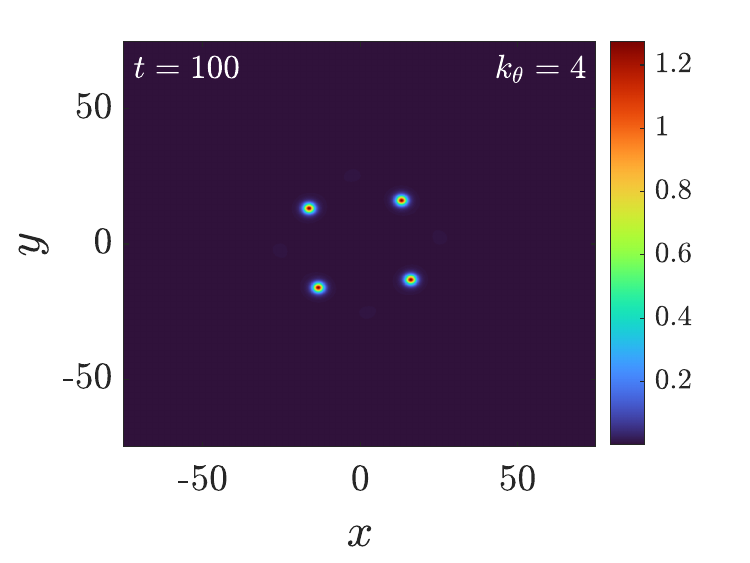} &
\includegraphics[width=\figfourwidth]{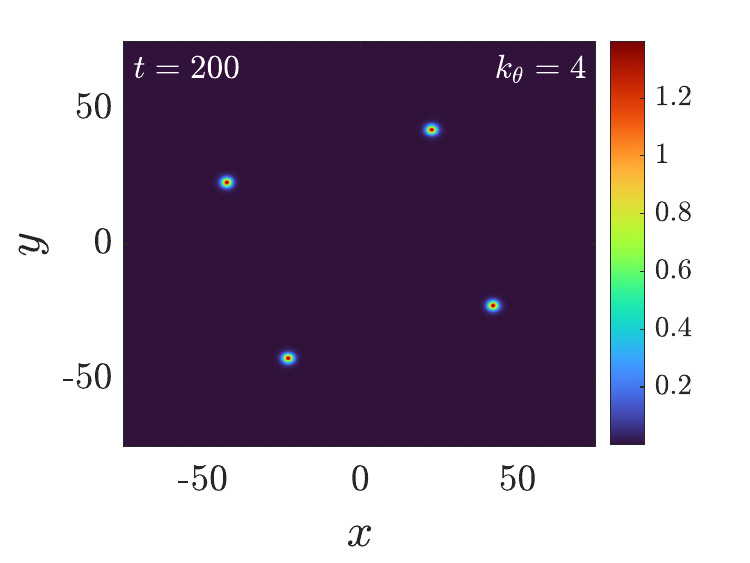} &
\includegraphics[width=\figfourwidth]{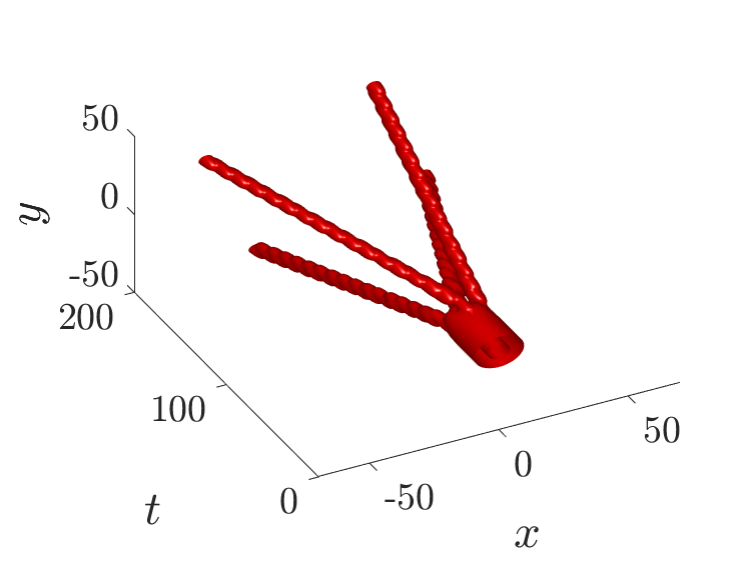} \\
\includegraphics[width=\figfourwidth]{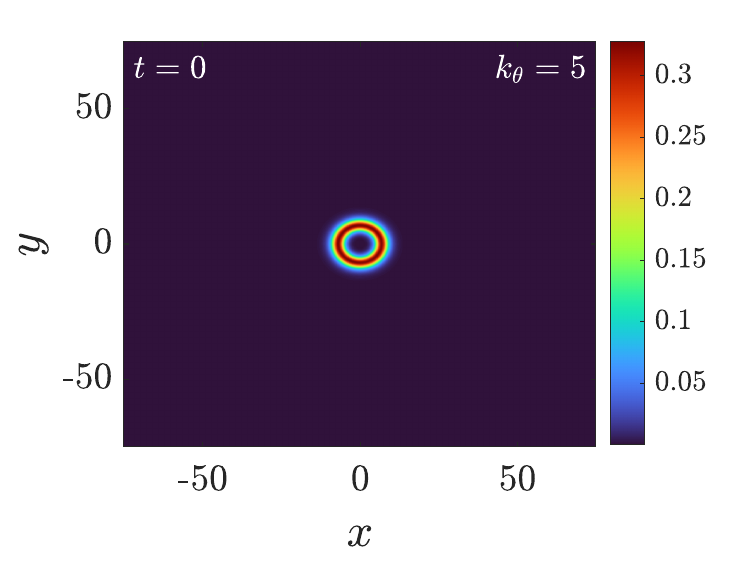} &
\includegraphics[width=\figfourwidth]{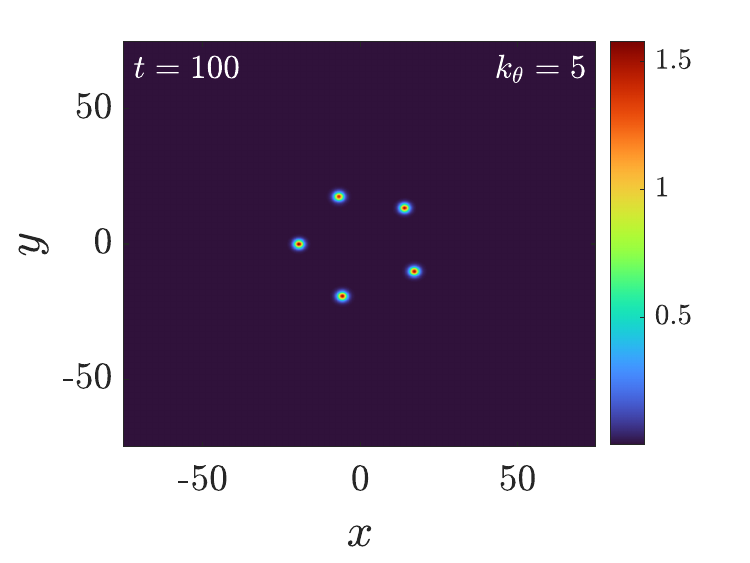} &
\includegraphics[width=\figfourwidth]{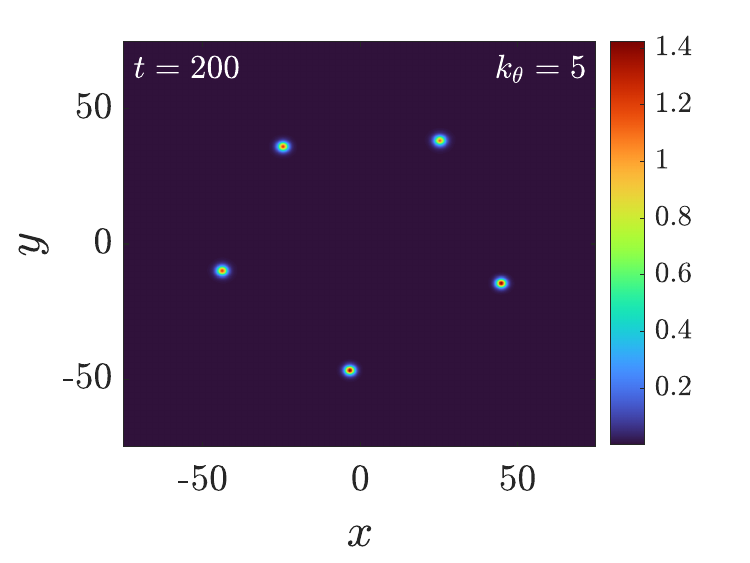} &
\includegraphics[width=\figfourwidth]{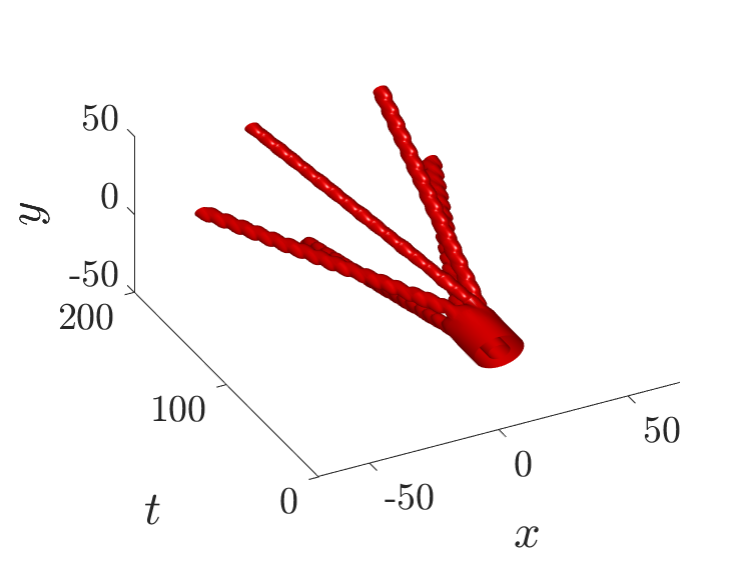} \\
\includegraphics[width=\figfourwidth]{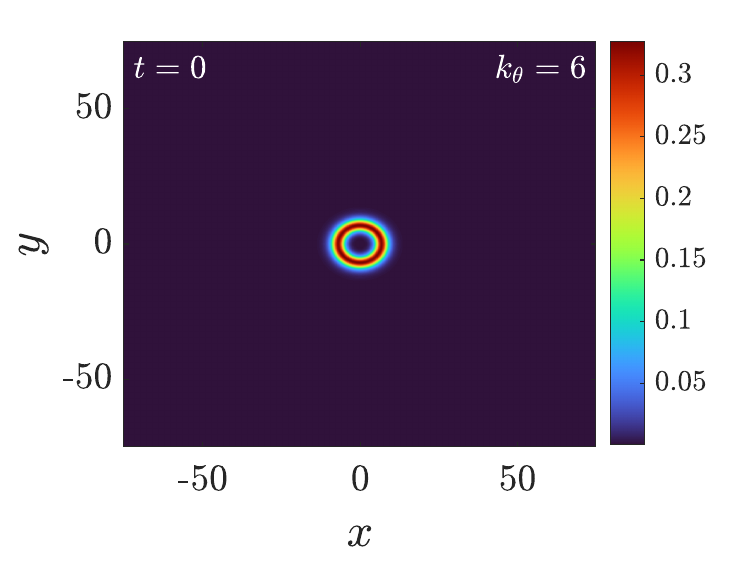} &
\includegraphics[width=\figfourwidth]{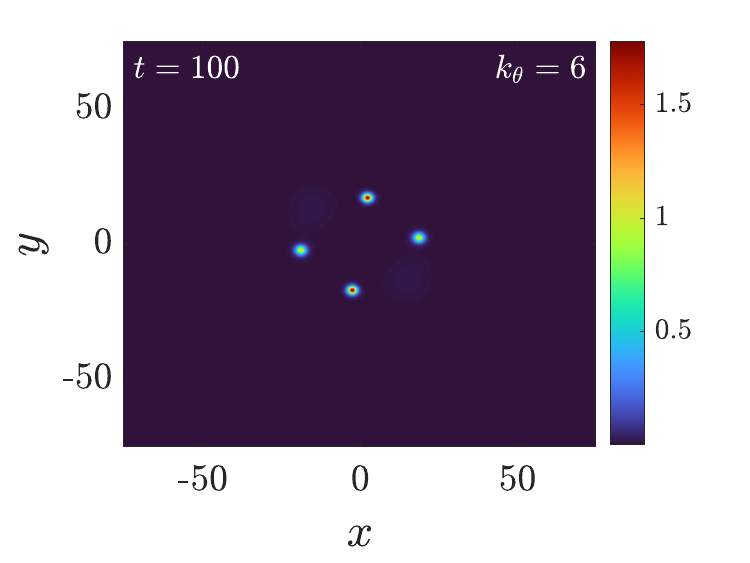} &
\includegraphics[width=\figfourwidth]{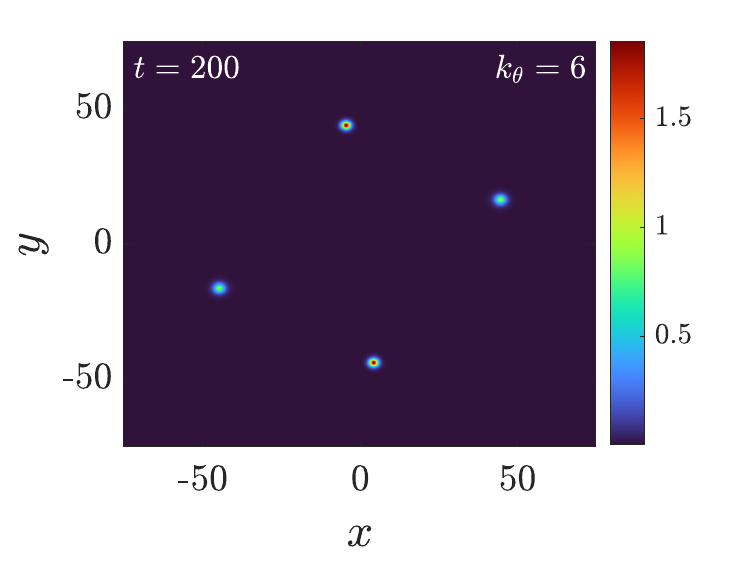} &
\includegraphics[width=\figfourwidth]{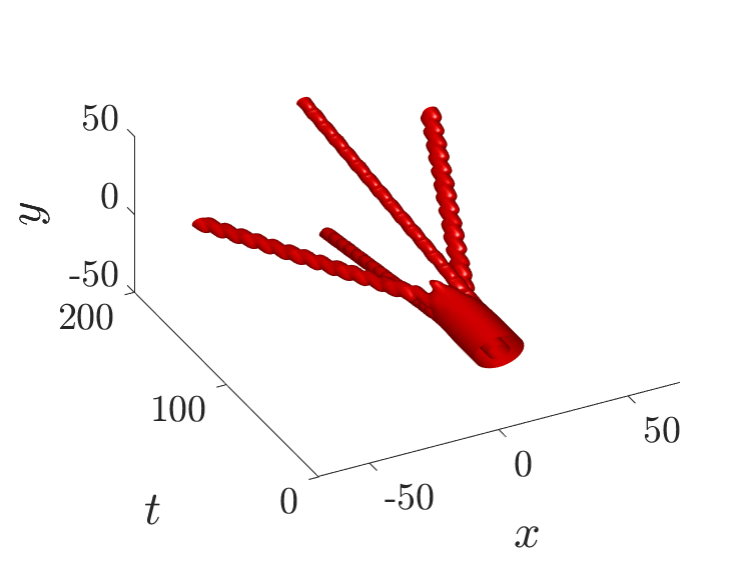} \\
\end{tabular}%
\caption{Power density plots showing the evolution of unstable topological solitons with $S=2$ and $\omega=0.1$ when they are perturbed along the direction of the azimuthal instability eigenmode with $k_\theta=4$ (top panels), $k_\theta=5$ (middle panels) and $k_\theta=6$ (bottom panels), which can be followed in more detail at \texttt{movie\_15.gif}, \texttt{movie\_16.gif} and \texttt{movie\_17.gif} respectively. Rightmost panels show the isosurfaces of the power density $|\psi(x,y,t)|^2$ when it is equal to 1/5th of the maximum density of the corresponding stationary topological soliton.}
\label{fig:simulS2b}
\end{figure}

For $S=2$ vortices, there are azimuthal instabilities for $k_\theta=2$ and $k_\theta=3$ for every $\omega$; for the rest of values of $k_\theta$, the maximum value for the existence of instabilities are $\omega=0.43$ ($k_\theta=1$), $\omega=0.66$ ($k_\theta=4$), $\omega=0.42$ ($k_\theta=5$) and $\omega=0,16$ ($k_\theta=6$). As in the $S=1$ case, we have analyzed the outcome of the vortex with $\omega=0.1$ (where all instabilities coexist) when it is perturbed with $k_\theta$ between 1 and 6 (see Figs.~\ref{fig:simulS2a} and \ref{fig:simulS2b}, together with the companion files from \texttt{movie\_12.gif} to \texttt{movie\_17.gif}). For even $k_\theta$, the number of generated radial solitons is always 4, while when $k_\theta=1$ or $3$, there are 3 generated solitons and only for $k_\theta=5$, the number of generated solitons is equal to $5$. We remark that the bias towards generation of four solitons is consistent with the $k_\theta = 4$ eigenvalue mode having the largest real part for this value of $\omega$.

\begin{figure}[!htbp]
\begin{tabular}{cccc}
\includegraphics[width=\figfourwidth]{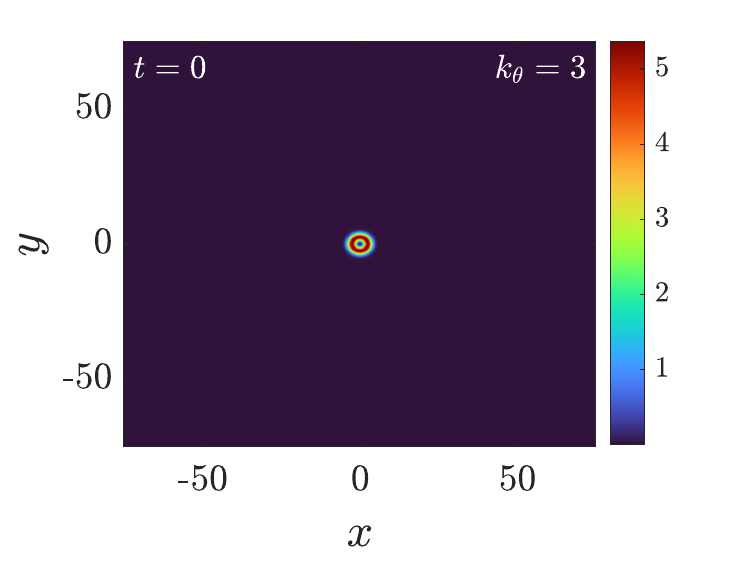} &
\includegraphics[width=\figfourwidth]{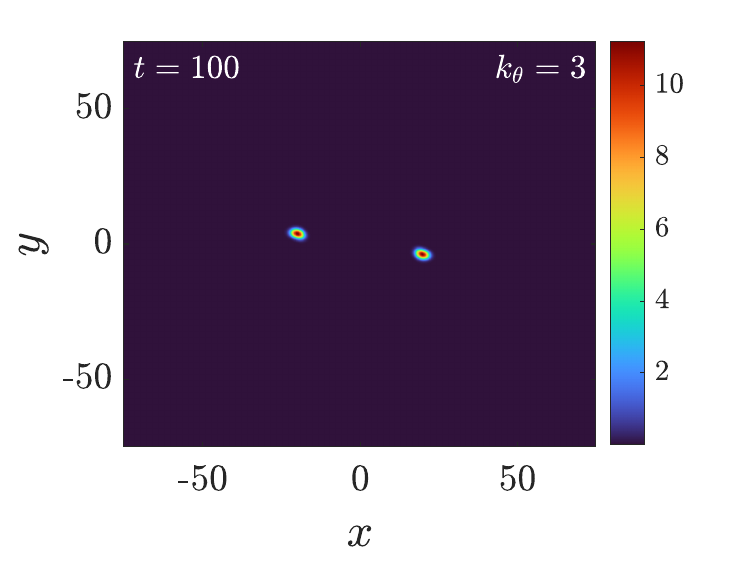} &
\includegraphics[width=\figfourwidth]{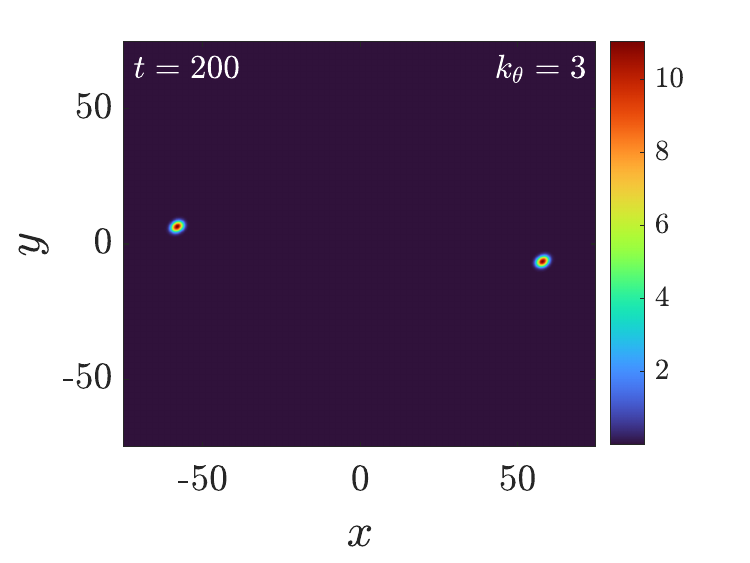} &
\includegraphics[width=\figfourwidth]{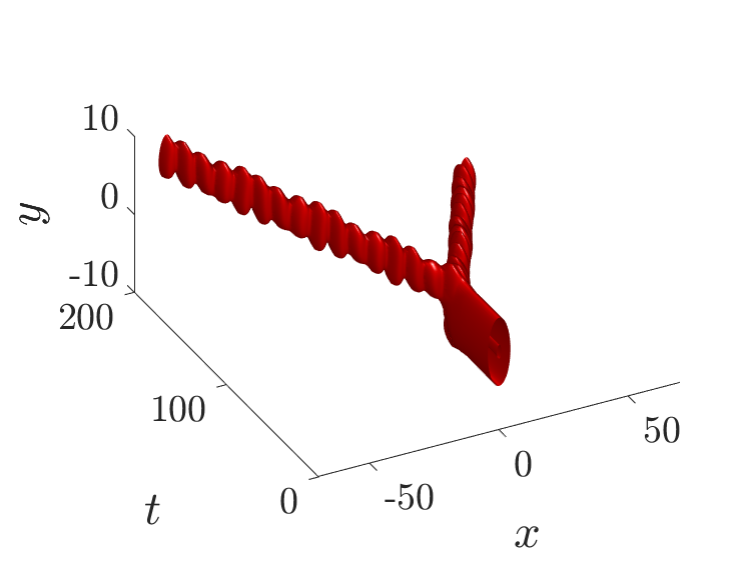} \\
\includegraphics[width=\figfourwidth]{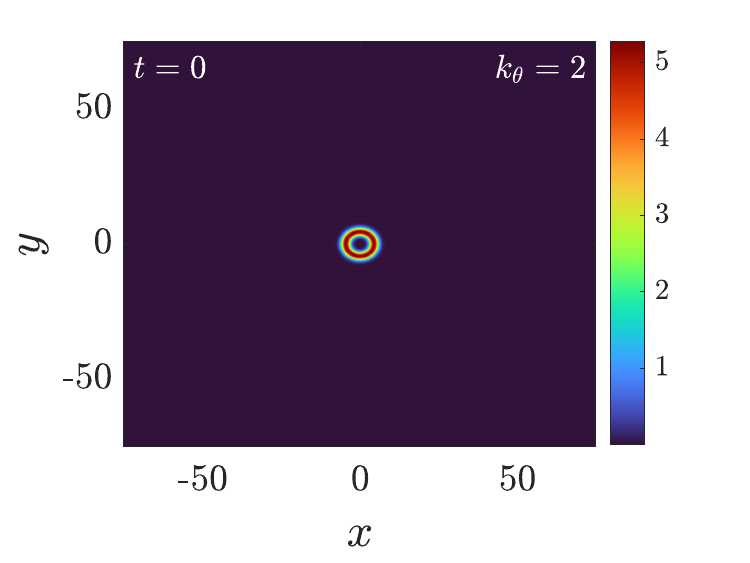} &
\includegraphics[width=\figfourwidth]{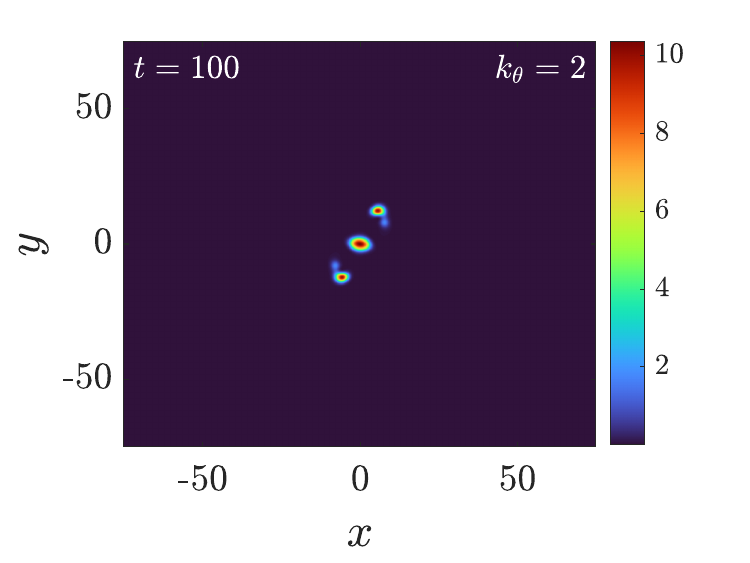} &
\includegraphics[width=\figfourwidth]{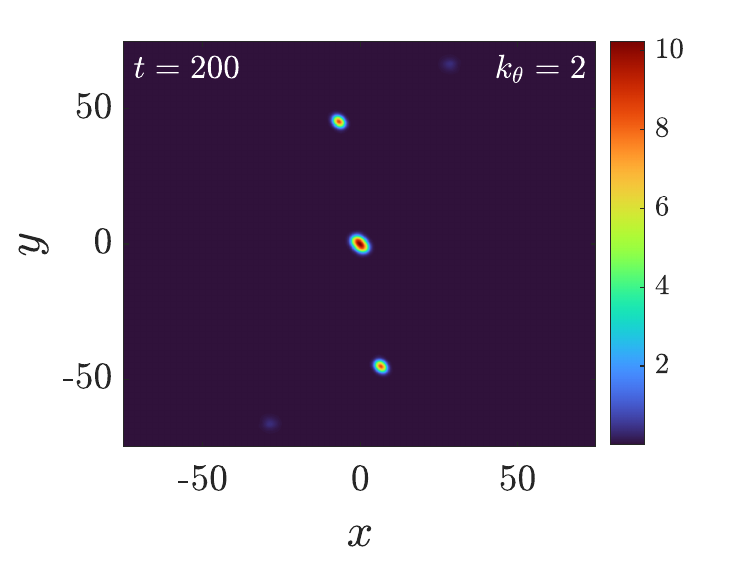} &
\includegraphics[width=\figfourwidth]{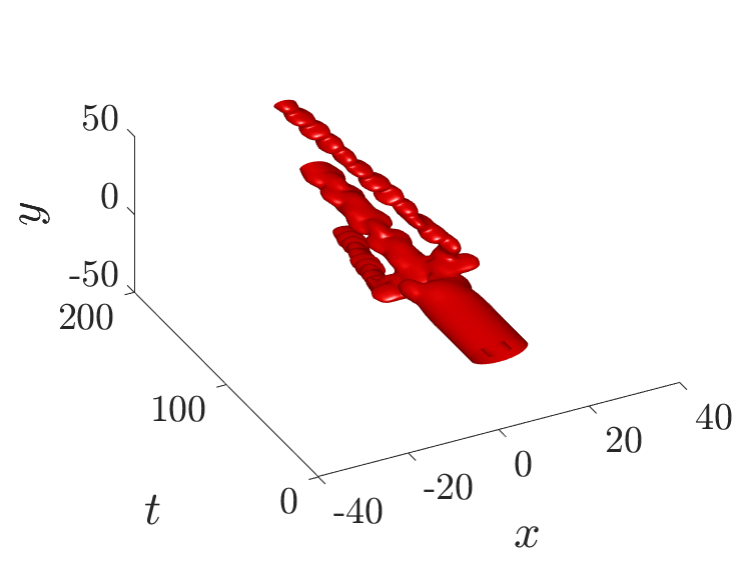} \\
\includegraphics[width=\figfourwidth]{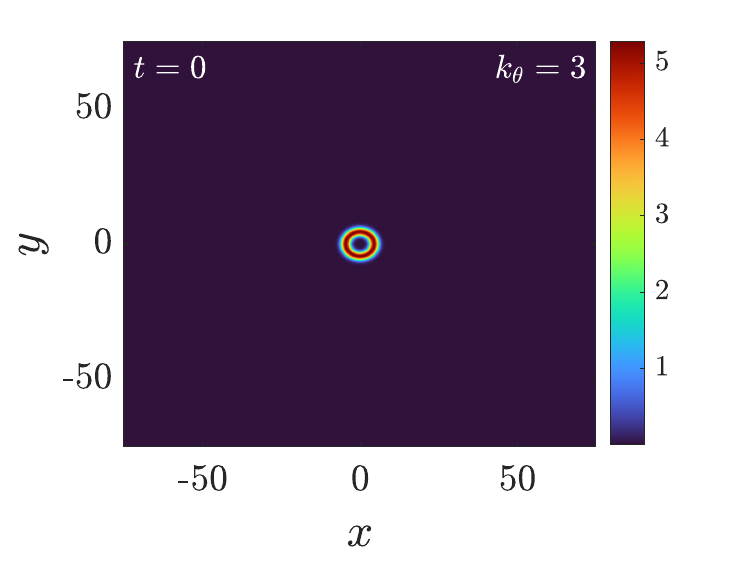} &
\includegraphics[width=\figfourwidth]{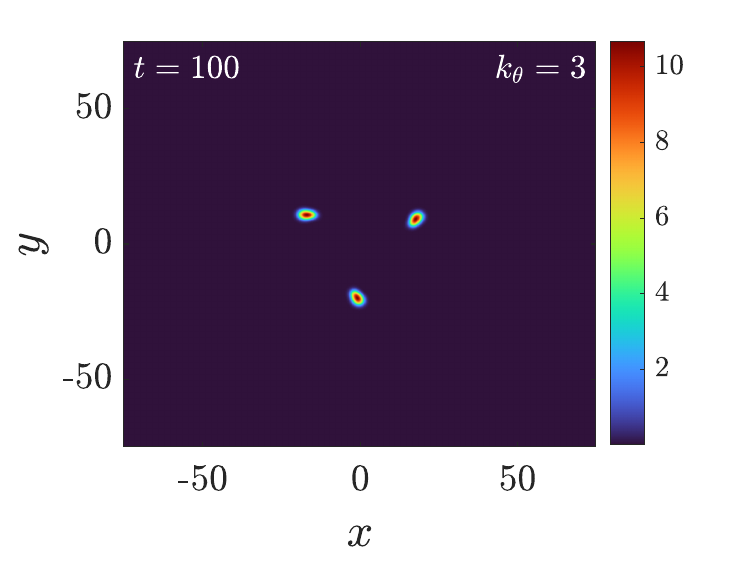} &
\includegraphics[width=\figfourwidth]{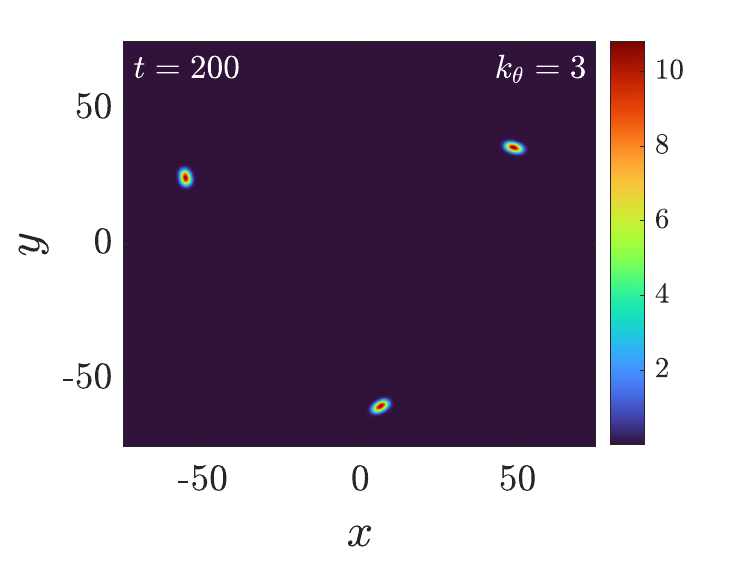} &
\includegraphics[width=\figfourwidth]{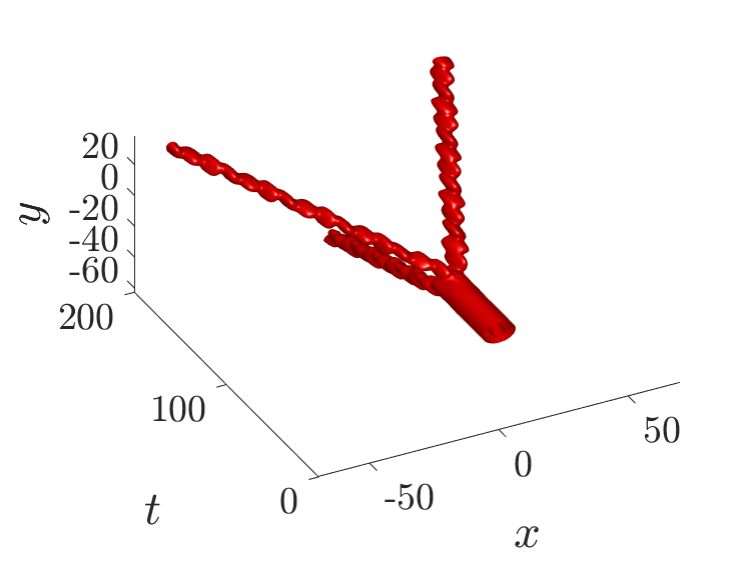} \\
\end{tabular}%
\caption{Power density plots showing the evolution of unstable topological solitons with $\omega=0.7$  when they are perturbed along the direction of the azimuthal instability eigenmode with $k_\theta=2$ (top and middle panels) and $k_\theta=3$ (bottom panels). The plots in top panels correspond to $S=1$ whereas the other ones hold for $S=2$. The dynamics of all these cases can be followed in more detail at \texttt{movie\_18.gif}, \texttt{movie\_19.gif} and \texttt{movie\_20.gif} respectively. Rightmost panels show the isosurfaces of the power density $|\psi(x,y,t)|^2$ when it is equal to 1/5th of the maximum density of the corresponding stationary topological soliton.}
\label{fig:simulS2c}
\end{figure}

We have considered $\omega=0.7$ for both $S=1$ and $S=2$ (see Fig.~\ref{fig:simulS2c} and companion files from \texttt{movie\_18.gif} to \texttt{movie\_20.gif}). In the $S=1$ case, the only instability corresponds to $k_\theta=2$ and perturbations to the corresponding eigenvector lead to the formation of two traveling single-hump solitary waves. For $S=2$, only instabilities with $k_\theta=2$ and $k_\theta=3$ are present and they typically lead to three radial solitons. Interestingly, in the former case, one of the solitons remains pinned to the center of the domain and the
other two splinters ``fly off'', whereas in the latter case all of them move along the domain. Remarkably, for both $S=1$ and $S=2$ vortices, the splinters 
have an elliptical shape deformation which allows them
to rotate as they move away from the original 
instability point.

\begin{figure}[!htbp]
\begin{tabular}{ccc}
\includegraphics[width=\figthreewidth]{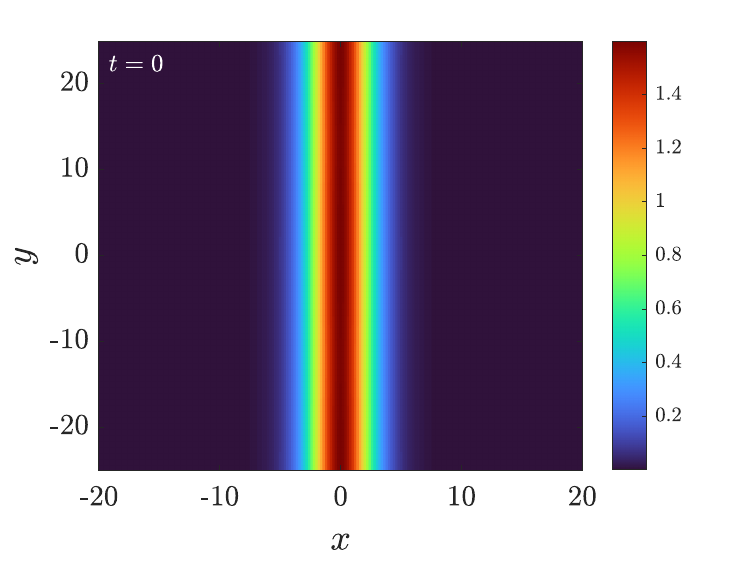} &
\includegraphics[width=\figthreewidth]{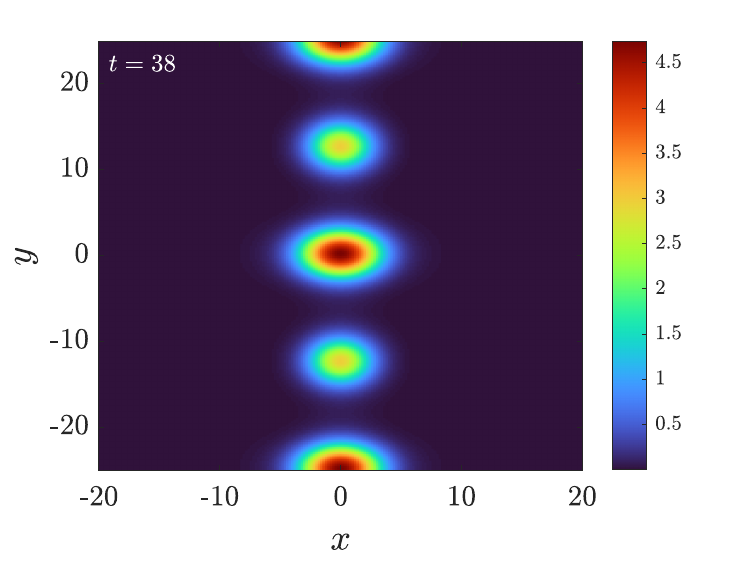} &
\includegraphics[width=\figthreewidth]{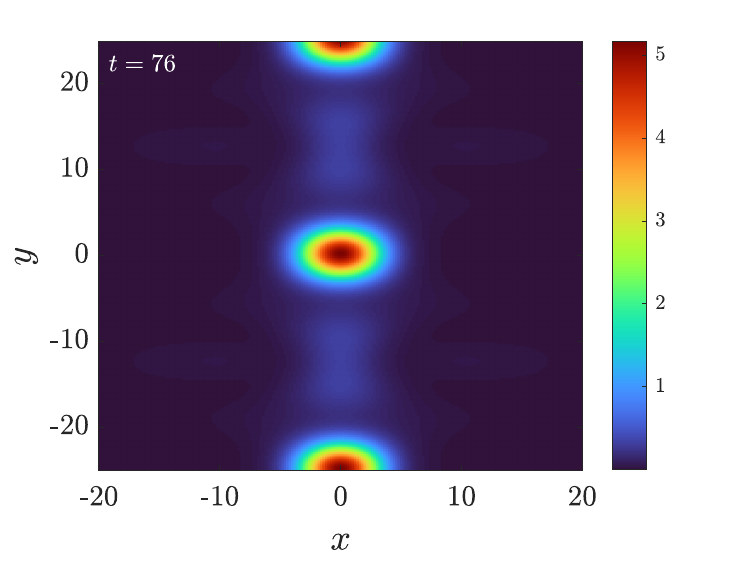} \\
\includegraphics[width=\figthreewidth]{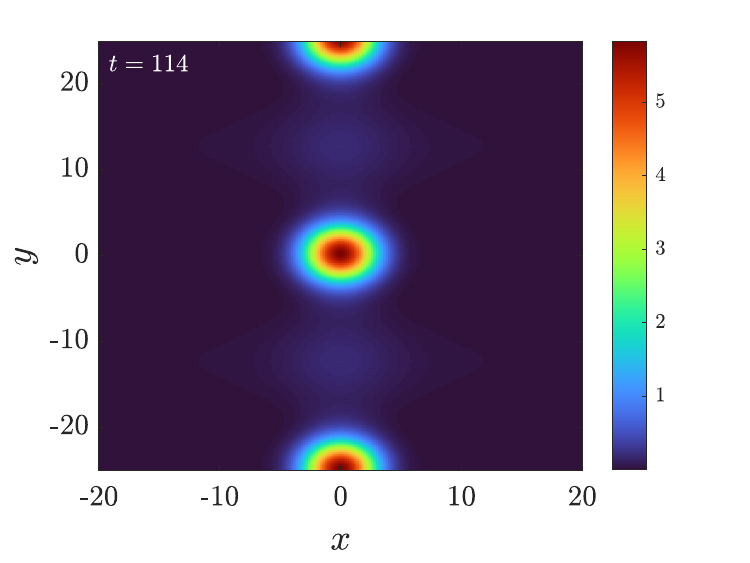} &
\includegraphics[width=\figthreewidth]{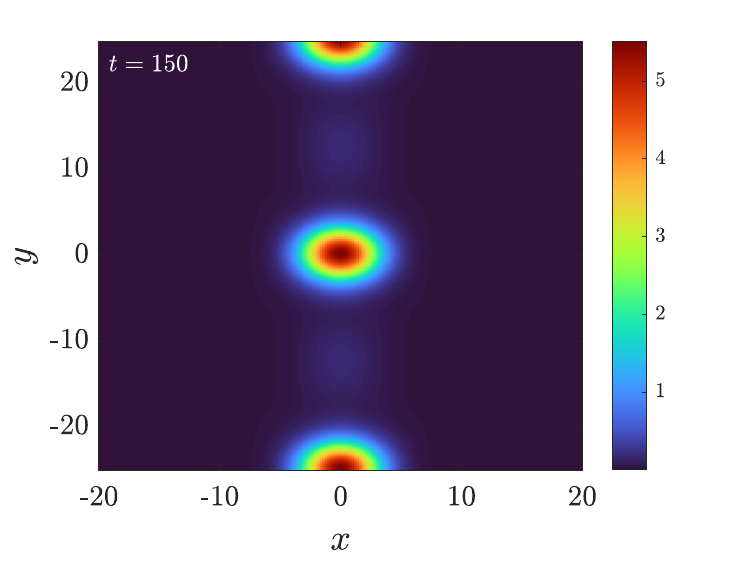} &
\includegraphics[width=\figthreewidth]{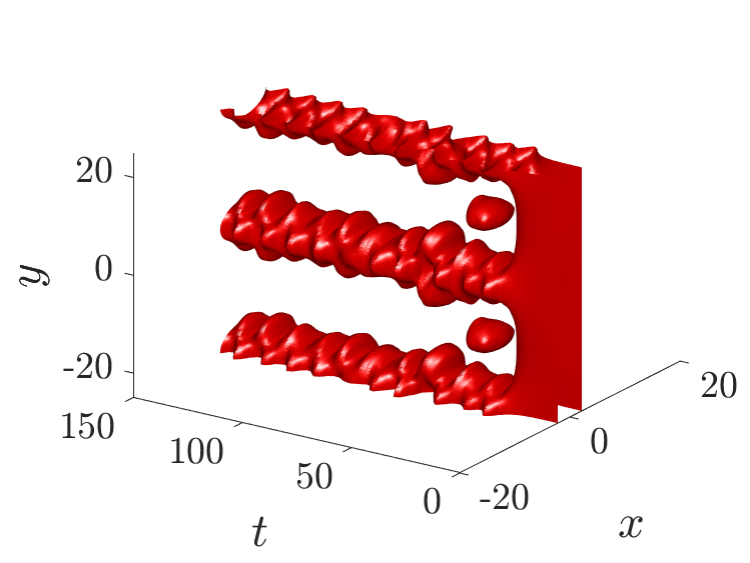} 
\end{tabular}%
\caption{Power density plots showing the evolution of an unstable line soliton with $\omega=0.5$ when it is perturbed along the direction of a transverse  instability eigenmode {with $k_y=2\pi/25\approx0.5027$,} which can be followed in more detail at \texttt{movie\_21.gif}. The last panel shows the isosurface of the power density $|\psi(x,y,t)|^2$ when it is equal to half of the maximum density of the corresponding stationary soliton.}
\label{fig:simulline}
\end{figure}

Finally, Fig.~\ref{fig:simulline} and companion movie file \texttt{movie\_21.gif} show the evolution of the transverse instability {with $k_y=4\pi/L_y$ for a line soliton with $\omega=0.5$, with $L_y=25$ being the length of the transverse direction}. One can observe the generation of localized non-topological solitons, emerging as a result of the breakup of the original line solitary wave into three radial solitons, with two of them eventually merging. Naturally,
the number of splinter solitons will depend on the most
unstable wavenumber for a given length of the spatial
domain.

\FloatBarrier

{ {\bf Acknowledgements:} 
This work was supported in part by the U.S. National Science Foundation under the awards PHY-2408988
(P.G.K.), DMS-2006887 (R.G.) and DMS-2307650 (R.G.). This research was partly conducted while P.G.K. was  visiting the Okinawa Institute of Science and
Technology (OIST) through the Theoretical Sciences Visiting Program (TSVP), the University of
Sydney through the visitor program of the Sydney Mathematical Research Institute (SMRI) and the Department of Mechanical Engineering at Seoul National
University through a Fulbright Fellowship. Their support is gratefully acknowledged.
Finally, this work was also  supported by a grant from the Simons Foundation [SFI-MPS-SFM-00011048, P.G.K.].
J.C.-M. acknowledges support from grants PID2022-143120OB-I00 and CEX2024-001517-M, both funded by MICIU/AEI/10.13039/501100011033 and ERDF/EU.}


\begin{thebibliography}{29}%
\makeatletter
\providecommand \@ifxundefined [1]{%
 \@ifx{#1\undefined}
}%
\providecommand \@ifnum [1]{%
 \ifnum #1\expandafter \@firstoftwo
 \else \expandafter \@secondoftwo
 \fi
}%
\providecommand \@ifx [1]{%
 \ifx #1\expandafter \@firstoftwo
 \else \expandafter \@secondoftwo
 \fi
}%
\providecommand \natexlab [1]{#1}%
\providecommand \enquote  [1]{``#1''}%
\providecommand \bibnamefont  [1]{#1}%
\providecommand \bibfnamefont [1]{#1}%
\providecommand \citenamefont [1]{#1}%
\providecommand \href@noop [0]{\@secondoftwo}%
\providecommand \href [0]{\begingroup \@sanitize@url \@href}%
\providecommand \@href[1]{\@@startlink{#1}\@@href}%
\providecommand \@@href[1]{\endgroup#1\@@endlink}%
\providecommand \@sanitize@url [0]{\catcode `\\12\catcode `\$12\catcode
  `\&12\catcode `\#12\catcode `\^12\catcode `\_12\catcode `\%12\relax}%
\providecommand \@@startlink[1]{}%
\providecommand \@@endlink[0]{}%
\providecommand \url  [0]{\begingroup\@sanitize@url \@url }%
\providecommand \@url [1]{\endgroup\@href {#1}{\urlprefix }}%
\providecommand \urlprefix  [0]{URL }%
\providecommand \Eprint [0]{\href }%
\providecommand \doibase [0]{https://doi.org/}%
\providecommand \selectlanguage [0]{\@gobble}%
\providecommand \bibinfo  [0]{\@secondoftwo}%
\providecommand \bibfield  [0]{\@secondoftwo}%
\providecommand \translation [1]{[#1]}%
\providecommand \BibitemOpen [0]{}%
\providecommand \bibitemStop [0]{}%
\providecommand \bibitemNoStop [0]{.\EOS\space}%
\providecommand \EOS [0]{\spacefactor3000\relax}%
\providecommand \BibitemShut  [1]{\csname bibitem#1\endcsname}%
\let\auto@bib@innerbib\@empty
\bibitem [{\citenamefont {Sulem}\ and\ \citenamefont {Sulem}(1999)}]{sulem}%
  \BibitemOpen
  \bibfield  {author} {\bibinfo {author} {\bibfnamefont {C.}~\bibnamefont
  {Sulem}}\ and\ \bibinfo {author} {\bibfnamefont {P.}~\bibnamefont {Sulem}},\
  }\href@noop {} {\emph {\bibinfo {title} {The nonlinear {Schr\"odinger}
  equation: self-focusing and wave collapse}}}\ (\bibinfo  {publisher}
  {Springer},\ \bibinfo {address} {New York},\ \bibinfo {year}
  {1999})\BibitemShut {NoStop}%
\bibitem [{\citenamefont {Fibich}\ and\ \citenamefont
  {Papanicolaou}(1999)}]{fibich}%
  \BibitemOpen
  \bibfield  {author} {\bibinfo {author} {\bibfnamefont {G.}~\bibnamefont
  {Fibich}}\ and\ \bibinfo {author} {\bibfnamefont {G.}~\bibnamefont
  {Papanicolaou}},\ }\href@noop {} {\bibfield  {journal} {\bibinfo  {journal}
  {SIAM Journal on Applied Mathematics}\ }\textbf {\bibinfo {volume} {60}},\
  \bibinfo {pages} {183} (\bibinfo {year} {1999})}\BibitemShut {NoStop}%
\bibitem [{\citenamefont {Kivshar}\ and\ \citenamefont
  {Agrawal}(2003)}]{Kivshar2003}%
  \BibitemOpen
  \bibfield  {author} {\bibinfo {author} {\bibfnamefont {Y.~S.}\ \bibnamefont
  {Kivshar}}\ and\ \bibinfo {author} {\bibfnamefont {G.~P.}\ \bibnamefont
  {Agrawal}},\ }\href {https://doi.org/10.1016/B978-0-12-410590-4.X5000-1}
  {\emph {\bibinfo {title} {Optical Solitons: From Fibers to Photonic
  Crystals}}}\ (\bibinfo  {publisher} {Academic Press},\ \bibinfo {year}
  {2003})\ pp.\ \bibinfo {pages} {1--540}\BibitemShut {NoStop}%
\bibitem [{\citenamefont {Kevrekidis}\ \emph {et~al.}(2015)\citenamefont
  {Kevrekidis}, \citenamefont {Frantzeskakis},\ and\ \citenamefont
  {Carretero-Gonz\'alez}}]{siambook}%
  \BibitemOpen
  \bibfield  {author} {\bibinfo {author} {\bibfnamefont {P.~G.}\ \bibnamefont
  {Kevrekidis}}, \bibinfo {author} {\bibfnamefont {D.~J.}\ \bibnamefont
  {Frantzeskakis}},\ and\ \bibinfo {author} {\bibfnamefont {R.}~\bibnamefont
  {Carretero-Gonz\'alez}},\ }\href@noop {} {\emph {\bibinfo {title} {The
  Defocusing Nonlinear Schr{\"o}dinger Equation}}}\ (\bibinfo  {publisher}
  {SIAM, Philadelphia},\ \bibinfo {year} {2015})\BibitemShut {NoStop}%
\bibitem [{\citenamefont {Ablowitz}(2011)}]{ablowitz2}%
  \BibitemOpen
  \bibfield  {author} {\bibinfo {author} {\bibfnamefont {M.}~\bibnamefont
  {Ablowitz}},\ }\href {https://doi.org/10.1017/cbo9780511998324} {\emph
  {\bibinfo {title} {Nonlinear Dispersive Waves, Asymptotic Analysis and
  Solitons}}}\ (\bibinfo  {publisher} {Cambridge University Press, Cambridge},\
  \bibinfo {year} {2011})\BibitemShut {NoStop}%
\bibitem [{\citenamefont {Kono}\ and\ \citenamefont {Skori{\'c}}(2010)}]{kono}%
  \BibitemOpen
  \bibfield  {author} {\bibinfo {author} {\bibfnamefont {M.}~\bibnamefont
  {Kono}}\ and\ \bibinfo {author} {\bibfnamefont {M.}~\bibnamefont
  {Skori{\'c}}},\ }\href@noop {} {\emph {\bibinfo {title} {Nonlinear Physics of
  Plasmas}}}\ (\bibinfo  {publisher} {Springer-Verlag, Heidelberg},\ \bibinfo
  {year} {2010})\BibitemShut {NoStop}%
\bibitem [{\citenamefont {Pitaevskii}\ and\ \citenamefont
  {Stringari}(2003)}]{Pitaevskii2003}%
  \BibitemOpen
  \bibfield  {author} {\bibinfo {author} {\bibfnamefont {L.~P.}\ \bibnamefont
  {Pitaevskii}}\ and\ \bibinfo {author} {\bibfnamefont {S.}~\bibnamefont
  {Stringari}},\ }\href@noop {} {\emph {\bibinfo {title} {Bose-{{Einstein}}
  Condensation}}},\ \bibinfo {series} {Oxford Science Publications}\ No.\
  \bibinfo {number} {116}\ (\bibinfo  {publisher} {{Clarendon Press}},\
  \bibinfo {address} {{Oxford ; New York}},\ \bibinfo {year}
  {2003})\BibitemShut {NoStop}%
\bibitem [{\citenamefont {Manton}\ and\ \citenamefont
  {Sutcliffe}(2004)}]{MantonSutcliffe2004_TopologicalSolitons}%
  \BibitemOpen
  \bibfield  {author} {\bibinfo {author} {\bibfnamefont {N.~S.}\ \bibnamefont
  {Manton}}\ and\ \bibinfo {author} {\bibfnamefont {P.}~\bibnamefont
  {Sutcliffe}},\ }\href {https://doi.org/10.1017/CBO9780511617034} {\emph
  {\bibinfo {title} {Topological Solitons}}}\ (\bibinfo  {publisher} {Cambridge
  University Press},\ \bibinfo {address} {Cambridge, UK},\ \bibinfo {year}
  {2004})\BibitemShut {NoStop}%
\bibitem [{\citenamefont {Zeng}\ \emph {et~al.}(2026)\citenamefont {Zeng},
  \citenamefont {Malomed}, \citenamefont {Mihalache}, \citenamefont {Li},\ and\
  \citenamefont {Zhu}}]{ZengMalomedMihalacheLiZhu2026Chaos_CQTroughs}%
  \BibitemOpen
  \bibfield  {author} {\bibinfo {author} {\bibfnamefont {L.}~\bibnamefont
  {Zeng}}, \bibinfo {author} {\bibfnamefont {B.~A.}\ \bibnamefont {Malomed}},
  \bibinfo {author} {\bibfnamefont {D.}~\bibnamefont {Mihalache}}, \bibinfo
  {author} {\bibfnamefont {J.}~\bibnamefont {Li}},\ and\ \bibinfo {author}
  {\bibfnamefont {X.}~\bibnamefont {Zhu}},\ }\href
  {https://doi.org/10.1063/5.0309512} {\bibfield  {journal} {\bibinfo
  {journal} {Chaos}\ }\textbf {\bibinfo {volume} {36}},\ \bibinfo {pages}
  {013138} (\bibinfo {year} {2026})},\ \bibinfo {note} {arXiv:2601.00187},\
  \Eprint {https://arxiv.org/abs/2601.00187} {arXiv:2601.00187 [nlin.PS]}
  \BibitemShut {NoStop}%
\bibitem [{\citenamefont {Caplan}\ \emph {et~al.}(2012)\citenamefont {Caplan},
  \citenamefont {Carretero-Gonz{\'a}lez}, \citenamefont {Kevrekidis},\ and\
  \citenamefont
  {Malomed}}]{CaplanCarreteroGonzalezKevrekidisMalomed2012_MatComSim}%
  \BibitemOpen
  \bibfield  {author} {\bibinfo {author} {\bibfnamefont {R.~M.}\ \bibnamefont
  {Caplan}}, \bibinfo {author} {\bibfnamefont {R.}~\bibnamefont
  {Carretero-Gonz{\'a}lez}}, \bibinfo {author} {\bibfnamefont {P.~G.}\
  \bibnamefont {Kevrekidis}},\ and\ \bibinfo {author} {\bibfnamefont {B.~A.}\
  \bibnamefont {Malomed}},\ }\href
  {https://doi.org/10.1016/j.matcom.2010.11.019} {\bibfield  {journal}
  {\bibinfo  {journal} {Mathematics and Computers in Simulation}\ }\textbf
  {\bibinfo {volume} {82}},\ \bibinfo {pages} {1150} (\bibinfo {year}
  {2012})},\ \bibinfo {note} {preprint version: arXiv:0910.5758}\BibitemShut
  {NoStop}%
\bibitem [{\citenamefont {Zezyulin}(2025)}]{Zezyulin2025_arXiv2512_05763}%
  \BibitemOpen
  \bibfield  {author} {\bibinfo {author} {\bibfnamefont {D.~A.}\ \bibnamefont
  {Zezyulin}},\ }\href@noop {} {\  (\bibinfo {year} {2025})},\ \bibinfo {note}
  {arXiv:2512.05763},\ \Eprint {https://arxiv.org/abs/2512.05763}
  {arXiv:2512.05763 [nlin.PS]} \BibitemShut {NoStop}%
\bibitem [{\citenamefont {Cuevas-Maraver}\ \emph {et~al.}(2023)\citenamefont
  {Cuevas-Maraver}, \citenamefont {Kevrekidis},\ and\ \citenamefont
  {Zhang}}]{hongkun}%
  \BibitemOpen
  \bibfield  {author} {\bibinfo {author} {\bibfnamefont {J.}~\bibnamefont
  {Cuevas-Maraver}}, \bibinfo {author} {\bibfnamefont {P.~G.}\ \bibnamefont
  {Kevrekidis}},\ and\ \bibinfo {author} {\bibfnamefont {H.-K.}\ \bibnamefont
  {Zhang}},\ }\href {https://doi.org/10.1103/PhysRevE.107.034217} {\bibfield
  {journal} {\bibinfo  {journal} {Phys. Rev. E}\ }\textbf {\bibinfo {volume}
  {107}},\ \bibinfo {pages} {034217} (\bibinfo {year} {2023})}\BibitemShut
  {NoStop}%
\bibitem [{\citenamefont {Kr\'{o}likowski}\ and\ \citenamefont
  {Holmstrom}(1997)}]{wieslaw}%
  \BibitemOpen
  \bibfield  {author} {\bibinfo {author} {\bibfnamefont {W.}~\bibnamefont
  {Kr\'{o}likowski}}\ and\ \bibinfo {author} {\bibfnamefont {S.~A.}\
  \bibnamefont {Holmstrom}},\ }\href {https://doi.org/10.1364/OL.22.000369}
  {\bibfield  {journal} {\bibinfo  {journal} {Opt. Lett.}\ }\textbf {\bibinfo
  {volume} {22}},\ \bibinfo {pages} {369} (\bibinfo {year} {1997})}\BibitemShut
  {NoStop}%
\bibitem [{\citenamefont {Meng}\ \emph {et~al.}(1997)\citenamefont {Meng},
  \citenamefont {Salamo}, \citenamefont {feng Shih},\ and\ \citenamefont
  {Segev}}]{Meng:97}%
  \BibitemOpen
  \bibfield  {author} {\bibinfo {author} {\bibfnamefont {H.}~\bibnamefont
  {Meng}}, \bibinfo {author} {\bibfnamefont {G.}~\bibnamefont {Salamo}},
  \bibinfo {author} {\bibfnamefont {M.}~\bibnamefont {feng Shih}},\ and\
  \bibinfo {author} {\bibfnamefont {M.}~\bibnamefont {Segev}},\ }\href
  {https://doi.org/10.1364/OL.22.000448} {\bibfield  {journal} {\bibinfo
  {journal} {Opt. Lett.}\ }\textbf {\bibinfo {volume} {22}},\ \bibinfo {pages}
  {448} (\bibinfo {year} {1997})}\BibitemShut {NoStop}%
\bibitem [{\citenamefont {Ferioli}\ \emph {et~al.}(2019)\citenamefont
  {Ferioli}, \citenamefont {Semeghini}, \citenamefont {Masi}, \citenamefont
  {Giusti}, \citenamefont {Modugno}, \citenamefont {Inguscio}, \citenamefont
  {Gallem{\`i}}, \citenamefont {Recati},\ and\ \citenamefont
  {Fattori}}]{FerioliSemeghiniMasiEtAl2019_PRL}%
  \BibitemOpen
  \bibfield  {author} {\bibinfo {author} {\bibfnamefont {G.}~\bibnamefont
  {Ferioli}}, \bibinfo {author} {\bibfnamefont {G.}~\bibnamefont {Semeghini}},
  \bibinfo {author} {\bibfnamefont {L.}~\bibnamefont {Masi}}, \bibinfo {author}
  {\bibfnamefont {G.}~\bibnamefont {Giusti}}, \bibinfo {author} {\bibfnamefont
  {G.}~\bibnamefont {Modugno}}, \bibinfo {author} {\bibfnamefont
  {M.}~\bibnamefont {Inguscio}}, \bibinfo {author} {\bibfnamefont
  {A.}~\bibnamefont {Gallem{\`i}}}, \bibinfo {author} {\bibfnamefont
  {A.}~\bibnamefont {Recati}},\ and\ \bibinfo {author} {\bibfnamefont
  {M.}~\bibnamefont {Fattori}},\ }\href
  {https://doi.org/10.1103/PhysRevLett.122.090401} {\bibfield  {journal}
  {\bibinfo  {journal} {Physical Review Letters}\ }\textbf {\bibinfo {volume}
  {122}},\ \bibinfo {pages} {090401} (\bibinfo {year} {2019})},\ \bibinfo
  {note} {arXiv:1812.09151},\ \Eprint {https://arxiv.org/abs/1812.09151}
  {arXiv:1812.09151 [cond-mat.quant-gas]} \BibitemShut {NoStop}%
\bibitem [{\citenamefont {Hu}\ \emph {et~al.}(2024)\citenamefont {Hu},
  \citenamefont {Fei}, \citenamefont {Chen},\ and\ \citenamefont
  {Zhang}}]{HuFeiChenZhang2024_arXiv2404_19295}%
  \BibitemOpen
  \bibfield  {author} {\bibinfo {author} {\bibfnamefont {Y.}~\bibnamefont
  {Hu}}, \bibinfo {author} {\bibfnamefont {Y.}~\bibnamefont {Fei}}, \bibinfo
  {author} {\bibfnamefont {X.-L.}\ \bibnamefont {Chen}},\ and\ \bibinfo
  {author} {\bibfnamefont {Y.}~\bibnamefont {Zhang}},\ }\href@noop {} {\
  (\bibinfo {year} {2024})},\ \bibinfo {note} {arXiv:2404.19295},\ \Eprint
  {https://arxiv.org/abs/2404.19295} {arXiv:2404.19295 [cond-mat.quant-gas]}
  \BibitemShut {NoStop}%
\bibitem [{\citenamefont {Piette}\ \emph {et~al.}(1995)\citenamefont {Piette},
  \citenamefont {Schroers},\ and\ \citenamefont
  {Zakrzewski}}]{PietteSchroersZakrzewski1995_NPB}%
  \BibitemOpen
  \bibfield  {author} {\bibinfo {author} {\bibfnamefont {B.~M. A.~G.}\
  \bibnamefont {Piette}}, \bibinfo {author} {\bibfnamefont {B.~J.}\
  \bibnamefont {Schroers}},\ and\ \bibinfo {author} {\bibfnamefont {W.~J.}\
  \bibnamefont {Zakrzewski}},\ }\href
  {https://doi.org/10.1016/0550-3213(95)00011-G} {\bibfield  {journal}
  {\bibinfo  {journal} {Nuclear Physics B}\ }\textbf {\bibinfo {volume}
  {439}},\ \bibinfo {pages} {205} (\bibinfo {year} {1995})},\ \bibinfo {note}
  {arXiv:hep-ph/9410256},\ \Eprint {https://arxiv.org/abs/hep-ph/9410256}
  {arXiv:hep-ph/9410256 [hep-ph]} \BibitemShut {NoStop}%
\bibitem [{\citenamefont {Lam}\ \emph {et~al.}(1977)\citenamefont {Lam},
  \citenamefont {Lippmann},\ and\ \citenamefont {Tappert}}]{tappert77}%
  \BibitemOpen
  \bibfield  {author} {\bibinfo {author} {\bibfnamefont {J.~F.}\ \bibnamefont
  {Lam}}, \bibinfo {author} {\bibfnamefont {B.}~\bibnamefont {Lippmann}},\ and\
  \bibinfo {author} {\bibfnamefont {F.}~\bibnamefont {Tappert}},\ }\href
  {https://doi.org/10.1063/1.861679} {\bibfield  {journal} {\bibinfo  {journal}
  {The Physics of Fluids}\ }\textbf {\bibinfo {volume} {20}},\ \bibinfo {pages}
  {1176} (\bibinfo {year} {1977})}\BibitemShut {NoStop}%
\bibitem [{\citenamefont {Max}(1976)}]{max76}%
  \BibitemOpen
  \bibfield  {author} {\bibinfo {author} {\bibfnamefont {C.~E.}\ \bibnamefont
  {Max}},\ }\href {https://doi.org/10.1063/1.861305} {\bibfield  {journal}
  {\bibinfo  {journal} {The Physics of Fluids}\ }\textbf {\bibinfo {volume}
  {19}},\ \bibinfo {pages} {74} (\bibinfo {year} {1976})}\BibitemShut {NoStop}%
\bibitem [{\citenamefont {Anderson}\ and\ \citenamefont
  {Bonnedal}(1979)}]{anderson79}%
  \BibitemOpen
  \bibfield  {author} {\bibinfo {author} {\bibfnamefont {D.}~\bibnamefont
  {Anderson}}\ and\ \bibinfo {author} {\bibfnamefont {M.}~\bibnamefont
  {Bonnedal}},\ }\href {https://doi.org/10.1063/1.862445} {\bibfield  {journal}
  {\bibinfo  {journal} {The Physics of Fluids}\ }\textbf {\bibinfo {volume}
  {22}},\ \bibinfo {pages} {105} (\bibinfo {year} {1979})}\BibitemShut
  {NoStop}%
\bibitem [{\citenamefont {Cohen}\ \emph {et~al.}(1991)\citenamefont {Cohen},
  \citenamefont {Lasinski}, \citenamefont {Langdon},\ and\ \citenamefont
  {Cummings}}]{cohen91}%
  \BibitemOpen
  \bibfield  {author} {\bibinfo {author} {\bibfnamefont {B.~I.}\ \bibnamefont
  {Cohen}}, \bibinfo {author} {\bibfnamefont {B.~F.}\ \bibnamefont {Lasinski}},
  \bibinfo {author} {\bibfnamefont {A.~B.}\ \bibnamefont {Langdon}},\ and\
  \bibinfo {author} {\bibfnamefont {J.~C.}\ \bibnamefont {Cummings}},\ }\href
  {https://doi.org/10.1063/1.859872} {\bibfield  {journal} {\bibinfo  {journal}
  {Physics of Fluids B: Plasma Physics}\ }\textbf {\bibinfo {volume} {3}},\
  \bibinfo {pages} {766} (\bibinfo {year} {1991})}\BibitemShut {NoStop}%
\bibitem [{Note1()}]{Note1}%
  \BibitemOpen
  \bibinfo {note} {Per a private communication of F. Tappert to one of us
  (DKC)}\BibitemShut {NoStop}%
\bibitem [{\citenamefont {Ruback}(1988)}]{RUBACK1988669}%
  \BibitemOpen
  \bibfield  {author} {\bibinfo {author} {\bibfnamefont {P.}~\bibnamefont
  {Ruback}},\ }\href
  {https://doi.org/https://doi.org/10.1016/0550-3213(88)90038-7} {\bibfield
  {journal} {\bibinfo  {journal} {Nuclear Physics B}\ }\textbf {\bibinfo
  {volume} {296}},\ \bibinfo {pages} {669} (\bibinfo {year}
  {1988})}\BibitemShut {NoStop}%
\bibitem [{\citenamefont {Foster}\ and\ \citenamefont
  {Krusch}(2015)}]{foster15}%
  \BibitemOpen
  \bibfield  {author} {\bibinfo {author} {\bibfnamefont {D.}~\bibnamefont
  {Foster}}\ and\ \bibinfo {author} {\bibfnamefont {S.}~\bibnamefont
  {Krusch}},\ }\href
  {https://doi.org/https://doi.org/10.1016/j.nuclphysb.2015.06.011} {\bibfield
  {journal} {\bibinfo  {journal} {Nuclear Physics B}\ }\textbf {\bibinfo
  {volume} {897}},\ \bibinfo {pages} {697} (\bibinfo {year}
  {2015})}\BibitemShut {NoStop}%
\bibitem [{\citenamefont {Koll\'ar}\ and\ \citenamefont {Pego}(2012)}]{Kollar}%
  \BibitemOpen
  \bibfield  {author} {\bibinfo {author} {\bibfnamefont {R.}~\bibnamefont
  {Koll\'ar}}\ and\ \bibinfo {author} {\bibfnamefont {R.}~\bibnamefont
  {Pego}},\ }\href@noop {} {\bibfield  {journal} {\bibinfo  {journal} {Applied
  Mathematics Research eXpress}\ }\textbf {\bibinfo {volume} {2012}},\ \bibinfo
  {pages} {1} (\bibinfo {year} {2012})}\BibitemShut {NoStop}%
\bibitem [{\citenamefont {Vakhitov}\ and\ \citenamefont
  {Kolokolov}(1973)}]{VakhitovKolokolov1973_VKCriterion}%
  \BibitemOpen
  \bibfield  {author} {\bibinfo {author} {\bibfnamefont {N.~G.}\ \bibnamefont
  {Vakhitov}}\ and\ \bibinfo {author} {\bibfnamefont {A.~A.}\ \bibnamefont
  {Kolokolov}},\ }\href {https://doi.org/10.1007/BF01031343} {\bibfield
  {journal} {\bibinfo  {journal} {Radiophysics and Quantum Electronics}\
  }\textbf {\bibinfo {volume} {16}},\ \bibinfo {pages} {783} (\bibinfo {year}
  {1973})},\ \bibinfo {note} {translated from Radiotekhnika i
  Elektronika}\BibitemShut {NoStop}%
\bibitem [{\citenamefont {Cuevas}\ and\ \citenamefont {Eilbeck}(2006)}]{chris}%
  \BibitemOpen
  \bibfield  {author} {\bibinfo {author} {\bibfnamefont {J.}~\bibnamefont
  {Cuevas}}\ and\ \bibinfo {author} {\bibfnamefont {J.}~\bibnamefont
  {Eilbeck}},\ }\href
  {https://doi.org/https://doi.org/10.1016/j.physleta.2006.04.095} {\bibfield
  {journal} {\bibinfo  {journal} {Physics Letters A}\ }\textbf {\bibinfo
  {volume} {358}},\ \bibinfo {pages} {15} (\bibinfo {year} {2006})}\BibitemShut
  {NoStop}%
\bibitem [{\citenamefont {Snyder}\ and\ \citenamefont
  {Sheppard}(1993)}]{snyder}%
  \BibitemOpen
  \bibfield  {author} {\bibinfo {author} {\bibfnamefont {A.~W.}\ \bibnamefont
  {Snyder}}\ and\ \bibinfo {author} {\bibfnamefont {A.~P.}\ \bibnamefont
  {Sheppard}},\ }\href {https://doi.org/10.1364/OL.18.000482} {\bibfield
  {journal} {\bibinfo  {journal} {Opt. Lett.}\ }\textbf {\bibinfo {volume}
  {18}},\ \bibinfo {pages} {482} (\bibinfo {year} {1993})}\BibitemShut
  {NoStop}%
\bibitem [{\citenamefont {Malomed}(2002)}]{MalomedV}%
  \BibitemOpen
  \bibfield  {author} {\bibinfo {author} {\bibfnamefont {B.~A.}\ \bibnamefont
  {Malomed}},\ }\href@noop {} {\bibfield  {journal} {\bibinfo  {journal}
  {Progress in Optics}\ }\textbf {\bibinfo {volume} {43}},\ \bibinfo {pages}
  {71} (\bibinfo {year} {2002})}\BibitemShut {NoStop}%
\end{thebibliography}
\end{document}